\documentclass{aastex63}
\usepackage{lineno}
\usepackage{gensymb}
\usepackage{graphicx}
\usepackage{multirow}
\usepackage{booktabs}
\usepackage{natbib}
\usepackage{longtable}
\hypersetup{hypertexnames=false}

\received{XXX}
\revised{XXX}
\accepted{XXX}
\submitjournal{ApJ}
\shortauthors{Bi et al.}
\begin{document}
\title{ Spectral Properties of X-Ray Flares in Gamma-Ray Bursts}

\author{Xiongwei Bi}
\affiliation{School of Physics and Astronomy, Yunnan University, and Key Laboratory of Astroparticle Physics of Yunnan Province, Kunming 650091, People's Republic of China}
\affiliation{Department of Physics, Honghe University, Mengzi 661199, Yunnan Province, People's Republic of China}

\author{Zejun Jiang}
\affiliation{School of Physics and Astronomy, Yunnan University, and Key Laboratory of Astroparticle Physics of Yunnan Province, Kunming 650091, People's Republic of China}

\author{Jirong Mao}$^{\star}$
\affiliation{Yunnan Observatories, Chinese Academy of Sciences, Kunming 650011, Yunnan Province, People's Republic of China}
\affiliation{Center for Astronomical Mega-Science, Chinese Academy of Sciences, 20A Datun Road, Chaoyang District, Beijing 100012, People's Republic of China}
\affiliation{Key Laboratory for the Structure and Evolution of Celestial Objects, Chinese Academy of Sciences, Kunming 650011, People's Republic of China}

\correspondingauthor{Zejun Jiang}
\email{zjjiang@ynu.edu.cn}
\correspondingauthor{Jirong Mao}
\email{jirongmao@mail.ynao.ac.cn}

\begin{abstract}
  We conducted a systematic spectral analysis of the X-ray flares of gamma-ray bursts (GRBs) observed by Swift/X-ray Telescope from 2005 April to 2015 March. The sample includes 521 flares from 206 GRBs. The GRBs cover the three subtypes: X-ray flashes (XRFs), X-ray-rich GRBs (XRRs), and classical GRBs (C-GRBs). We used a simple power-law model to fit the spectra during the flares, as well as to fit the preflare and postflare afterglow emissions, and obtained the parameters such as photon index and intrinsic hydrogen column density ($N_{\rm H}$). By comparing the spectral characteristics of flares, prompt emission, and afterglow emission, and analyzing the temporal evolution of the X-ray flare spectra, we obtained the following main conclusions. (1) The spectrum during the flare phase is significantly harder than that of the preflare and postflare phases, while there is no obvious difference in the spectrum between the preflare and postflare phases. (2) The intrinsic $N_{\rm H}$ shows no significant change between the preflare and flare phases, but significantly decreases after the flare. (3) The overall X-ray flare spectra are softer than those of the prompt emission. (4) For GRBs with multiple flares, the photon index softens with increasing flare peak time. (5) The X-ray flare spectra in XRFs are softer than those in C-GRBs. These results support the origin of X-ray flares being from the internal shock radiation driven by the late activity of the GRB central engine.

 \end{abstract}

\keywords{gamma-ray burst}

\section{Introduction}
X-ray flares are a common phenomenon observed in the X-ray afterglows of gamma-ray bursts (GRBs; \citealt{Chincarini2007,Chincarini2010,Bernardini2011}). In the sample of GRBs observed by the Swift satellite, approximately one-third of the bursts exhibited X-ray flares during the afterglow phase \citep{Chincarini2010}. From the manifestations of the flares, the X-ray flares of different GRBs exhibit diverse characteristics: some GRBs show only a single flare, while others display multiple successive flares, and sometimes with multiple flare components overlapping with each other \citep[e.g.,][]{Chincarini2010,Falcone2007,Burrows2005,Margutti2011a,Yi2016,Liu2019}.

The physical origin of X-ray flares remains debated. Due to the similar curves of light variation and characteristics of the spectral evolution of X-ray flares to prompt emission pulses, many researchers generally believe that X-ray flares and  prompt emission may have the same physical origin, and both are related to prolonged activity of the central engine of GRBs \citep[e.g.,][]{Burrows2005,Zhang2006,Nousek2006,Romano2006,Chincarini2007,Chincarini2010,Dereli-Begue2025,Ma2025}. Therefore, X-ray flares are regarded as an important probe for exploring the physical properties of the central engine of GRBs.

Observationally, X-ray flares exhibit the following characteristics. (1) The profile of light variation can usually be described by an asymmetric structure that has a rapid rise and rapid decline, and the duration of the decline is approximately twice that of the rise \citep{Chincarini2010,Hascoet2017}. (2) The duration of the flare is related to its isotropic equivalent X-ray energy $E_{\rm X,iso}$ and peak time; that is, flares with higher $E_{\rm X,iso}$ have shorter durations, whereas later-peaking flares have longer durations \citep{Chincarini2010, Yi2016}. (3) The peak luminosity $L_p$ of flares declines with increasing peak time, indicating that later-peaking flares are less energetic (i.e., lower $E_{\rm X,iso}$; \citealt{Chincarini2010, Yi2016}). The spectral evolution of X-ray flares is observed to exhibit diverse features, and their physical origin remains unclear. Much research has studied the spectral characteristics of X-ray flares \citep{Burrows2005,Pagani2006,Chincarini2007,Chincarini2010,Falcone2007,Margutti2010,Abdo2011,Liu2019,Lu2022}.

Some theoretical models have been proposed, such as the fragmentation model of collapsing stars \citep{King2005}, the accretion disk fragmentation model \citep{Perna2006}, the intermittent accretion model driven by magnetic fields \citep{Proga2006}, the magnetic reconnection model after the merger of millisecond pulsars \citep{Dai2006}, the magnetic dissipation model of decelerating shells \citep{Giannios2006}, and the internal shock models \citep{Burrows2005,Fan2005,Lazzati2007}. X-ray flares undoubtedly carry rich physical information about the central engine, providing us with a unique perspective for understanding the physical processes of GRBs.

Spectral analysis of GRBs is one of the most important methods for understanding their physical origins. In this paper, based on the X-ray observation data from Swift/XRT, we use a simple power-law model to fit the spectra of X-ray flares and construct a comprehensive spectral data sample of these flares. By comparing and analyzing the spectral properties of X-ray flares with afterglow emission and with prompt emission, we can test and restrict some physical models of GRBs.
This paper is organized as follows:
Section 2 is the sample selection and spectral fitting of X-ray flares.
Section 3 is the statistical analysis of the spectra of X-ray flares.
Section 4 is a discussion.
Section 5 is the summary of our work.

\section{Sample Selection and Spectral Fitting} \label{sec:sample}
\subsection{Sample Selection}
In this paper, the selection of samples of X-ray flares is mainly based on the data of X-ray flares that have been verified by  \cite{Chincarini2010} and \cite{Yi2016}. According to the temporal parameters of X-ray flares provided by those authors, we conducted time-sliced spectral analyses for the three phases: the flare phase, the preflare phase, and the postflare phase.

Our final sample comprises 521 X-ray flares originating from 206 GRBs observed between 2005 April 6 and 2015 March 23. Within this GRB sample, 98 bursts exhibit a single flare, 57 bursts display two flares, and the remaining 51 bursts show three or more flares. The sample includes four short-duration GRBs (S-GRBs) and 202 long-duration GRBs (L-GRBs), among which a total of 99 GRBs have known redshifts.

In addition, we referred to the third Swift/BAT Gamma-Ray Burst Catalog published by \cite{Lien2016} and collected the prompt emission spectral parameters of these 206 GRBs, including the power-law photon index ($\alpha_{\rm PL}$) and the cutoff power-law photon index ($\alpha_{\rm CPL}$). According to the classification criteria for GRB subclasses proposed by \cite{Bi2018}, our sample contains 17 X-ray flashes (XRFs), 122 X-ray-rich GRBs (XRRs), and 67 classical GRBs (C-GRBs). The relevant data of the GRB X-ray samples are summarized in Table 1.

\subsection{Spectral Fitting}
The X-ray observations of GRBs in our sample were obtained from the public website of the Neil Gehrels Swift Observatory\footnote{\url{https://www.swift.ac.uk/xrt_curves/}}, which provides data from both the Burst Alert Telescope (BAT) and the X-ray Telescope (XRT). Detailed descriptions of the Swift/XRT data reduction procedures, light-curve fitting, spectral fitting, example results, and the usage of the online database are presented by \cite{Evans2007,Evans2009}. The Swift/XRT team also maintains an X-ray spectral repository\footnote{\url{https://www.swift.ac.uk/xrt_spectra/}} for GRBs, offering time-averaged spectra and their best-fit parameters for each burst. This repository website further provides an automated tool for time-sliced spectral fitting, allowing users to select specific time intervals for analysis. In this work, we primarily employed this tool to perform time-sliced spectral fits for the X-ray flares and their preflare and postflare phases.

For the definition of the flare duration, we follow the method of \citet{Chincarini2010} and \citet{Yi2016}. The start time ($T_{\mathrm{start}}$) and end time ($T_{\mathrm{end}}$) of an X-ray flare are taken as the intersections of power-law fits to the rising and decaying phases of the flare with the underlying continuum, respectively. The peak time $T_{\mathrm{peak}}$ is the time of the maximum flux, and the duration is then $T_{\mathrm{end}}-T_{\mathrm{start}}$. This approach cleanly separates the flare emission from the underlying afterglow \citep[e.g.,][]{Falcone2007,Chincarini2010,Yi2016}.

We provide a schematic diagram to explain the spectral fitting of the X-ray flare in Figure~1.
\begin{figure*}[!htb]
  \centering
  \includegraphics[height=8cm,width=12cm]{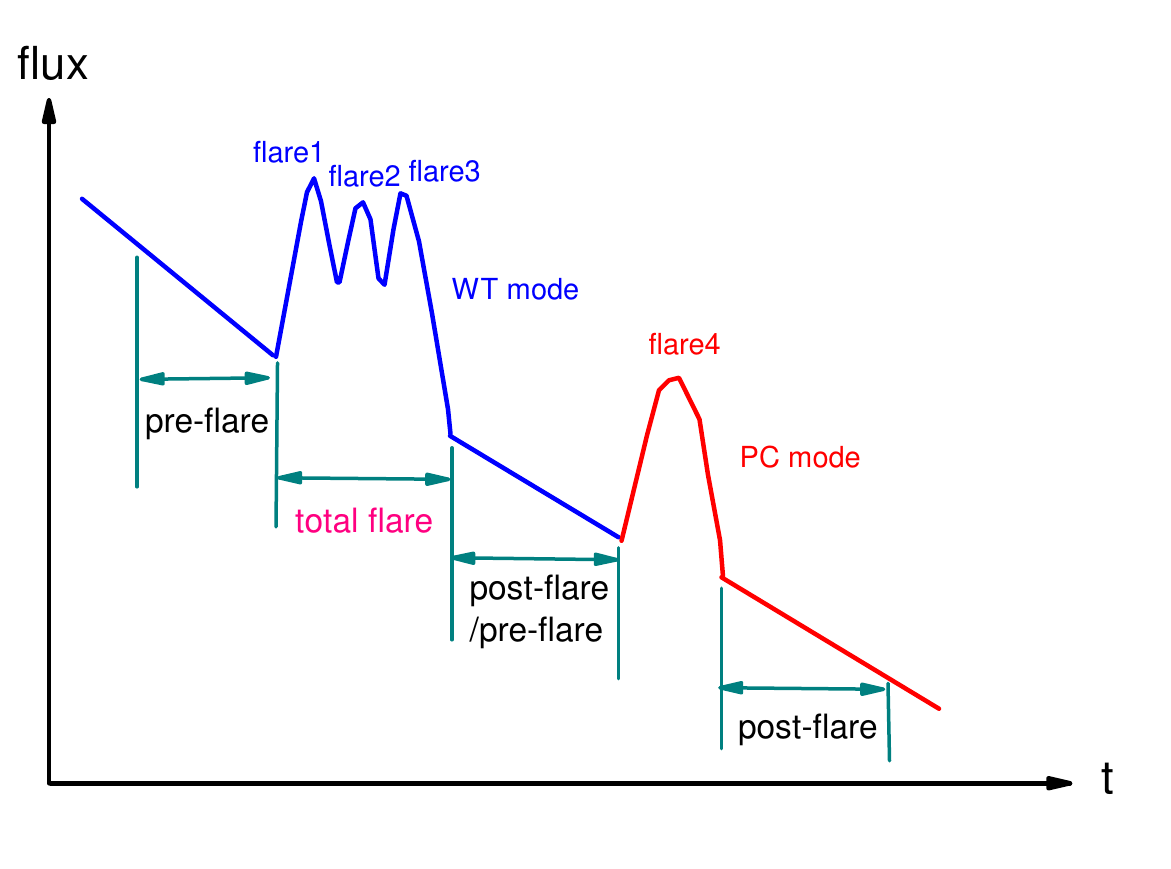}
   \caption{Schematic diagram of spectral fitting for X-ray flares and their preflare and postflare phases. The data acquired in WT mode and PC mode are represented in blue and red, respectively.}
  \label{fig:1}
\end{figure*}

Figure~1 illustrates our strategy for spectral fitting of X-ray flares. Taking an X-ray light curve containing multiple flares as an example, the figure marks three consecutive and overlapping flares (flare 1, flare 2, and flare 3). For such multiflare events, we performed individual spectral fits for each flare as well as a combined fit for the entire flare sequence (total-flare fitting), with the results labeled as ``total-flare'' in Table~1. Time-sliced spectral analysis was also conducted for the preflare and postflare intervals characterized by relatively smooth and continuous afterglow emission (regions marked as preflare and postflare in the figure). Based on the observational modes of the X-ray data, we carried out fits separately for data acquired in windowed timing (WT) mode and photon counting (PC) mode, which are shown in the figure in blue and red, respectively.
\begin{figure*}[!htb]
  \centering
  \includegraphics[height=6cm,width=8cm]{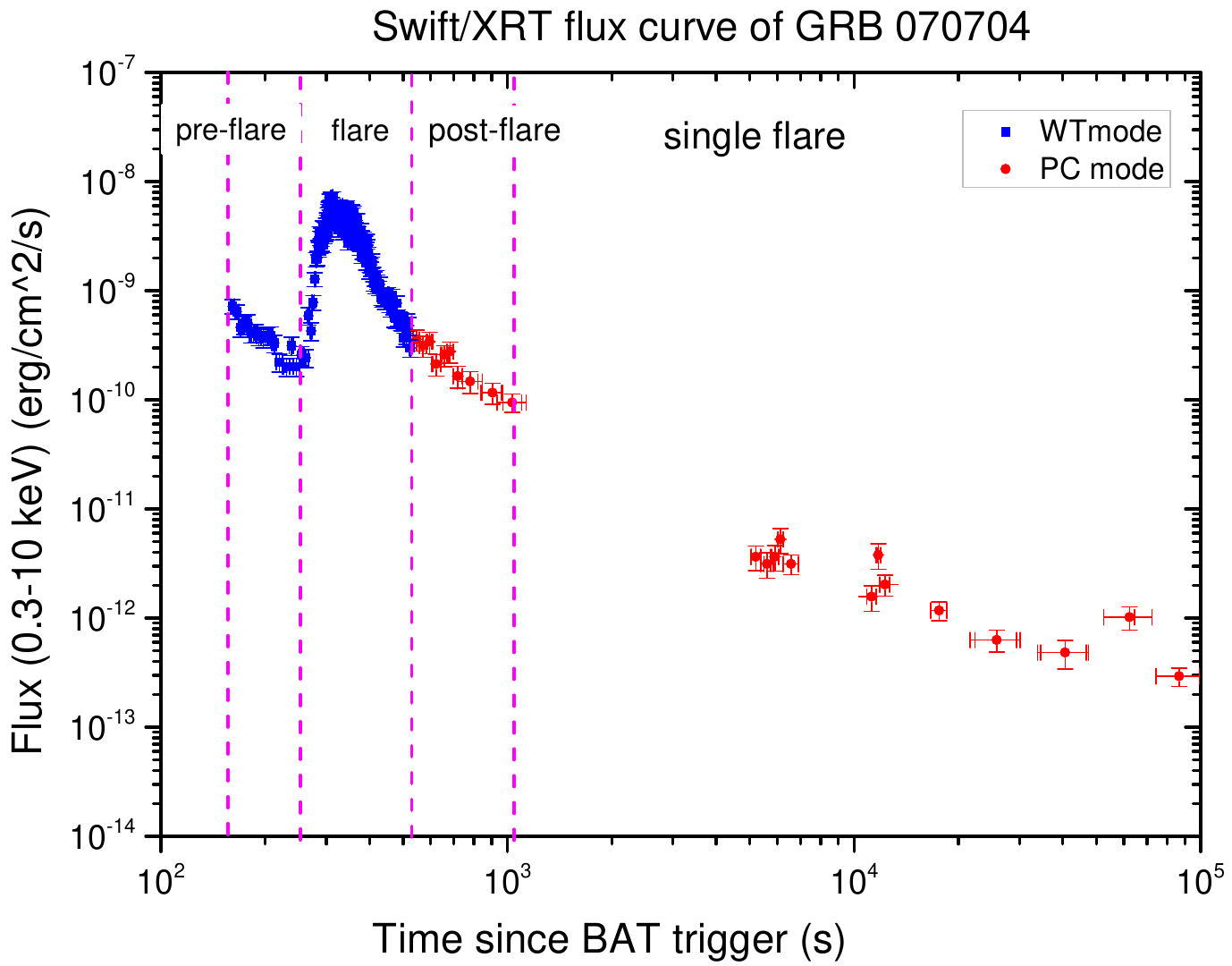}
  \includegraphics[height=6cm,width=8cm]{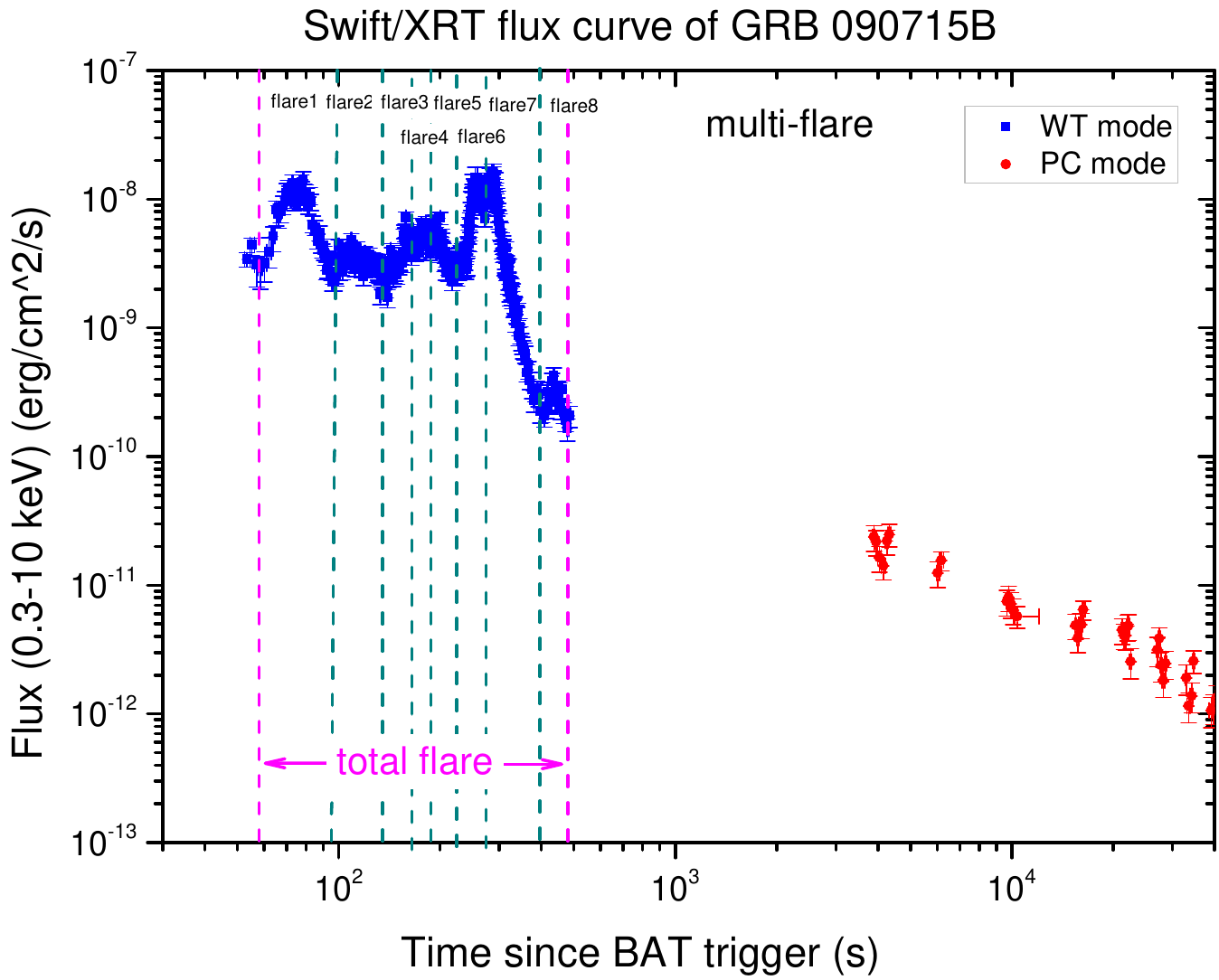}
   \caption{Examples of time-sliced spectral analysis for X-ray flares. Left: GRB 070704, a burst exhibiting only a single flare. Right: GRB 090715B, a burst exhibiting multiple consecutive flares. The data acquired in WT mode and PC mode are represented in blue and red, respectively.}
  \label{fig:2}
\end{figure*}

Figure~2 presents two representative X-ray light curves illustrating the selection of time intervals for time-sliced spectral analysis of different types of flares. The left panel shows the case of GRB 070704, a burst exhibiting only a single flare, where the time intervals for the flare, preflare, and postflare phases are marked. The right panel displays the case of GRB 090715B, a burst exhibiting multiple consecutive flares. In this case, the time intervals for the eight consecutive overlapping flares are individually marked.

Our time-resolved spectral analysis followed the strategy shown in Figures~1 and~2. For each GRB, we examined three distinct light-curve intervals: preflare, flare and postflare phases. Spectral extractions were performed via the UKSSDC online tool, which implements the standard XRT reduction routines \citep{Evans2007,Evans2009}. We simply repeat the treatment described by \citet{Evans2007,Evans2009} below.

We defined a circular extraction region for source events, centered on the afterglow position. In PC mode, this radius depended on the count rate, ranging between 5 and 30 pixels (i.e. $2\farcs36$ per pixel) -- see Table~1 of \citet{Evans2007}. For WT data, we instead used a rectangular box of $20 \times 40$ pixels. Background sampling came from a nearby source-free circle, with any interloping sources flagged and removed via \texttt{XIMAGE}. For PC observations, our default background annulus had radii of 60~pixels (inner, $142\arcsec$) and 110~pixels (outer, $260\arcsec$); we adjusted its location whenever the CCD edge posed a problem.

Pileup corrections are built into the UKSSDC pipeline, relying on a rate-dependent point-spread function (PSF) masking routine that discards the bright central core. In PC mode, the pipeline compares the observed PSF with the standard PSF \citep{Moretti2005} to identify pileup, and then applies an annular source region (inner radius $R_p$) when required. In WT mode, such an annular box is triggered above $150\ \mathrm{counts\,s^{-1}}$. To minimize residual systematics from pileup, we enforced the recommended event-grade selections: we kept grades 0--4 for PC mode and retained only grade~0 for WT mode.

For the actual fitting, we used XSPEC v12.15.1 \citep{Arnaud2025} and adopted an absorbed power-law form: \texttt{phabs * zphabs * powerlaw} \citep{Evans2009}. The Galactic column (\texttt{phabs}) was not fitted; instead, we fixed it to the 21 cm value from \citet{Kalberla2005} and \citet{Willingale2013}. The host-galaxy absorption (\texttt{zphabs}) was also held fixed, using the known spectroscopic redshift for each burst. The remaining parameter, the intrinsic neutral column density $N_{\rm H,int}$, was left free and fitted independently for each time-resolved spectrum.

We chose the Cash statistic ($C$-stat; \citealt{Cash1979}) in XSPEC for our fitting work. Including a separate background spectrum triggers automatic switching to the $W$-statistic in XSPEC. That is actually the recommended route for Poisson-limited, low-count X-ray data \citep{Arnaud2025}. Before we could compute the $C$-stat, we had to group each spectrum channel by channel. We used the FTOOL \texttt{grppha} with the \texttt{group min 1} setting \citep{Evans2009} to make sure every bin had at least one count. This mild grouping does two things: it preserves the energy resolution as much as possible, and it clears out all those empty zero-count bins.
A good number of our time-resolved spectra are pretty faint, especially at late times or during those dull intervals. There, many channels only register a few counts each. That is exactly why we turn to the $C$-statistic \citep{Cash1979} --- it is a maximum-likelihood estimator built for Poisson data, and it gives us unbiased parameters even when the counts are really low.

We checked the quality of our spectral fits in two ways. First, we visually compared the data with the best-fit model by examining the residuals. Second, we looked at the empirical ratio of the \(C\)-statistic to the degrees of freedom (dof). Admittedly, the Cash statistic does not behave exactly like \(\chi^2\) in terms of a strict probability distribution. Still, both \citet{Humphrey2009} and \citet{Kaastra2017} have shown that it can serve as a reliable guide for judging the goodness of fit in X-ray work. Combining the residual behavior with the $C$-stat/dof value gives us a workable, practical criterion for judging model adequacy.
Over the full $0.3$--$10$~keV band, the residuals (data/model) fluctuate randomly around unity. We see no persistent deviations or organized structures. At the same time, the $C$-stat/dof ratios all come out close to 1.0. This, empirically, tells us that the absorbed power-law model describes our observed X-ray spectra reasonably well.

In Figure~3, we present the histogram of the $C$-stat/dof distributions derived from the X-ray spectral fits for the flare, preflare, and postflare phases of our selected sample. The vertical dashed line at 1.0 marks the ideal value for a statistically acceptable fit according to the empirical $C$-stat/dof criterion. The individual $C$-stat/dof values are listed in Table~1.

\begin{figure*}[!htb]
  \centering
  \includegraphics[width=0.3\linewidth]{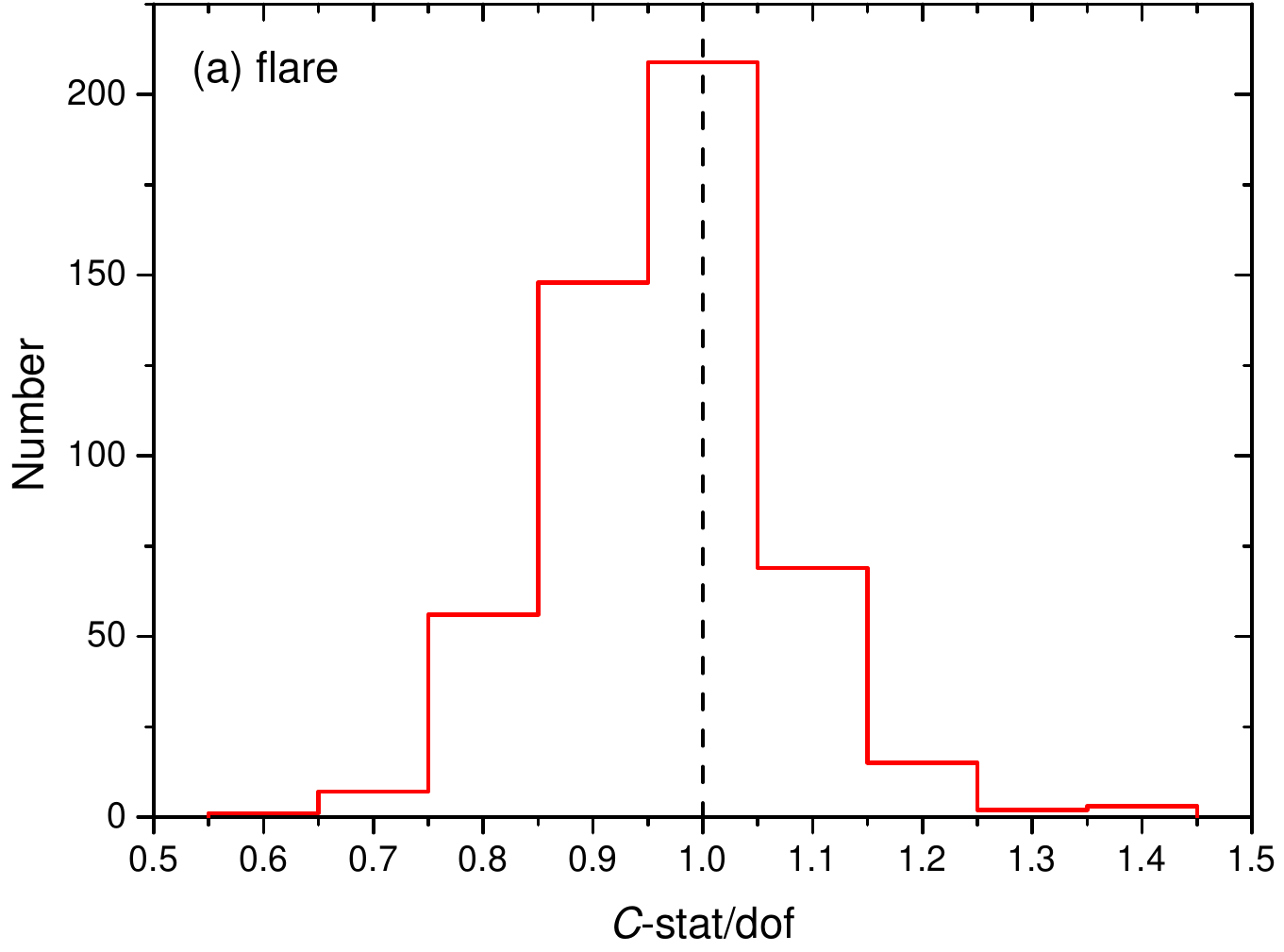}
  \includegraphics[width=0.3\linewidth]{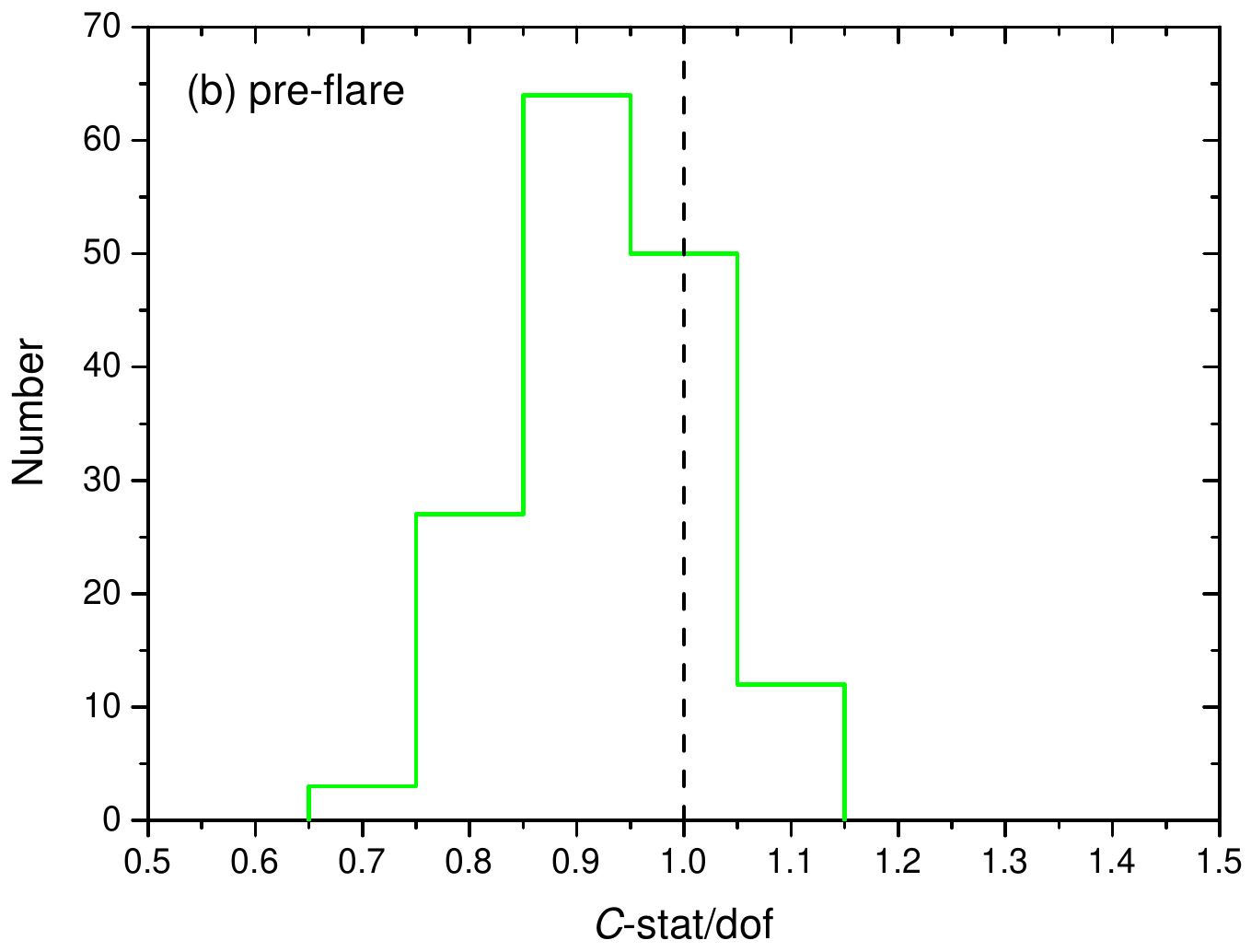}
  \includegraphics[width=0.3\linewidth]{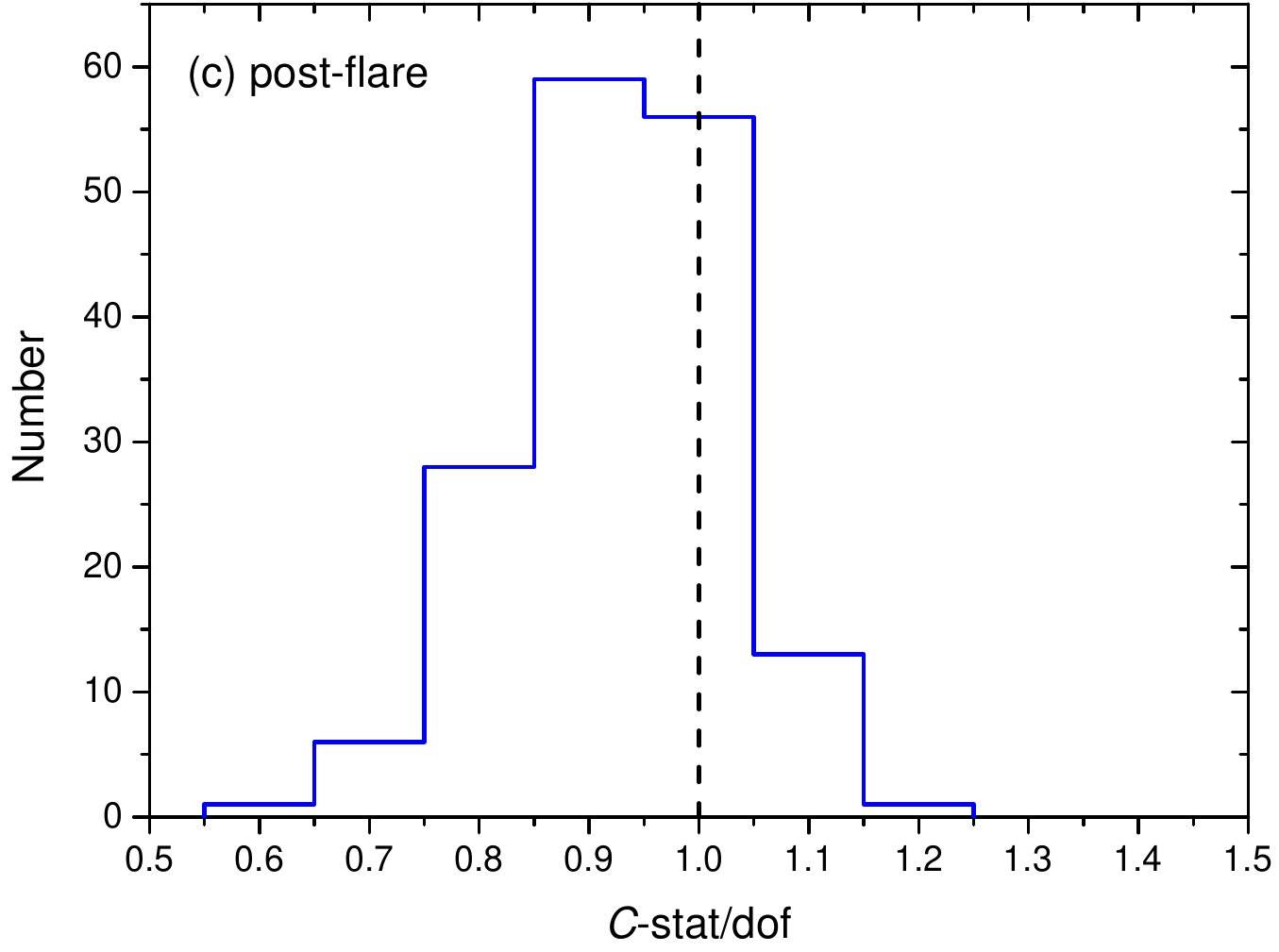}
  \caption{Distribution of the $C$-stat/dof values from the X-ray spectral fits for all samples: (a) flare, (b) preflare, and (c) postflare phases. The vertical dashed line marks the ideal value of 1.0.}
\end{figure*}

We illustrate the time-resolved spectral fitting procedure using the Swift/XRT observation of GRB~070704 as an example. Spectra were extracted for three epochs: preflare ($159$--$270$~s), flare ($280$--$500$~s), and postflare ($550$--$1078$~s). Each spectrum was fitted in the $0.3$--$10$~keV band with an absorbed power-law model, where the Galactic absorption was fixed at $N_{\rm H,Gal}=8.54\times10^{21}$~cm$^{-2}$ and the intrinsic neutral absorption $N_{\rm H,int}$ was left free. The best-fit parameters and fit statistics are summarized in Table~1. The derived $N_{\rm H,int}$ values are $(1.00^{+0.43}_{-0.37})\times10^{22}~\mathrm{cm}^{-2}$ for the preflare phase, $(8.8^{+1.1}_{-1.0})\times10^{21}~\mathrm{cm}^{-2}$ for the flare phase, and $(4.7^{+5.3}_{-4.3})\times10^{21}~\mathrm{cm}^{-2}$ for the postflare phase. The corresponding photon indices are $1.98 \pm 0.28$, $1.76 \pm 0.07$, and $1.47 \pm 0.35$, respectively. The absorbed power-law model provides acceptable fits for all three phases: $C$-stat/dof $= 0.84$ (preflare, $\text{dof} = 344$), $C$-stat/dof $= 0.95$ (flare, $\text{dof} = 788$), and $C$-stat/dof $= 0.78$ (postflare, $\text{dof} = 194$). Figure~4 presents the unfolded spectra and the data-to-model ratio for the three phases. In all cases, the residuals are randomly scattered around 1.0 with no systematic trends.

\begin{figure*}[!htb]
  \centering
  \includegraphics[width=0.3\linewidth]{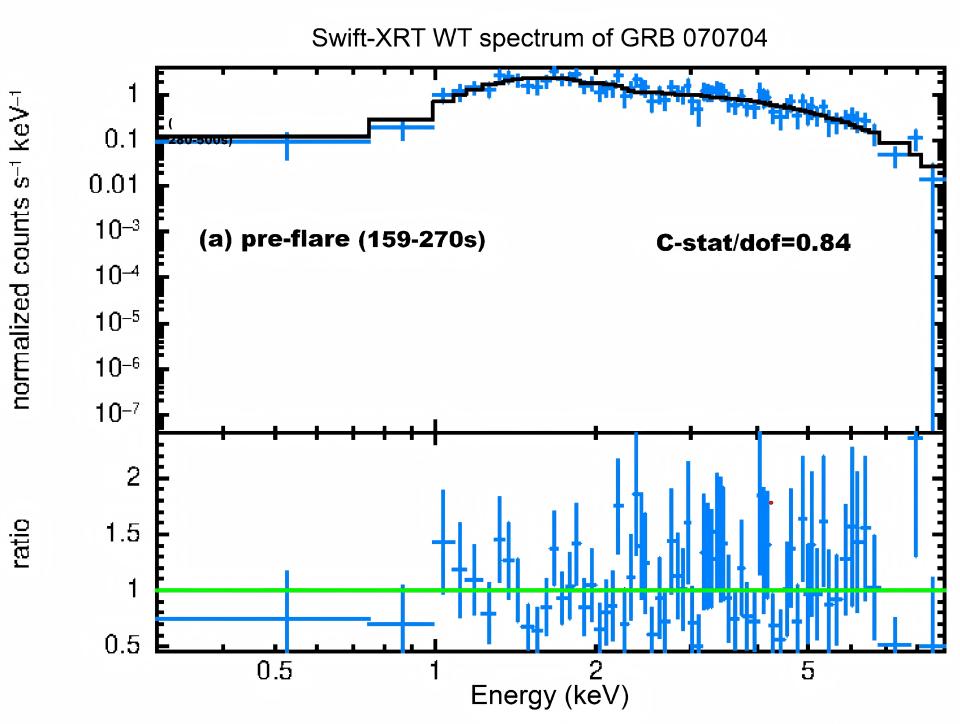}
  \includegraphics[width=0.3\linewidth]{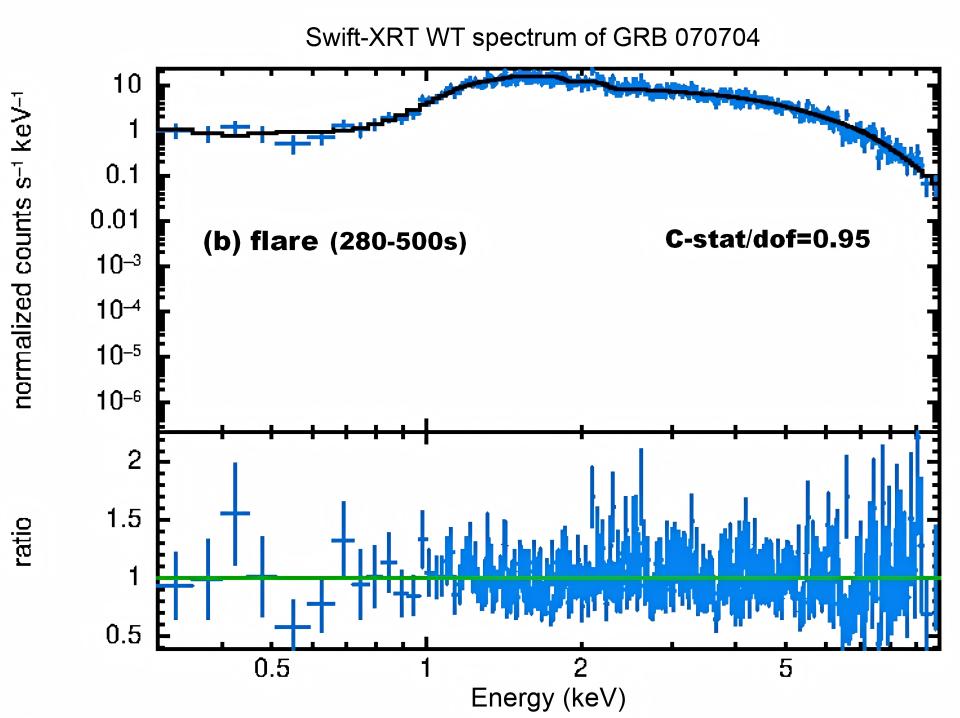}
  \includegraphics[width=0.3\linewidth]{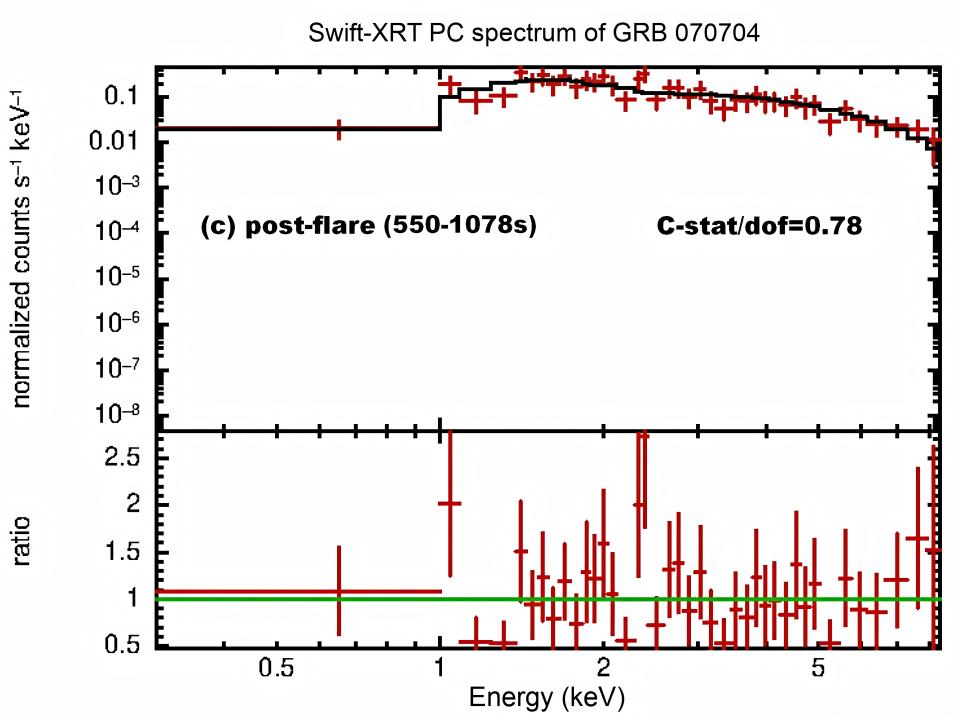}
  \caption{X-ray spectral fits for GRB~070704 during the (a) preflare, (b) flare, and (c) postflare phases. Upper panels: unfolded spectra (data points with $1\sigma$ error bars) and the best-fit absorbed power-law model (black solid line). Lower panels: data-to-model ratio. The horizontal solid line marks the ideal ratio of 1.0. The $C$-stat/dof values (0.84, 0.95, 0.78) are indicated in each panel. The fitting energy range is 0.3--10~keV.}
\end{figure*}

\subsection{Spectral Fitting Results}
We performed spectral fitting for all 521 X-ray flares, as well as for the preflare and postflare phases. The best-fit parameters and related physical quantities are summarized in Table~1.

The columns of Table~1 are organized as follows: Column (1) gives the GRB name; Column (2) indicates the GRB subtype (XRF, XRR, or C-GRB); Column (3) lists the redshift \(z\); Column (4) presents \(T_{90}\), the time interval over which 5\% to 95\% of the total fluence is accumulated; Column (5) presents the power-law photon index \(\alpha_{\rm PL}\) from the simple power-law model for the prompt emission; Column (6) presents the photon index \(\alpha_{\rm CPL}\) from the cutoff power-law model. The subsequent columns report the fitting results for the flare, preflare, and postflare phases. For each phase, the following parameters are provided: the observing mode (Obs., either WT mode or PC mode), the time interval used in the fit, the photon index ($\Gamma$), the intrinsic hydrogen column density \(N_{\rm H}(\rm intrinsic)\) in units of \(10^{20}~\rm cm^{-2}\), and the $C$-statistic value along with the degrees of freedom ($C$-stat/dof). The flare number (No.) is indicated, where ``tot.'' denotes a combined fit for multiple consecutive flares (total-flare), and an asterisk (\(*\)) marks flares for which data from both WT and PC modes overlap and are spectrally fitted separately.

\section{Statistical Analysis} \label{sec:analysis}

\subsection{Spectral Statistics of X-ray Flares}
Our X-ray flare sample is mainly based on the flares identified by \citet{Chincarini2010} and \citet{Yi2016}. According to their criteria, a flare is recognized by its temporal morphology: a well-defined rise, peak, and decay, clearly separable from the underlying afterglow continuum. Minor fluctuations or unresolved features are not considered as valid flares. The light-curve profile must be well fitted by an empirical function, the Norris 05 function \citep{Chincarini2010}, or a smooth broken power law \citep{Yi2016}. Overlapping flares are included only when individual pulses can be reliably separated.

For GRBs with multiple consecutive flares, we perform spectral fits for each individual flare and a global fit for the whole flare sequence, following the schemes shown in Figures 1 and 2. The start and end times of individual flares are determined using the flare boundaries from \citet{Chincarini2010} and \citet{Yi2016}, combined with GRB light curves; all flare time intervals are nonoverlapping.

We select GRB 090715B, a GRB exhibiting multiple consecutive flares, as an example; its light curve is displayed in Figure 2. This source shows eight separate X-ray flares. We carry out spectral fits for each individual flare and a global fit covering all eight consecutive flares (total-flare). The fitting results are shown in Figure 5.

\begin{figure*}[!htb]
  \centering
  \includegraphics[width=0.3\linewidth]{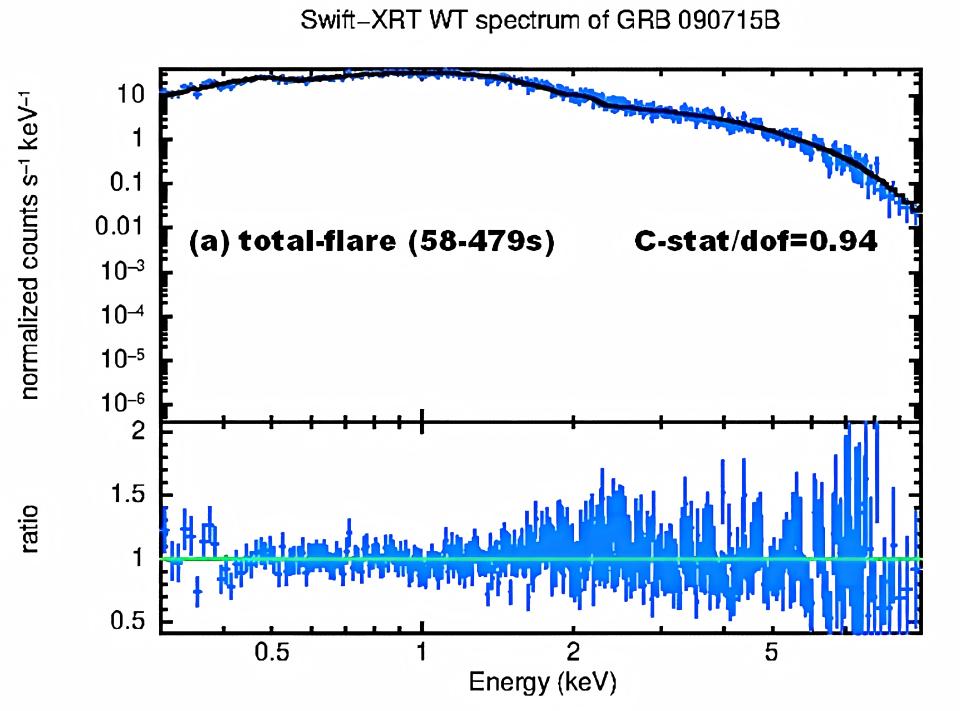}
  \includegraphics[width=0.3\linewidth]{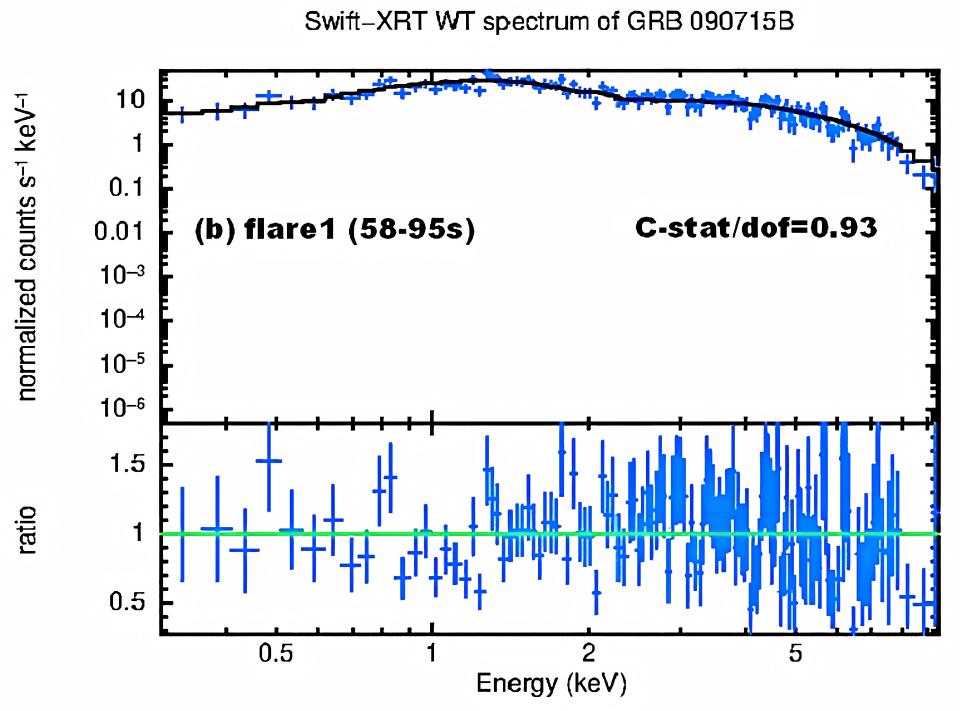}
  \includegraphics[width=0.3\linewidth]{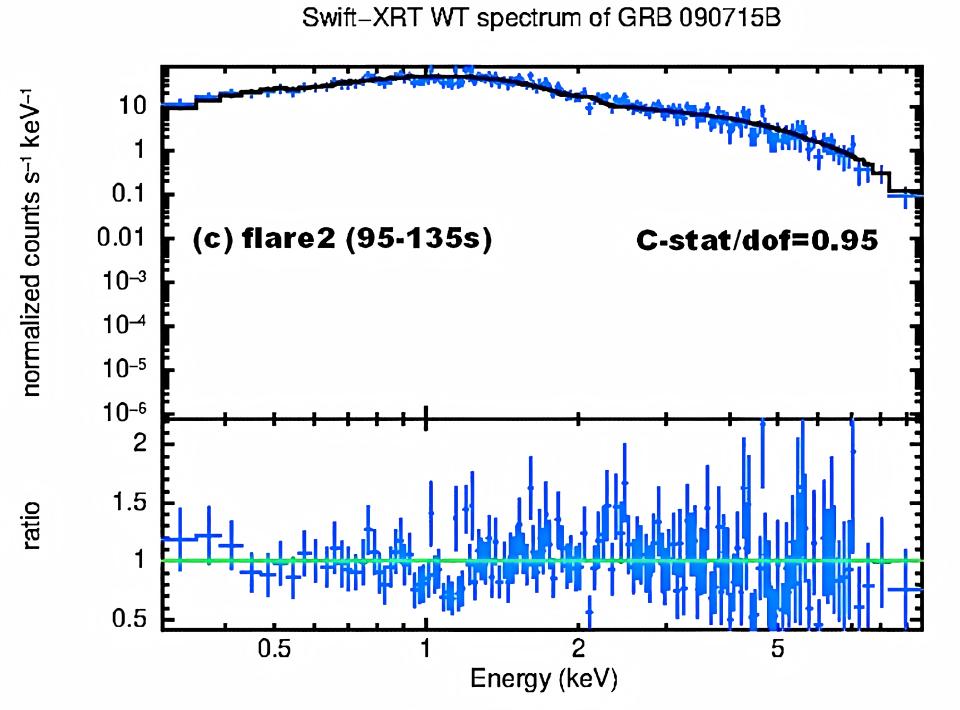}
  \includegraphics[width=0.3\linewidth]{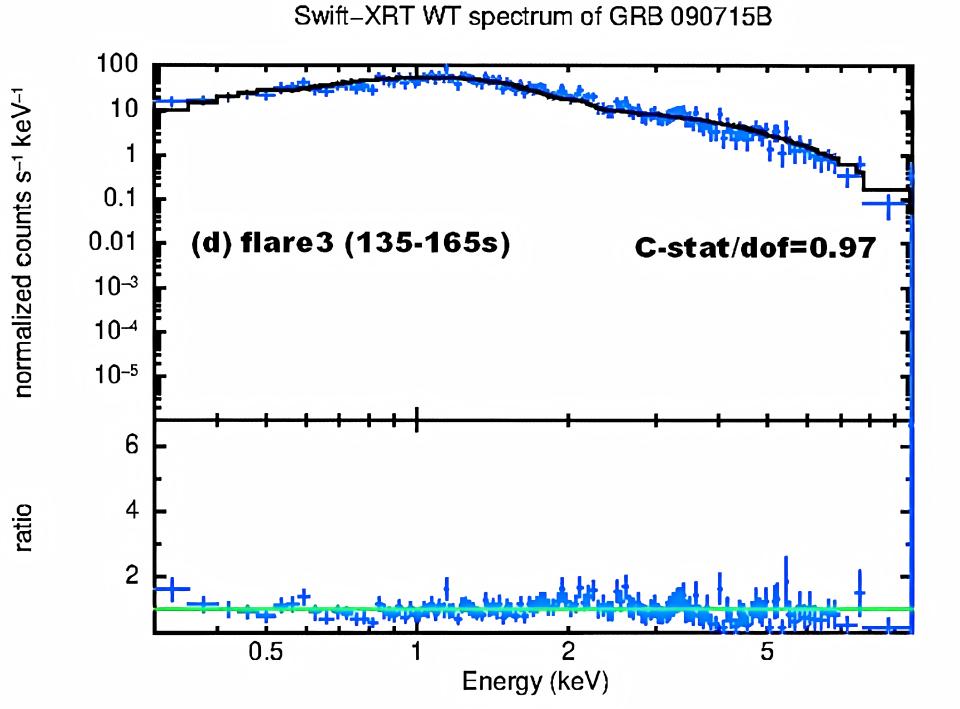}
  \includegraphics[width=0.3\linewidth]{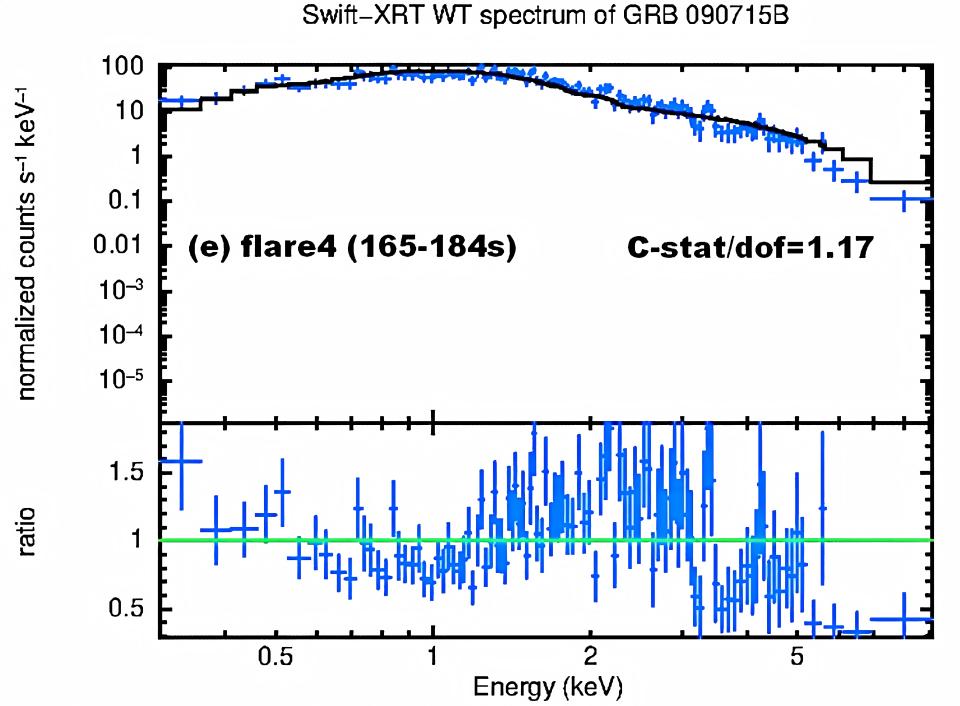}
  \includegraphics[width=0.3\linewidth]{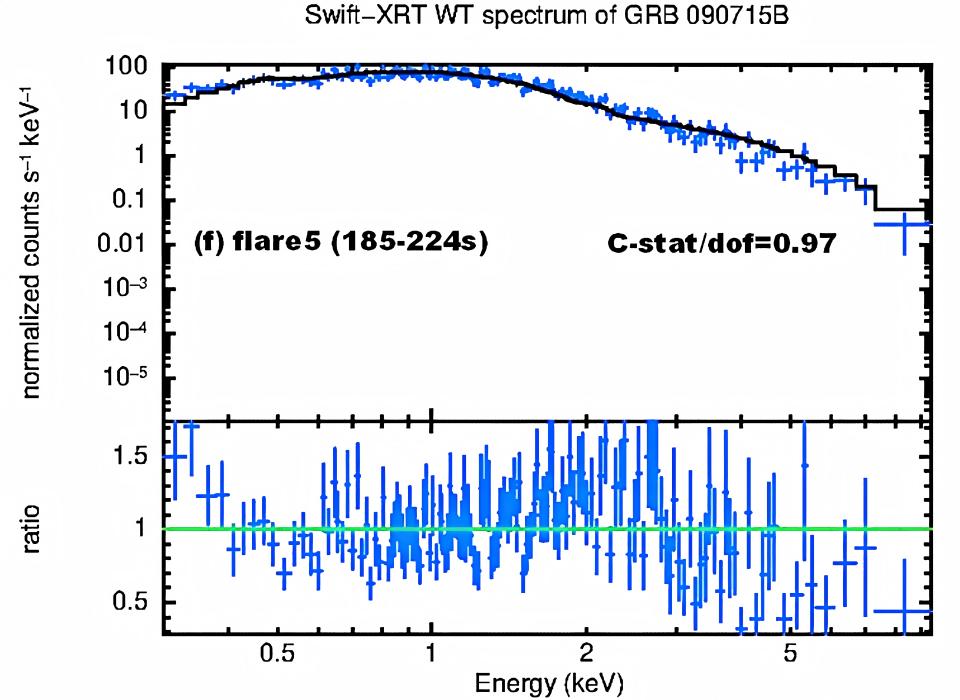}
  \includegraphics[width=0.3\linewidth]{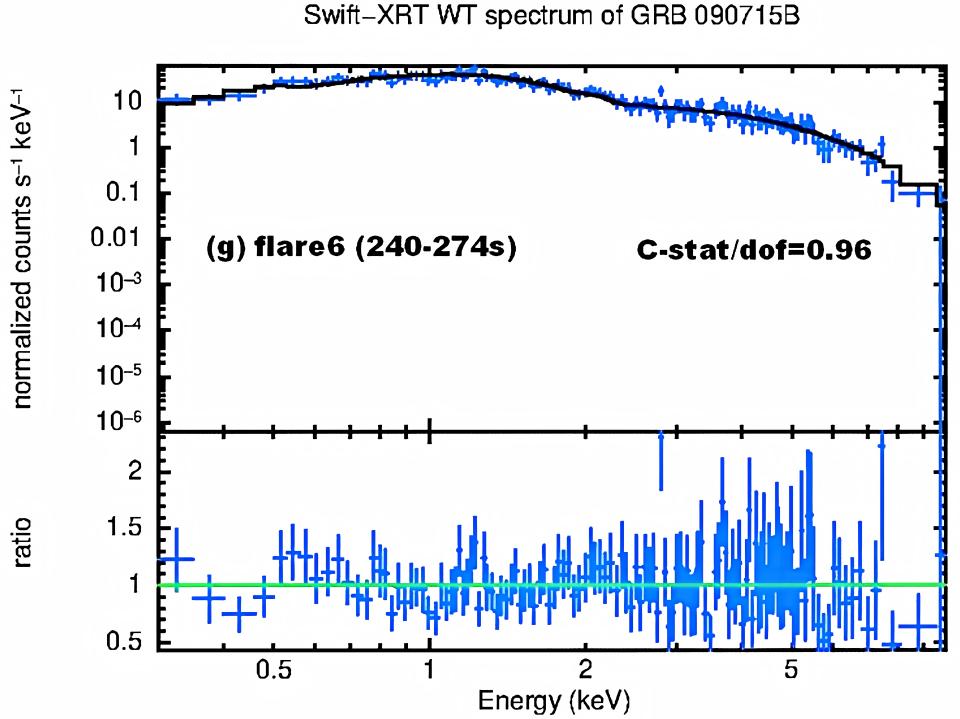}
  \includegraphics[width=0.3\linewidth]{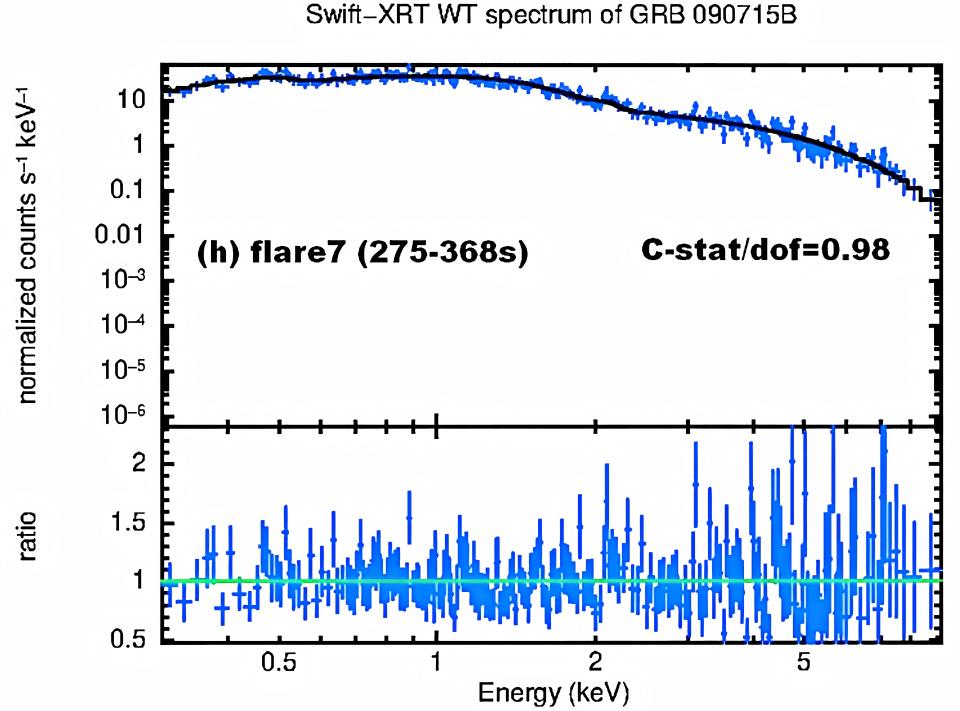}
  \includegraphics[width=0.3\linewidth]{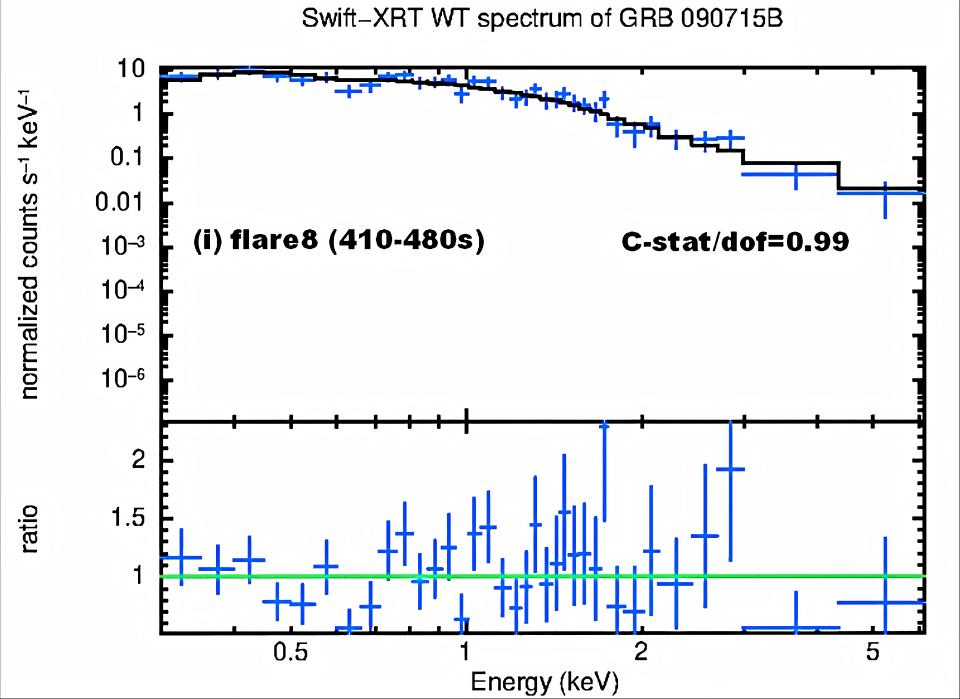}
  \caption{X-ray spectral fits for GRB~090715B. (a) Total-flare phase covering 58--479 s; (b)--(i) flares 1--8 respectively. Upper panels: unfolded spectra (data points with $1\sigma$ error bars) and the best-fit absorbed power-law model (black solid line). Lower panels: data-to-model ratio. The horizontal solid line marks the ideal ratio of 1.0. The $C$-stat/dof value for each fit is indicated in each panel. The fitting energy range is 0.3--10~keV.}
\end{figure*}

In our sample, we perform joint spectral fits on 74 light-curve segments containing consecutive flares from 73 GRB sources. For GRB~141031A, the WT-mode and PC-mode data overlapped during its consecutive flares, so we carried out two separate joint spectral analyses using WT-mode and PC-mode data, respectively. The 73 GRBs that exhibited multiple consecutive flares contained a total of 315 individual flares. We conducted statistical analyses on the fitting results from both the joint spectra and individual-flare spectra, and the distributions of $C$-stat/dof are shown in Figure~6. Figure~6(a) presents the $C$-stat/dof distribution for the 74 joint spectral fits of multiflare sequences, spanning 0.75--1.45. Figure~6(b) shows the $C$-stat/dof distribution for the 315 individual-flare spectral fits, covering 0.65--1.45. As shown in the figure, we obtained spectral fits of reasonable quality for both individual flares and consecutive multiflare sequences.

\begin{figure*}[!htb]
  \centering
  \includegraphics[width=0.4\linewidth]{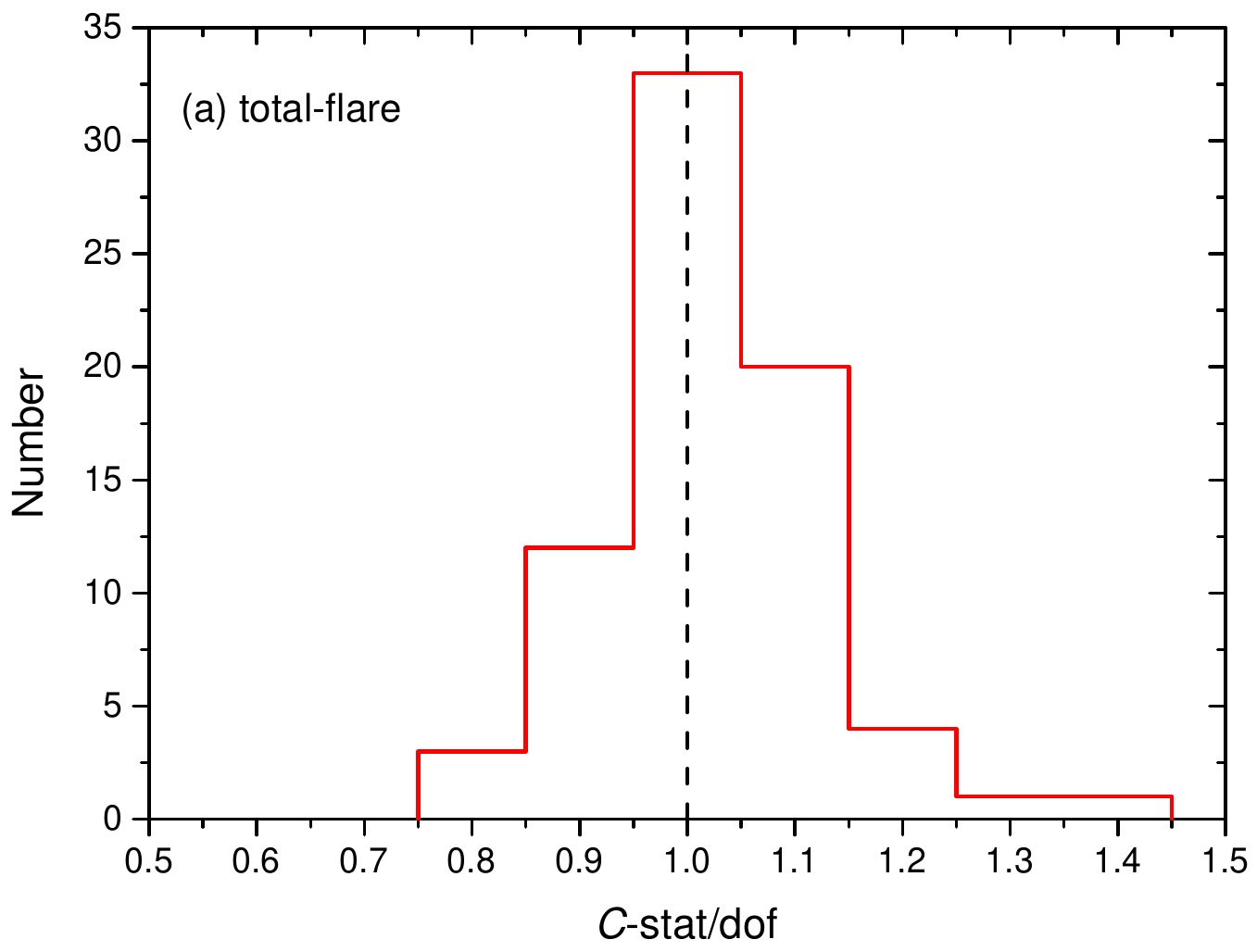}
  \includegraphics[width=0.4\linewidth]{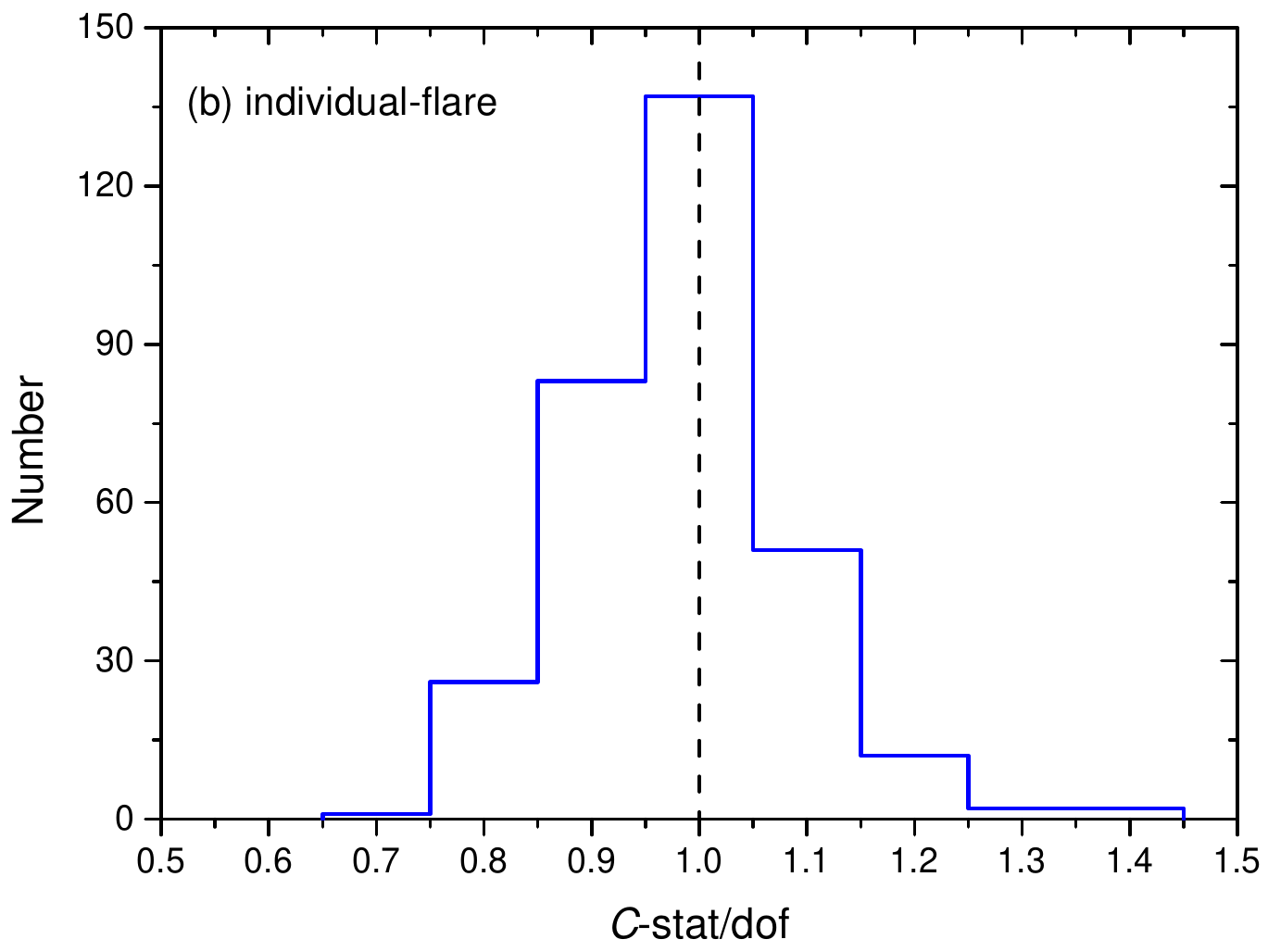}
   \caption{Distribution of the $C$-stat/dof values from the X-ray spectral fits: (a) total-flare sample consisting of 74 joint spectral fits, (b) individual-flare sample consisting of 315 individual flares. The vertical dashed line marks the ideal value of 1.0.}
\end{figure*}

We conducted a statistical analysis of the spectra of X-ray flares. In our sample, we obtained 521 X-ray flares from 206 GRBs. Among the sample, 98 GRBs exhibited a single flare, 57 GRBs showed two flares, and 51 GRBs displayed three or more flares. We performed joint spectral fits on 74 light-curve segments with consecutive X-ray flares from 73 GRB sources. In our sample, the total number of X-ray flares observed in WT and PC modes is 521, with 406 in WT mode and 115 in PC mode. We analyzed the photon index distributions of X-ray flares for the combined WT+PC mode, WT mode, and PC mode. The resulting histograms are shown in Figure 7.

\begin{figure*}[!htb]
  \centering
  \includegraphics[width = 0.7\linewidth]{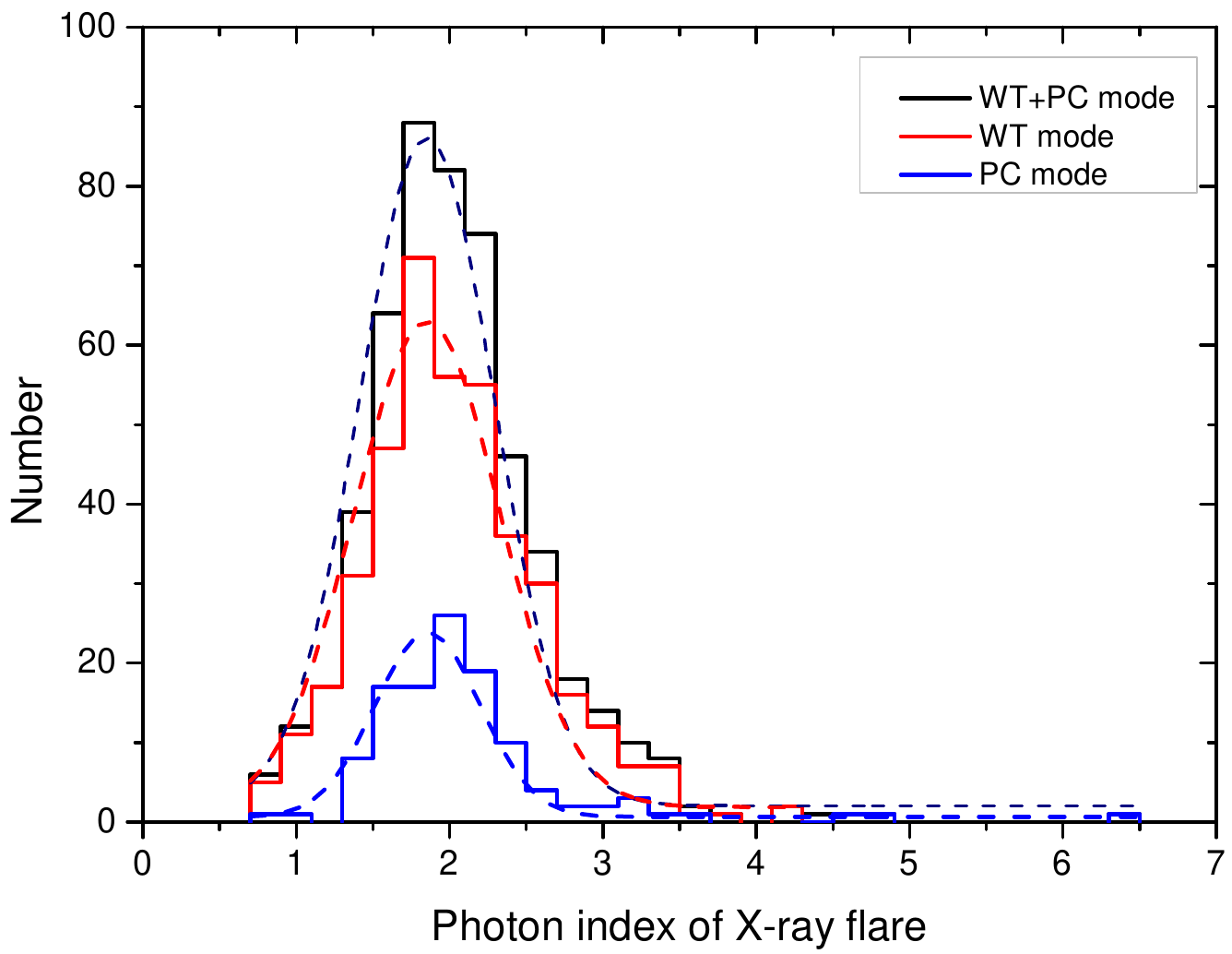}
  \caption{Histogram of the photon index distribution of X-ray flares. The black solid line represents data from both WT and PC modes, the red solid line represents data from WT mode, and the blue solid line represents data from PC mode. The dashed lines show the corresponding Gaussian fits.} \label{fig:7}
\end{figure*}

In Figure~7, we performed Gaussian fits to the histograms. For the WT+PC mode, the photon indices range from 0.63 to 6.30, with a mean value of 1.94 and a Gaussian peak centered at \(x_c = 1.85 \pm 0.01\). For the WT mode, the photon indices range from 0.67 to 4.36, with a mean of 1.92 and a Gaussian peak at \(x_c = 1.85 \pm 0.02\). For the PC mode, the photon indices range from 0.73 to 6.30, with a mean of 2.02 and a Gaussian peak at \(x_c = 1.86 \pm 0.02\).\footnote{In our sample of X-ray flares, GRB 050714B exhibits the largest photon index, with a value of 6.30, which is substantially higher than those of the other bursts. A review of the literature indicates that \citet{Page2005} performed spectral analysis on the early X-ray afterglow of GRB 050714B and found its spectrum to be extremely soft, with a photon index of approximately 6. \citet{Butler2007} conducted a comprehensive spectral analysis of the early X-ray afterglow data in the Swift/XRT GRB sample and found that while most spectra (90\%) can be well fitted by a simple power-law model, there exist notable outliers with anomalously soft spectra that may contain a thermal component. Four bursts---GRB 060218, GRB 060202, GRB 050822, and GRB 050714B---exhibit such unusually soft spectra, with GRB 050714B having a photon index of 5.90. Our spectral analysis results for the X-ray flares of GRB 050714B are generally consistent with those of \citet{Page2005} and \citet{Butler2007}, further confirming that GRB 050714B is a GRB with an anomalously soft spectrum.} We applied the Kolmogorov--Smirnov test (K-S test) to compare the photon index distributions between the WT and PC modes, yielding a $p$-value of \(p = 0.1956\). The K-S test indicates that there is no significant difference between the photon index distributions obtained in the two observing modes.

\subsection{Temporal Evolution of the Photon Index in X-Ray Flares}
To investigate the temporal evolution of the photon index in X-ray flares, we selected GRBs exhibiting multiple flares. The sample composition is as follows: 14 GRBs exhibit four X-ray flares (GRB 050712, GRB 050904, GRB 060124, GRB 070721B, GRB 071031, GRB 081028, GRB 090417B, GRB 090429A, GRB 100614A, GRB 111016A, GRB 111209A, GRB 140817A, GRB 141031A, GRB 150323C); seven GRBs exhibit five X-ray flares (GRB 070107, GRB 081210, GRB 100725B, GRB 100901A, GRB 110709B, GRB 130606A, GRB 130925A); one GRB exhibits six X-ray flares (GRB 080310); two GRBs exhibit seven X-ray flares (GRB 070129 and GRB 070616); one GRB exhibits eight X-ray flares (GRB 090715B); two GRBs exhibit nine X-ray flares (GRB 100728A and GRB 110119A); one GRB exhibits ten X-ray flares (GRB 051117A); and one GRB exhibits twelve X-ray flares (GRB 100212A).

\begin{figure*}[!htb]
  \centering
  \includegraphics[width=0.4\linewidth]{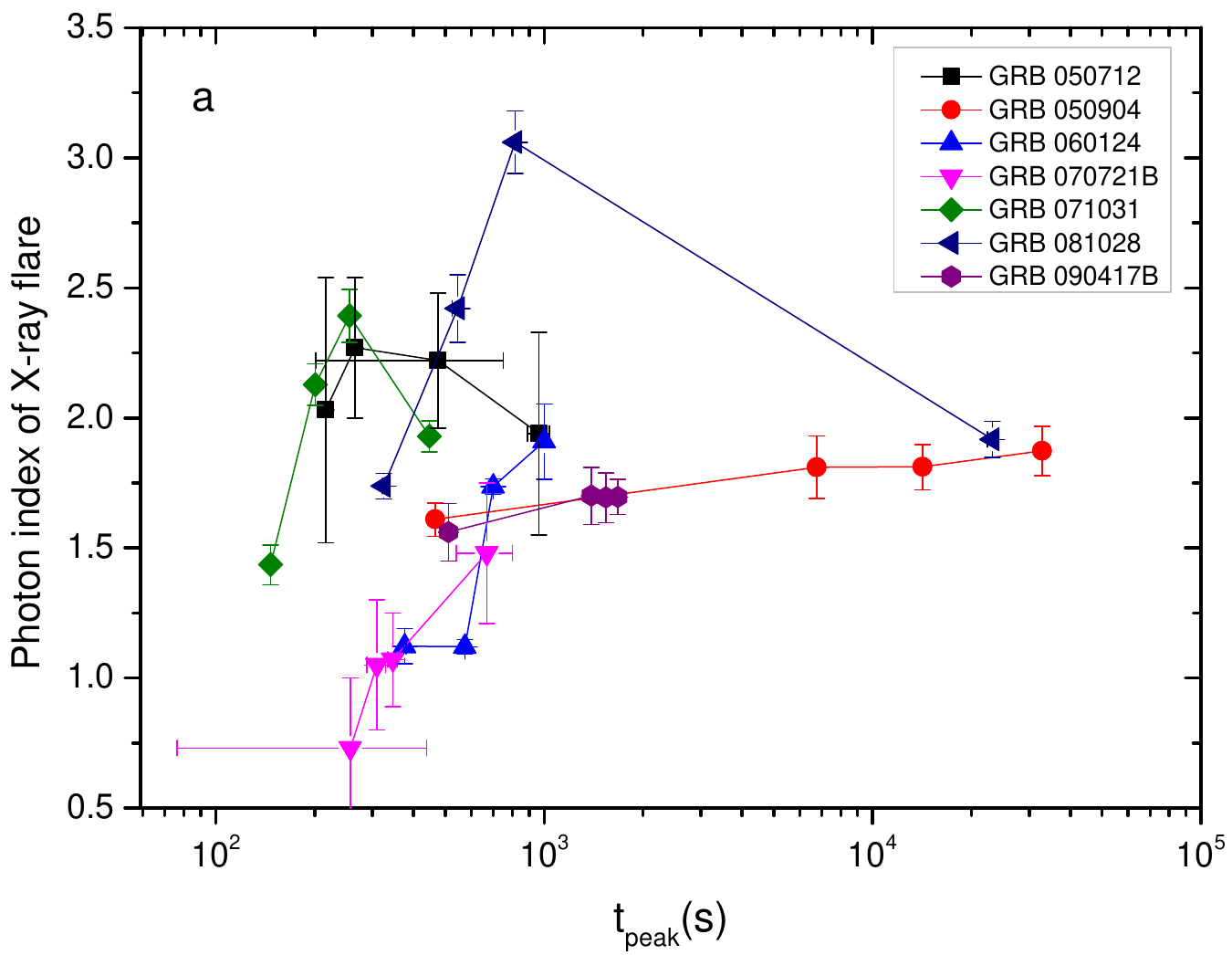}
  \includegraphics[width=0.4\linewidth]{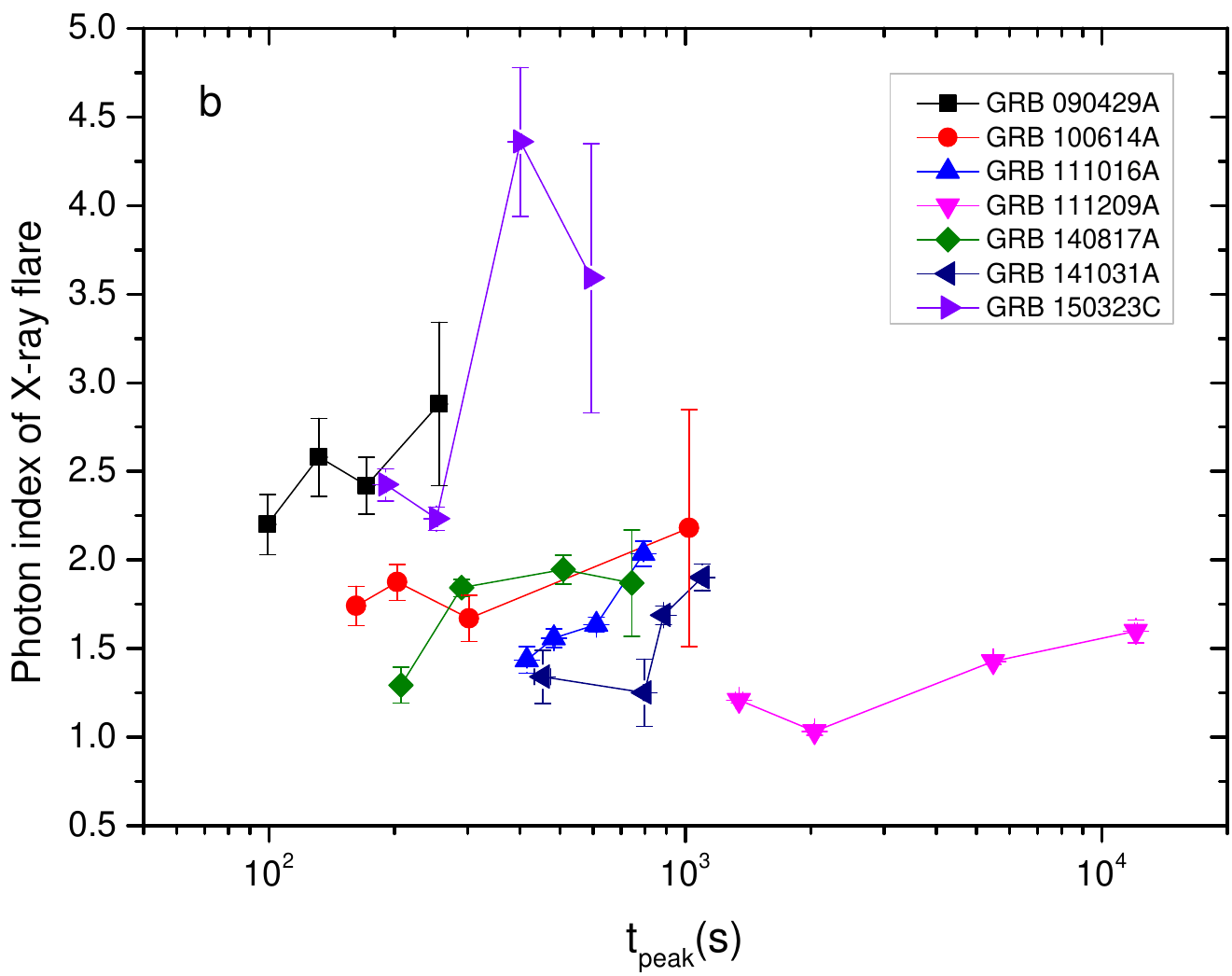}
  \includegraphics[width=0.4\linewidth]{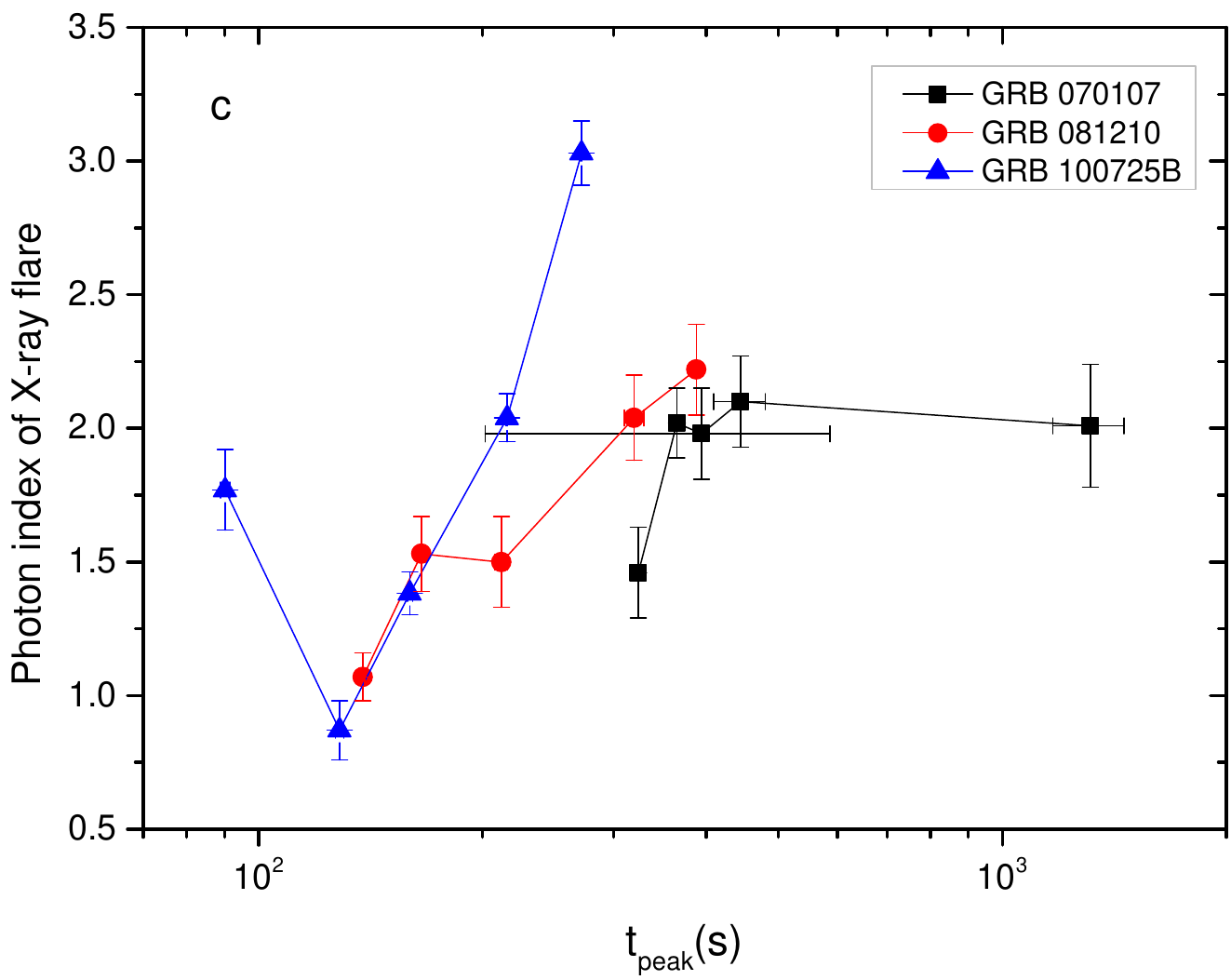}
  \includegraphics[width=0.4\linewidth]{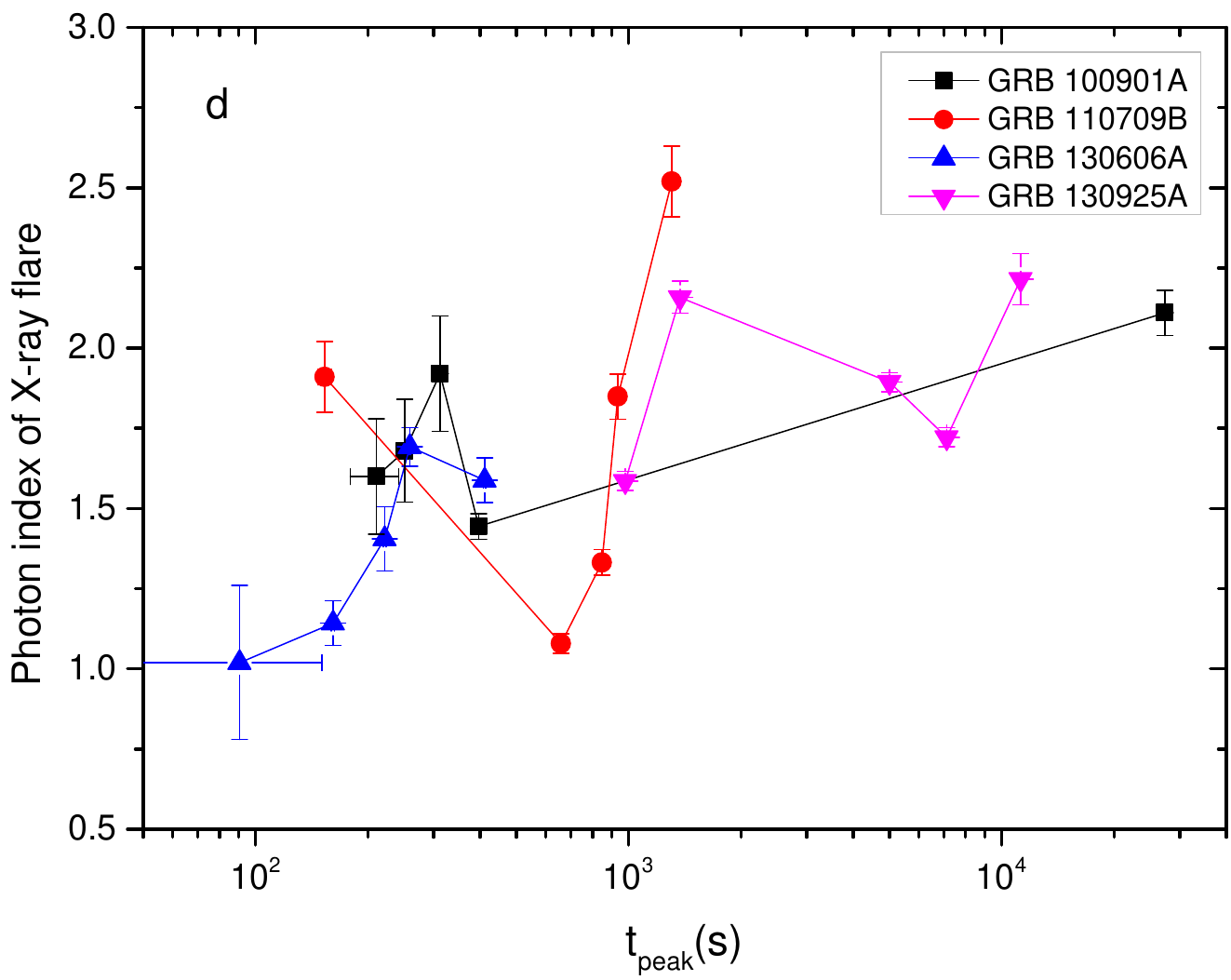}
  \includegraphics[width=0.4\linewidth]{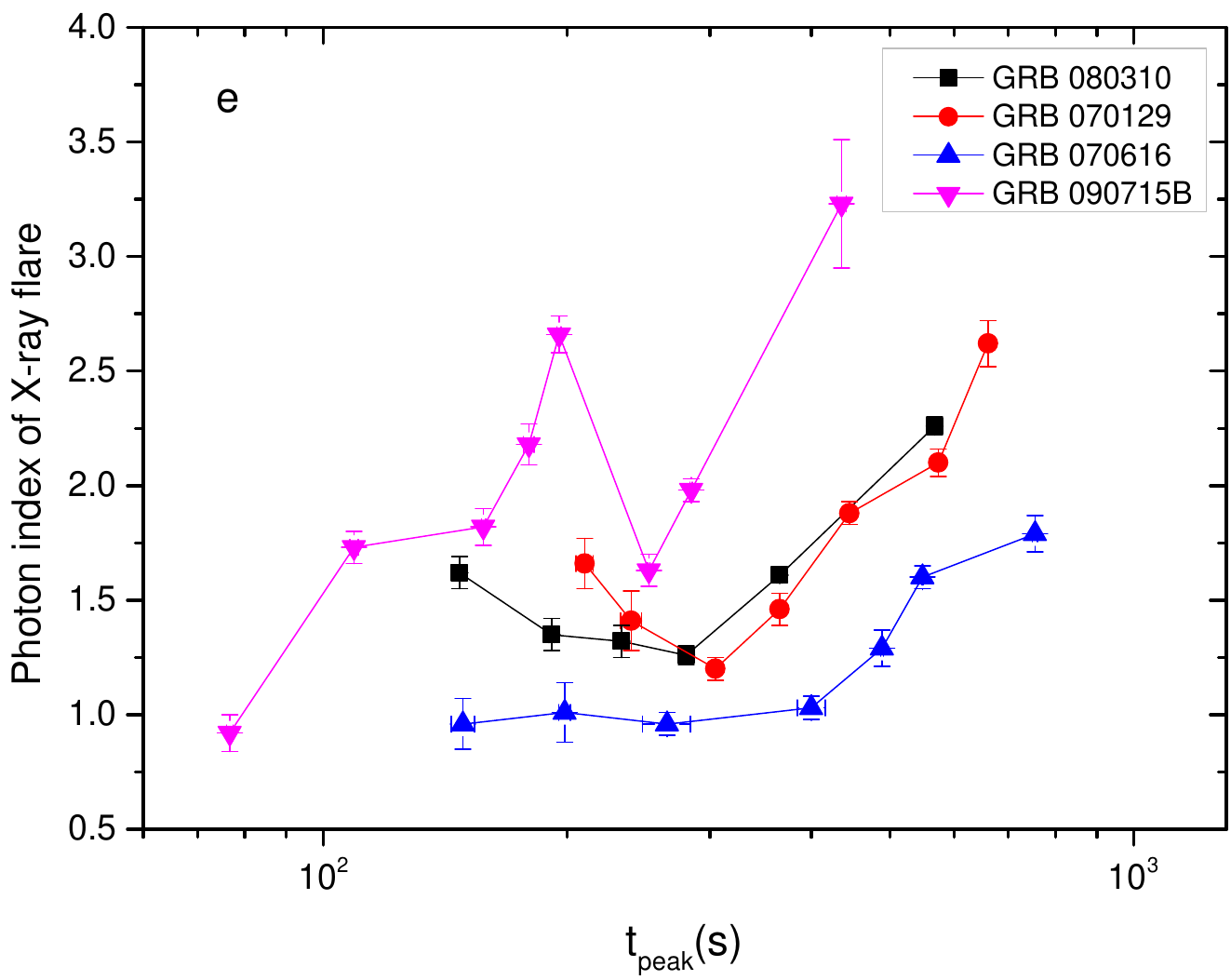}
  \includegraphics[width=0.4\linewidth]{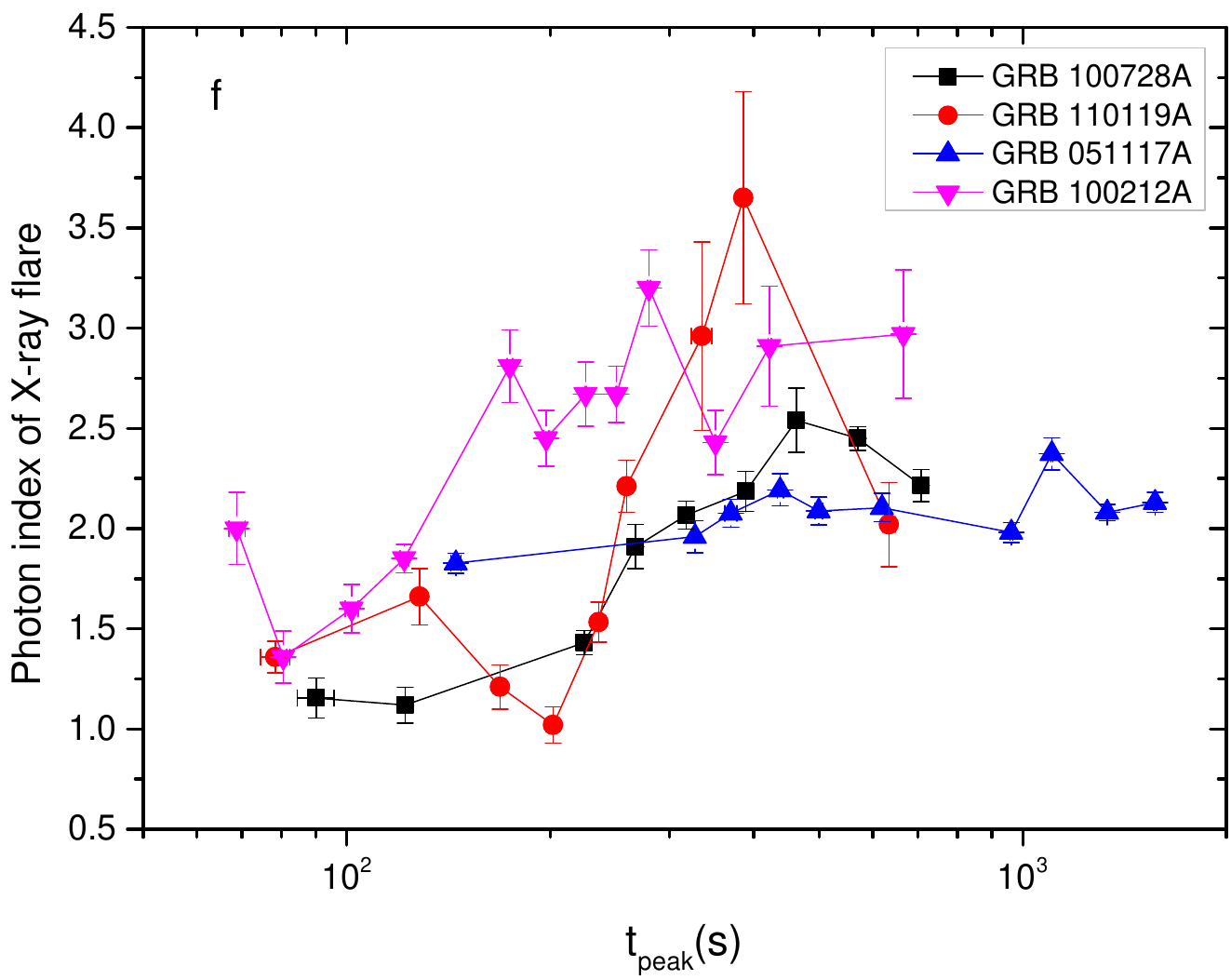}
  \caption{Correlation between the photon index and the peak time of X-ray flares. Panels present GRBs that exhibit (a), (b) four X-ray flares; (c), (d) five X-ray flares; (e) six, seven, and eight X-ray flares; (f) nine, ten, and twelve X-ray flares.}
  \label{fig:8}
\end{figure*}

Figure 8 presents the correlation between the photon index and the peak time (\(t_{\text{peak}}\)) of multiple X-ray flares occurring in the same GRB. Figures 8(a) and 8(b) show the cases in which each GRB  has four X-ray flares. Figures 8(c) and 8(d) show the cases in which each GRB has five X-ray flares. Figure 8(e) displays the cases that each GRB has six, seven, and eight X-ray flares, respectively. Figure 8(f) displays the cases that each GRB has nine, ten, and twelve X-ray flares, respectively. As illustrated, the photon index of X-ray flares increases progressively with peak time, indicating a spectral softening trend as the peak time increases. Our analysis further demonstrates that, for GRBs with multiple flare episodes, the photon index of X-ray flare exhibits a distinct temporal evolution pattern: later flares tend to have softer spectra than earlier ones, suggesting a progressive spectral softening over time.

\subsection{Statistical Analysis of Photon Indices in Preflare, Flare, and Postflare Phases}
To investigate the spectral evolution characteristics of X-ray flares, we conducted a comparative statistical analysis of the photon indices in the preflare, flare, and postflare phases. For GRBs with multiple consecutive flares in our sample, we adopted two approaches for statistical analysis. First, we analyzed the overall emission of multiple consecutive flares (total-flare sample). Second, we analyzed all individual flares (all-flare sample). The K-S test and the t-test were employed to statistically examine the variations in photon index distributions among the preflare, flare, and postflare phases.

In the histogram distributions, Gaussian fits were applied separately to the photon index distributions of the preflare, flare, and postflare phases, and the differences in the Gaussian peak centers were compared. Additionally, K-S tests were performed on the photon index distributions among these three phases. In this statistical analysis, we combined all observation data from both WT and PC modes for an integrated analysis.

Figure 9 presents the histograms of spectral index distributions along with Gaussian fits for the total-flare sample, which includes bursts exhibiting multiple flares. The central values derived from the Gaussian fits are: $x_c = 1.82 \pm 0.02$ for flare; $x_c = 1.88 \pm 0.02$ for preflare; and $x_c = 1.94 \pm 0.03$ for postflare. The K-S test yields the following $p$-values: $p = 0.452$ for flare versus preflare; $p = 0.002$ for flare versus postflare; $p = 0.063$ for preflare versus postflare. At a significance level of 0.05, no significant differences are found between the flare and preflare distributions, nor between the preflare and postflare distributions. However, a statistically significant difference exists between the flare and postflare distributions, manifested as a notably larger spectral index in the postflare phase than that during the flare, indicating a trend toward spectral softening after the flare.

\begin{figure*}[!htb]
  \centering
  \includegraphics[width=0.7\linewidth]{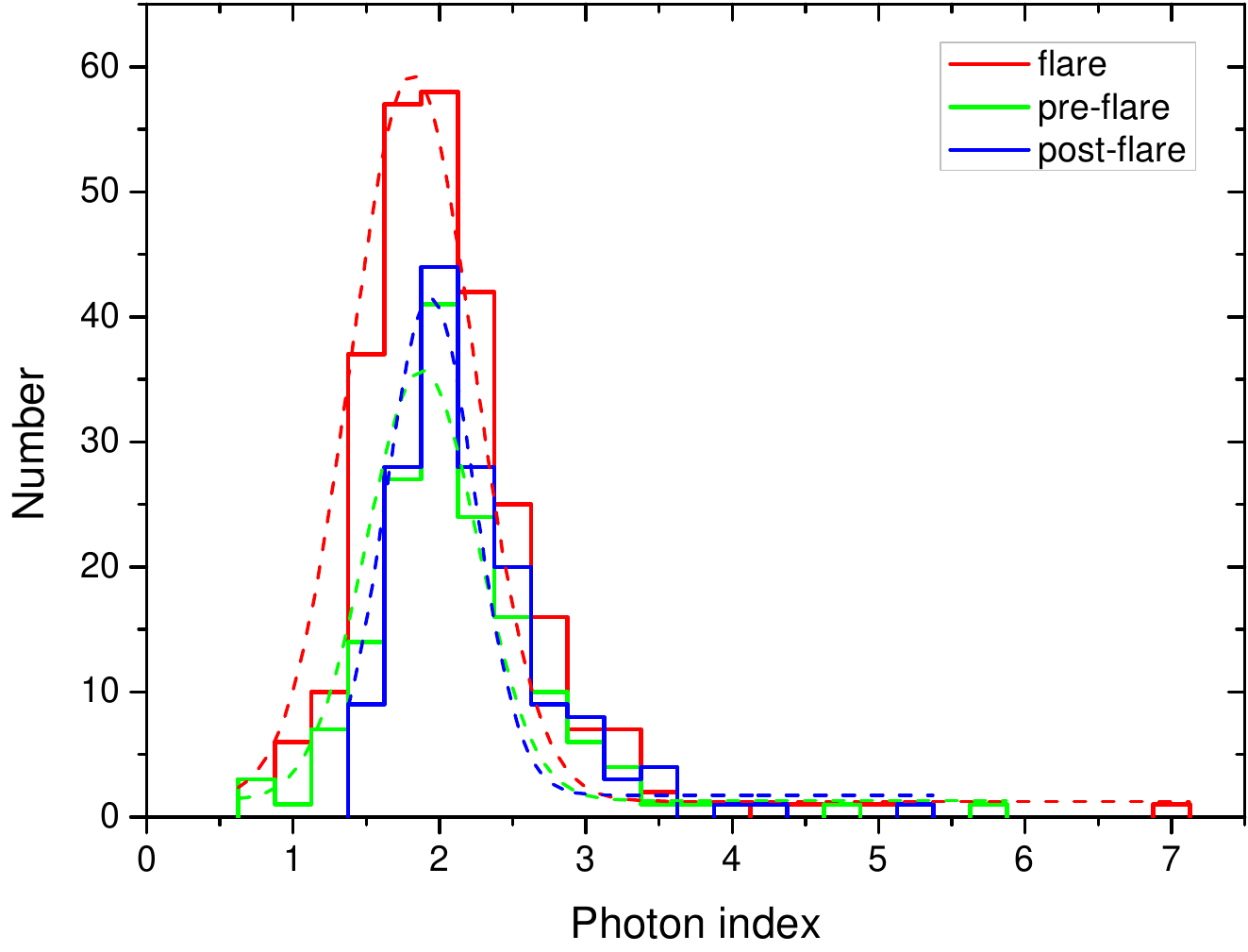}
  \caption{Histograms of photon index distribution for the flare, preflare, and postflare phases (total-flare sample). The red, green, and blue solid lines represent the data for the flare, preflare, and postflare phases, respectively, and the dashed lines are the Gaussian fits.}\label{fig9}
\end{figure*}

To further investigate the evolution of the spectral indices across the preflare, flare, and postflare phases, we performed one-to-one matching of the sample data among the three phases. The paired scatter plots for flare versus preflare, flare versus postflare, and preflare versus postflare are presented in Figure 10.

\begin{figure*}[!htb]
  \centering
  \includegraphics[width=0.5\linewidth]{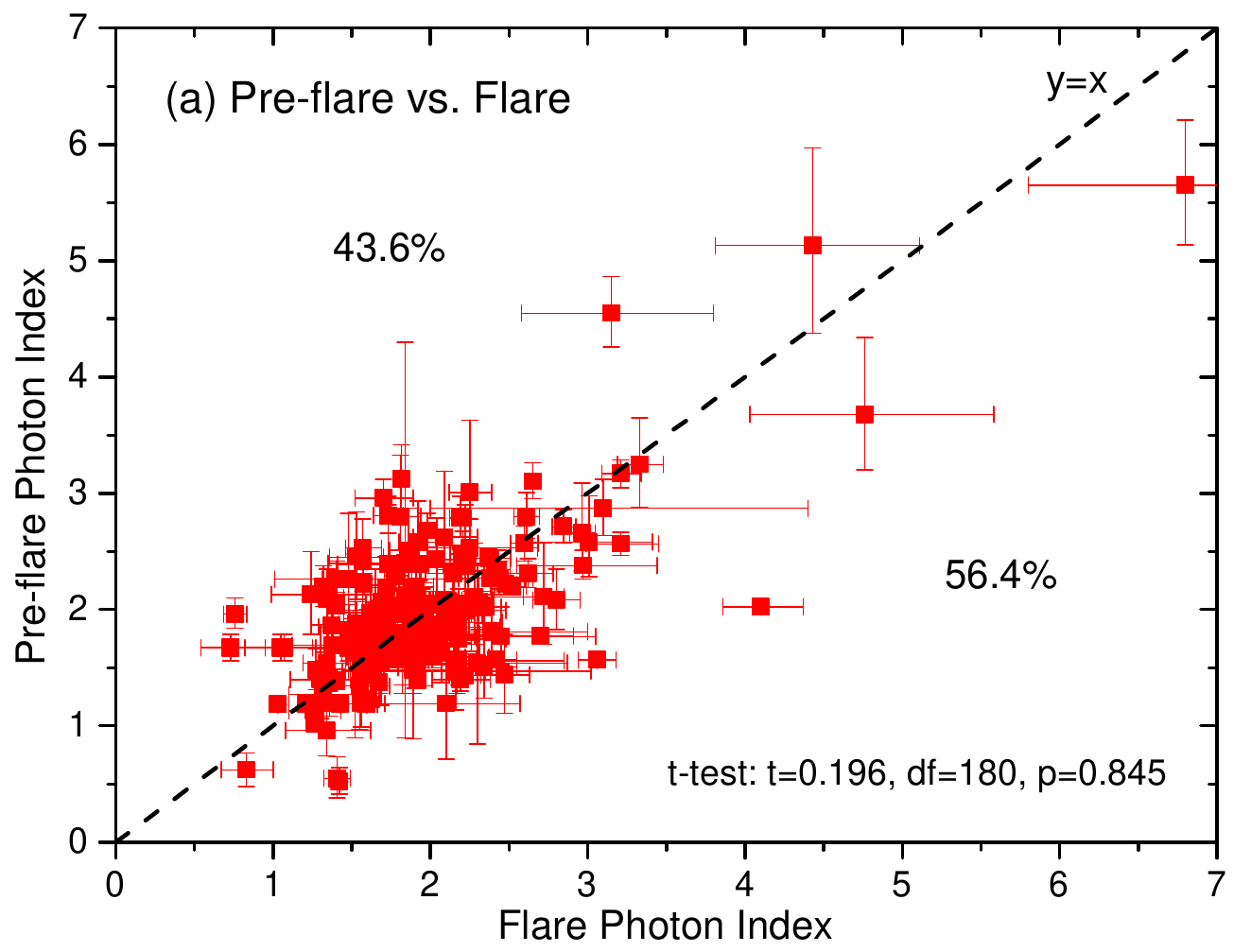}
  \includegraphics[width=0.5\linewidth]{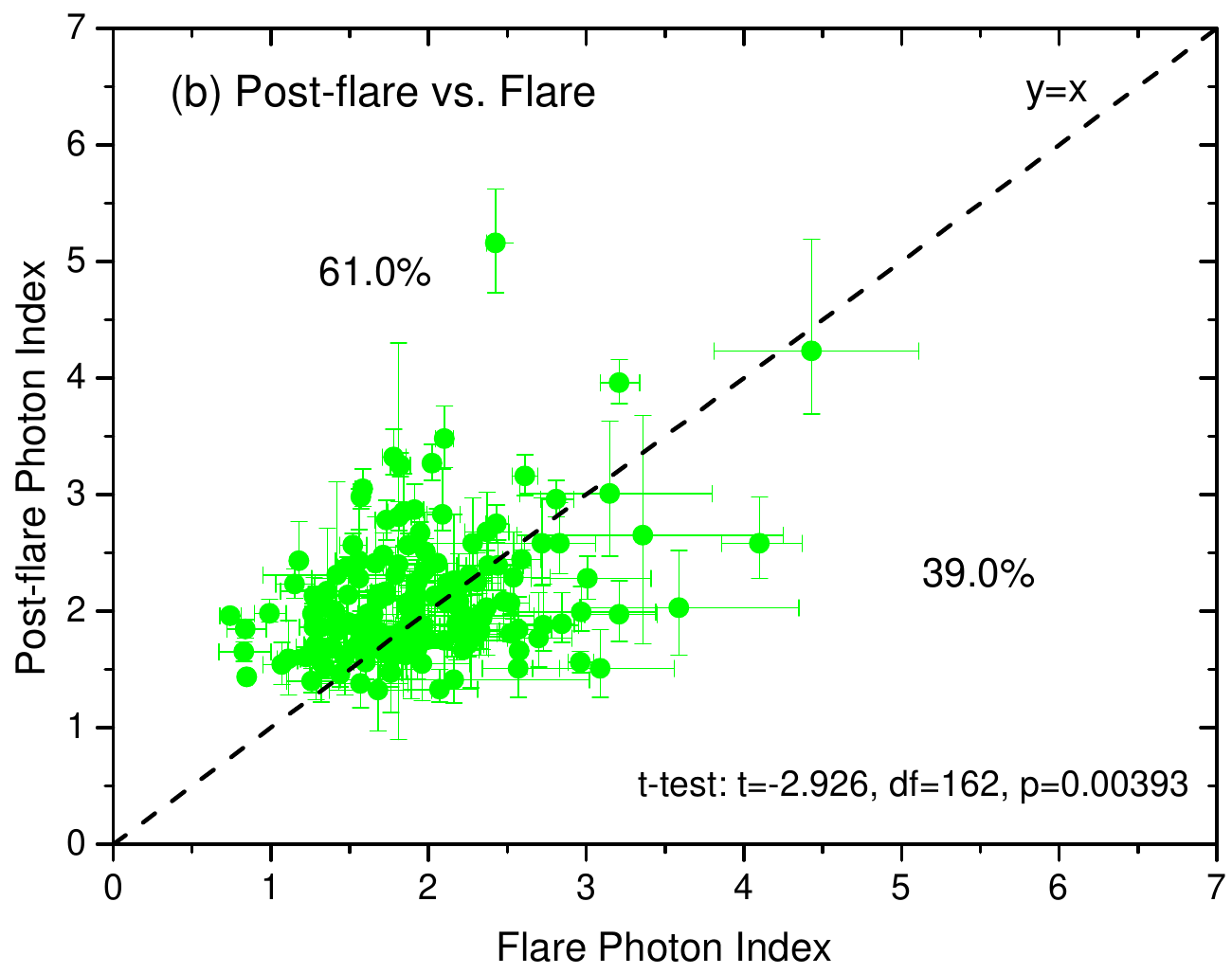}
  \includegraphics[width=0.5\linewidth]{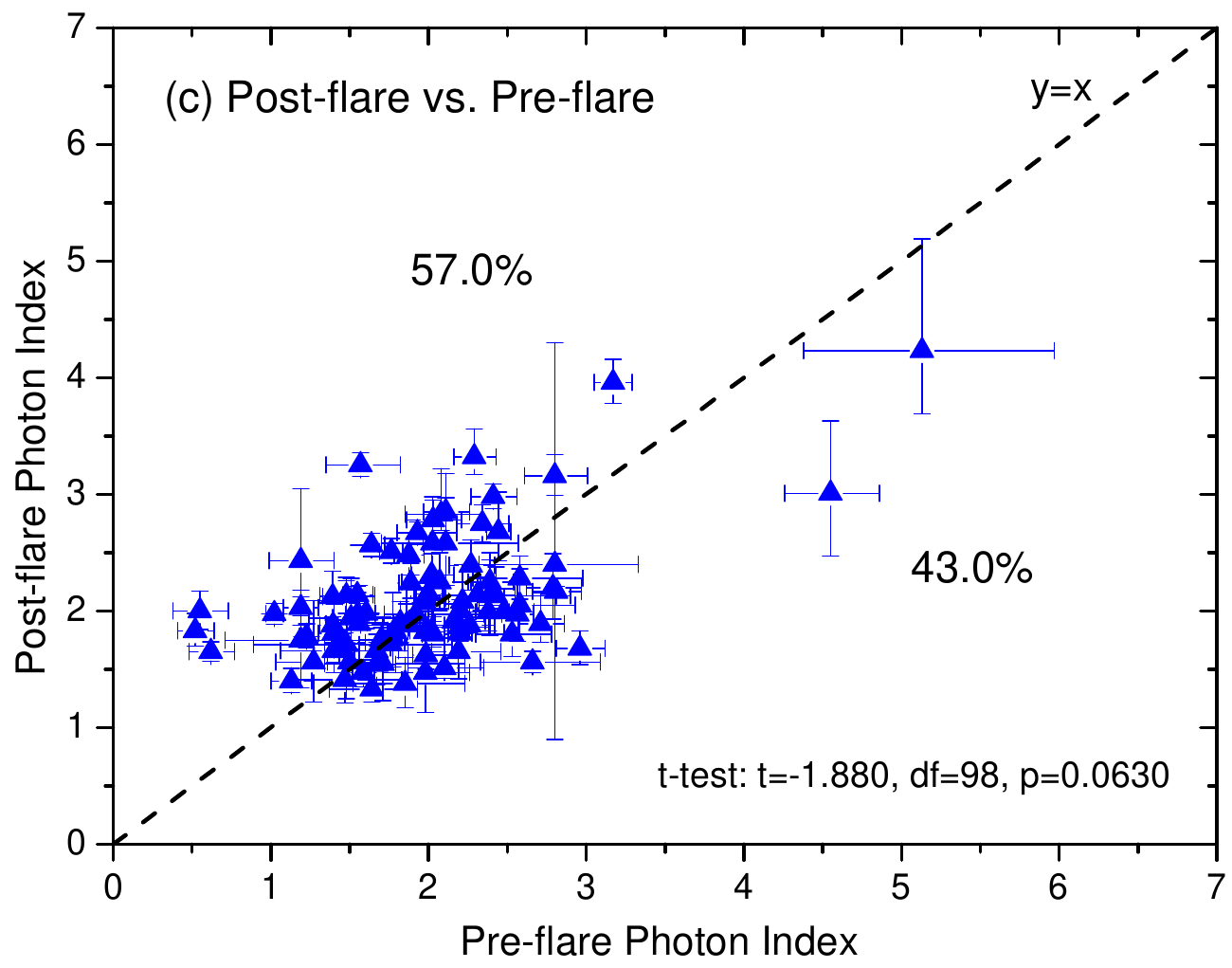}
  \caption{Comparison of photon index in the preflare, flare, and postflare phases (total-flare sample). The dashed line represents the $y = x$ identity line. (a) Flare versus preflare, (b) flare versus postflare, and (c) preflare versus postflare.}
  \label{fig:10}
\end{figure*}
To quantitatively compare the differences in spectral indices across different phases of X-ray flares, we performed paired-samples t-tests on the photon indices of the preflare, flare, and postflare phases. The results are annotated in each panel of Figure~10. The black dashed line in each panel represents the identity line ($y = x$), which serves as a reference for comparing the relative magnitudes of the spectral indices between two phases. The percentages indicate the fraction of samples in which the spectral index in one phase is larger or smaller than that in the other phase. In the comparison between preflare phase and flare phase (Figure~10(a)), the paired t-test reveals no significant difference between the two phases on the photon index ($t = 0.196$, $df = 180$, $p = 0.845$). In the comparison between flare phase and postflare phase (Figure~10(b)), the paired t-test indicates a statistically significant difference between the two phases on the photon index ($t = -2.926$, $df = 162$, $p = 0.00393$). In the comparison between preflare phase and postflare phase (Figure~6(c)), the paired t-test shows no significant difference between the two phases on the photon index ($t = -1.880$, $df = 98$, $p = 0.0630$).

Combining the results of the K-S tests and paired t-tests, we find that the photon index undergoes significant hardening (i.e., decreases) during the flare phase compared to the postflare phase. In contrast, the differences between preflare and flare, as well as between preflare and postflare, are not statistically significant at the 0.05 significance level.

To further investigate the variations in spectral indices across different phases of X-ray flares, we extended our analysis to the all-flare sample, applying the same methodology used for the total-flare sample. Figure~11 presents the histograms of spectral index distributions along with Gaussian fits for the all-flare sample across different phases. The central values derived from the Gaussian fits are: $x_c = 1.84 \pm 0.02$ for flare, $x_c = 1.89 \pm 0.02$ for preflare, and $x_c = 1.93 \pm 0.03$ for postflare. The K-S test yields the following $p$-values: $p = 0.360$ for flare versus preflare, $p = 0.001$ for flare versus postflare, $p = 0.108$ for preflare versus postflare. These results indicate a statistically significant difference in the spectral index distribution between the flare and postflare phases, consistent with the findings from the total-flare sample.

\begin{figure*}[!htb]
  \centering
  \includegraphics[width=0.7\linewidth]{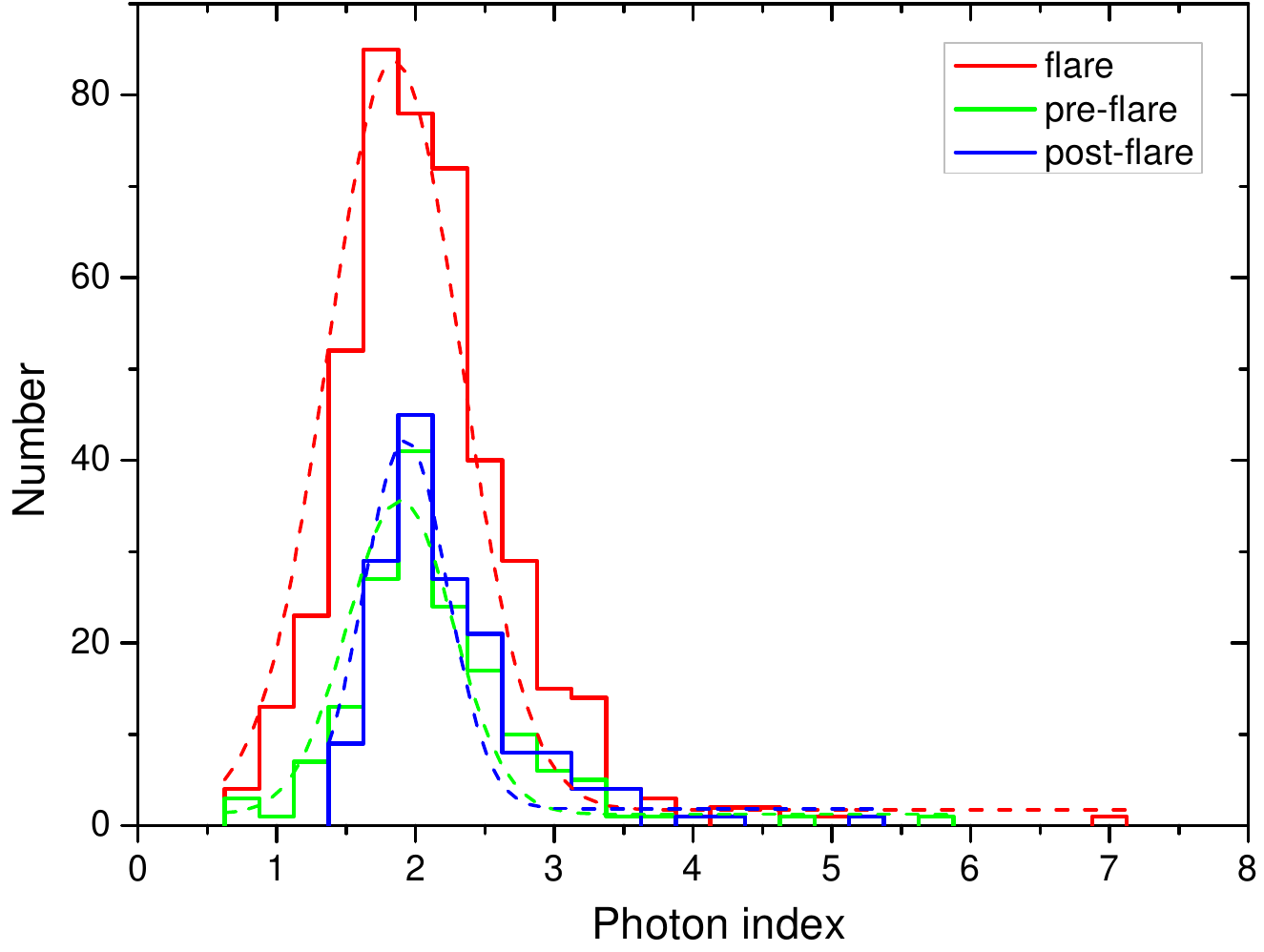}
   \caption{Histograms of photon index distributions for the flare, preflare, and postflare phases (all-flare sample). The red, green, and blue solid lines represent the data for the flare, preflare, and postflare phases, respectively, and the dashed lines are the Gaussian fits.}\label{fig11}
\end{figure*}

%\clearpage
\begin{figure*}[!htb]
  \centering
  \includegraphics[width=0.5\linewidth]{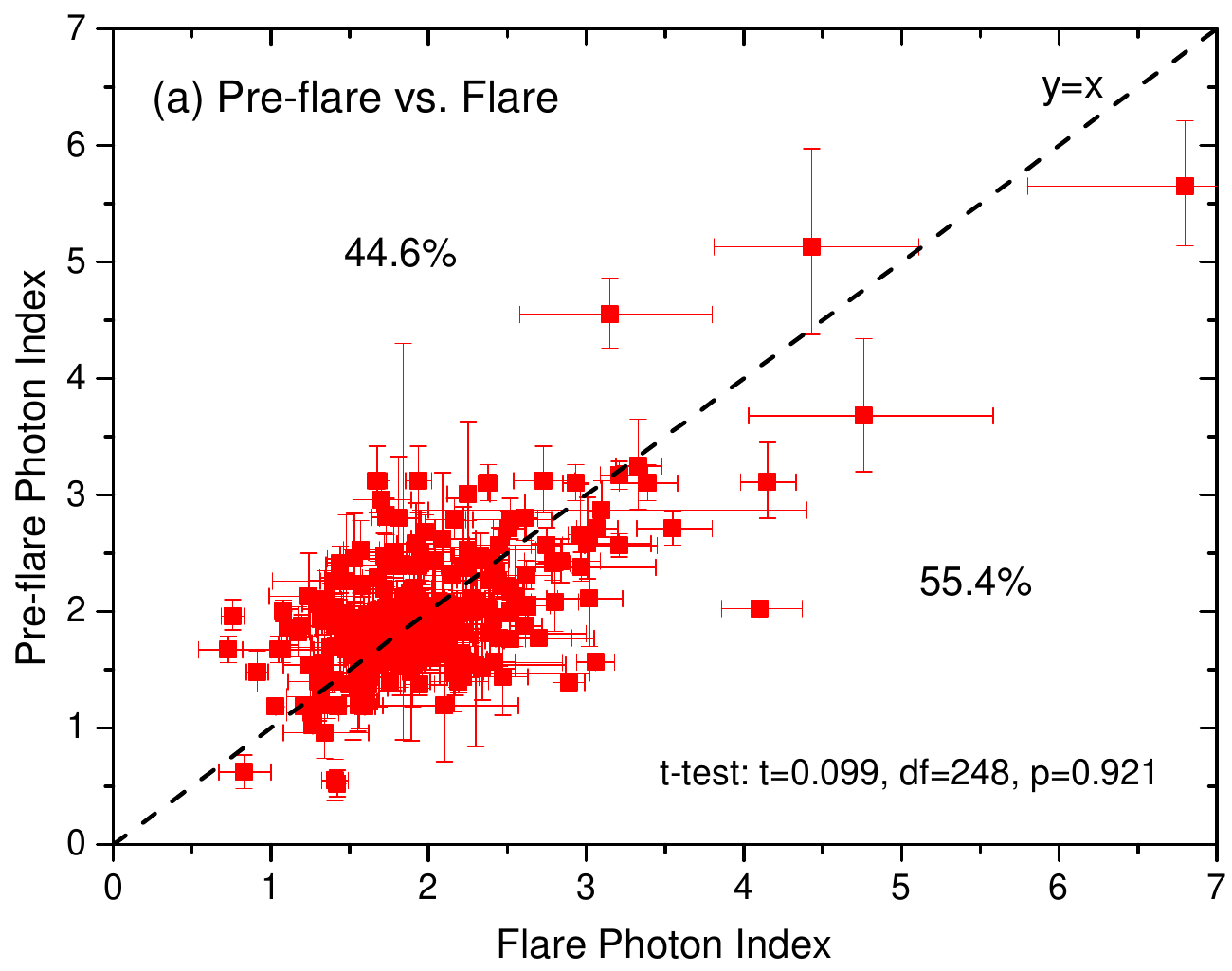}
  %\hspace{-1.5ex}
  \includegraphics[width=0.5\linewidth]{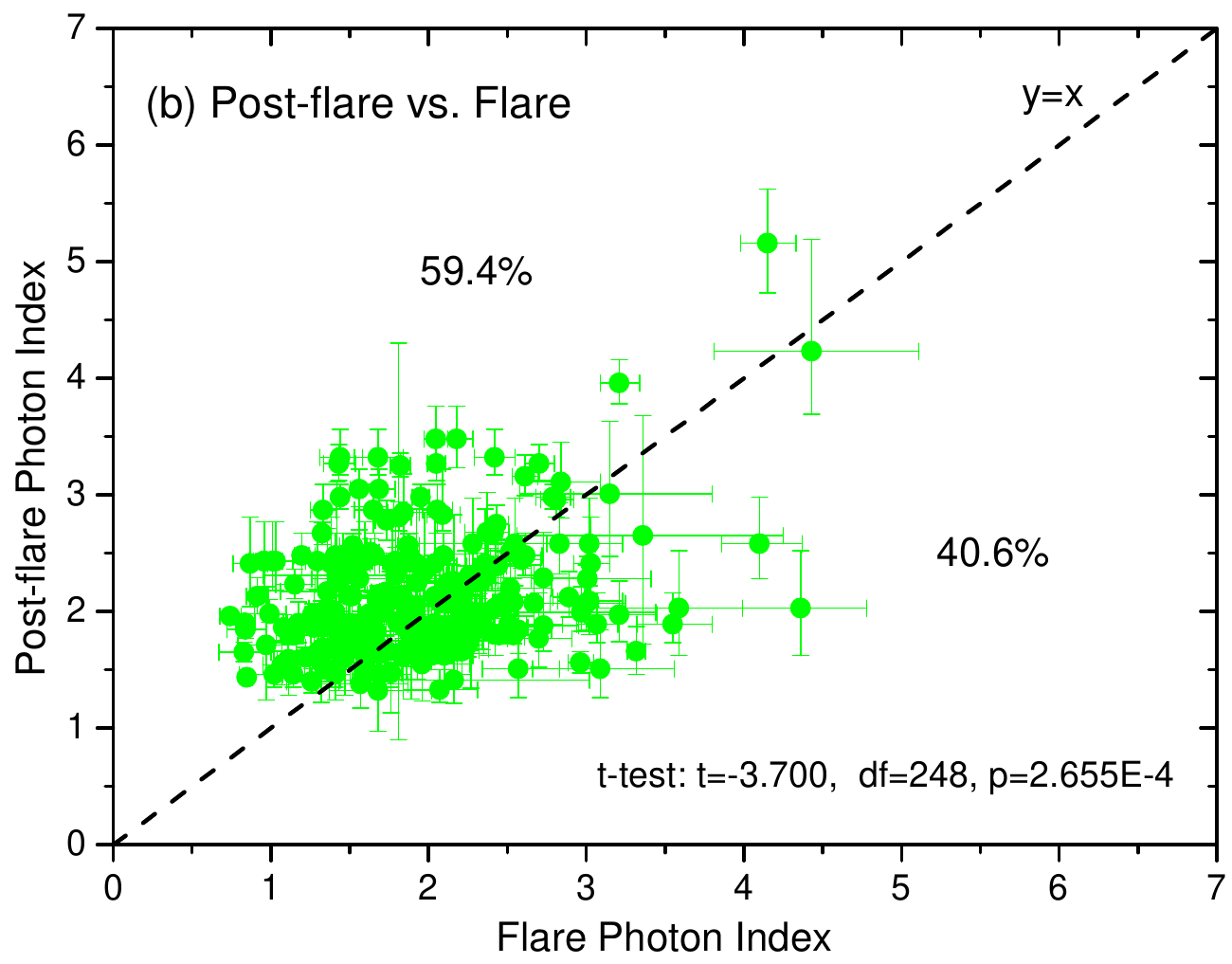}
  \includegraphics[width=0.5\linewidth]{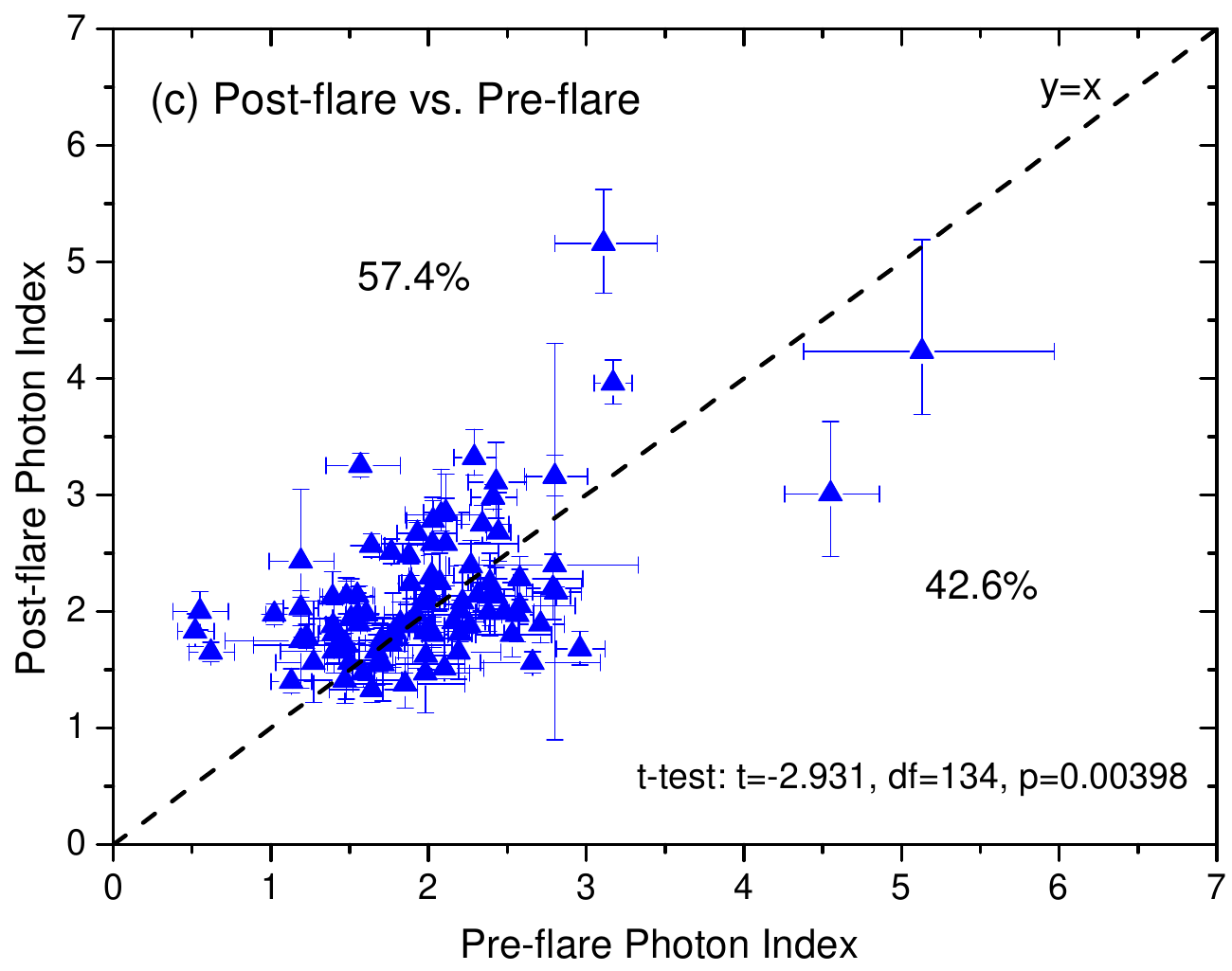}
  \caption{Comparison of photon index in the preflare, flare, and postflare phases (all-flare sample). The dashed line represents the $y = x$ identity line. (a) Flare versus preflare, (b) flare versus postflare, and (c) preflare versus postflare.}
  \label{fig:12}
\end{figure*}

Figure 12 presents one-to-one comparison of photon indices in the preflare, flare, and postflare phases for the all-flare sample. The annotations in each panel follow the same format as in Figure 10. The dashed line represents the $y=x$ identity line.
Paired t-tests were employed to compare the photon indices among the preflare, flare, and postflare phases. In the comparison between preflare phase and flare phase (Figure~12(a)), the paired t-test reveals no significant difference between the two phases on the photon index ($t = 0.099$, $df = 248$, $p = 0.921$). In the comparison between flare phase and postflare phase (Figure~12(b)), the paired t-test indicates a highly significant difference between the two phases on the photon index ($t = -3.700$, $df = 248$, $p = 2.655 \times 10^{-4}$). In the comparison between preflare phase and postflare phase (Figure~12(c)), the paired t-test shows a significant difference between the two phases on the photon index ($t = -2.931$, $df = 134$, $p = 0.00398$).

Comparing the statistical results of the spectral index distributions between the total-flare sample and the all-flare sample, both samples exhibit similar distribution characteristics. Consistent conclusions are drawn regarding the evolution of spectral indices across different flare phases: there is no significant difference in the photon index between the flare and preflare phases, while the photon index in the postflare phase is generally larger than that in the flare phase, implying an overall spectral softening. The main discrepancy between the two samples lies in the comparison between preflare and postflare phases. In the total-flare sample, there is no significant difference in the photon index between the preflare and postflare phases. In contrast, the all-flare sample shows a significant difference, where the postflare photon index is generally larger than that in the preflare phase.

In summary, the statistical analyses of both samples indicate that the spectral index during the flare phase differs significantly from that in the preflare or in the postflare phase, primarily manifesting as a hardening (decrease in photon index) during the flare phase.

\subsection{Statistical Analysis of Hydrogen Column Density in Preflare, Flare, and Postflare Phases}
To investigate the characteristics of the variations in intrinsic hydrogen column density ($N_{\rm H}$) variations during the preflare, flare, and postflare phases, we conducted a comparative statistical analysis of the $N_{\rm H}$ values in our sample. Analyses were performed separately for two samples: the total-flare sample and the all-flare sample. We employed K-S tests on the statistical histograms and paired t-tests. In this statistical analysis, we performed a combined analysis of all observational data from both WT and PC modes. We note that several spectral fits covering preflare, flare, and postflare phases only provide upper limits of $N_{\rm H}$. Although these upper limits are listed in Table~1, we do not include such spectral-fit results in our comparative statistical analysis of $N_{\rm H}$.

Figure~13 presents the distribution histograms of the hydrogen column density ($N_{\rm H}$ in units of cm$^{-2}$) in the preflare phase, flare phase, and postflare phase for the total-flare sample. The central peak positions derived from Gaussian fits to the histograms are: $\log(N_{\rm H}) = 21.71 \pm 0.04$ for flare phase, $\log(N_{\rm H}) = 21.71 \pm 0.08$ for preflare phase, and $\log(N_{\rm H}) = 21.58 \pm 0.04$ for postflare phase. The K-S test results are as follows: $p = 0.948$ for flare versus preflare, $p = 0.108$ for flare versus postflare, $p = 0.177$ for preflare versus postflare.

\begin{figure*}[!htb]
 \centering
  \includegraphics[width=0.7\linewidth]{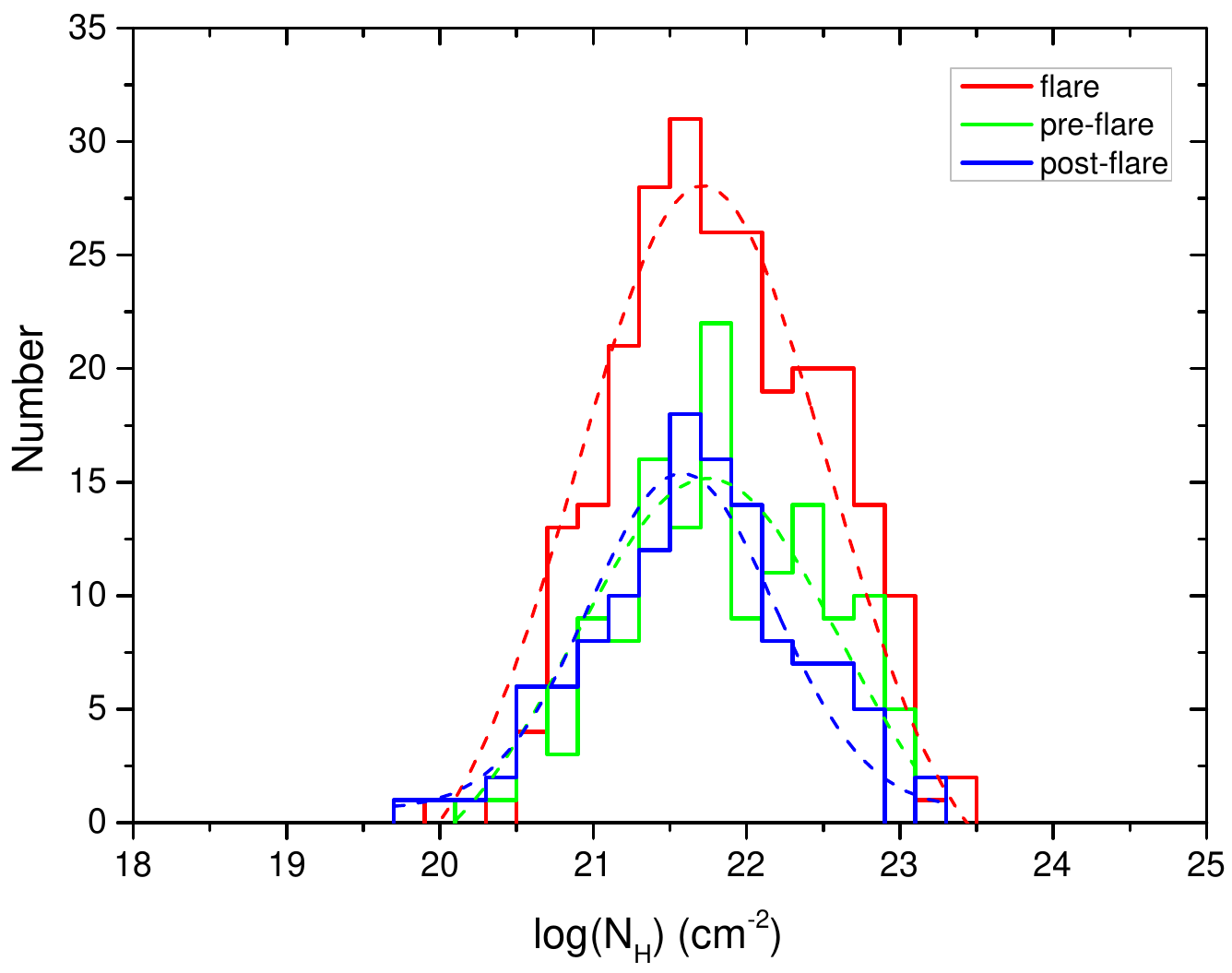}
   \caption{Histograms of $N_{\rm H}$ distributions in the preflare, flare, and postflare phase (total-flare sample). The red, green, and blue solid lines represent the data for the flare, preflare, and postflare phases, respectively, and the dashed lines are the Gaussian fits.}\label{fig13:NH_hist_total_flare}
\end{figure*}

\begin{figure*}[!htb]
  \centering
  \includegraphics[width=0.5\linewidth]{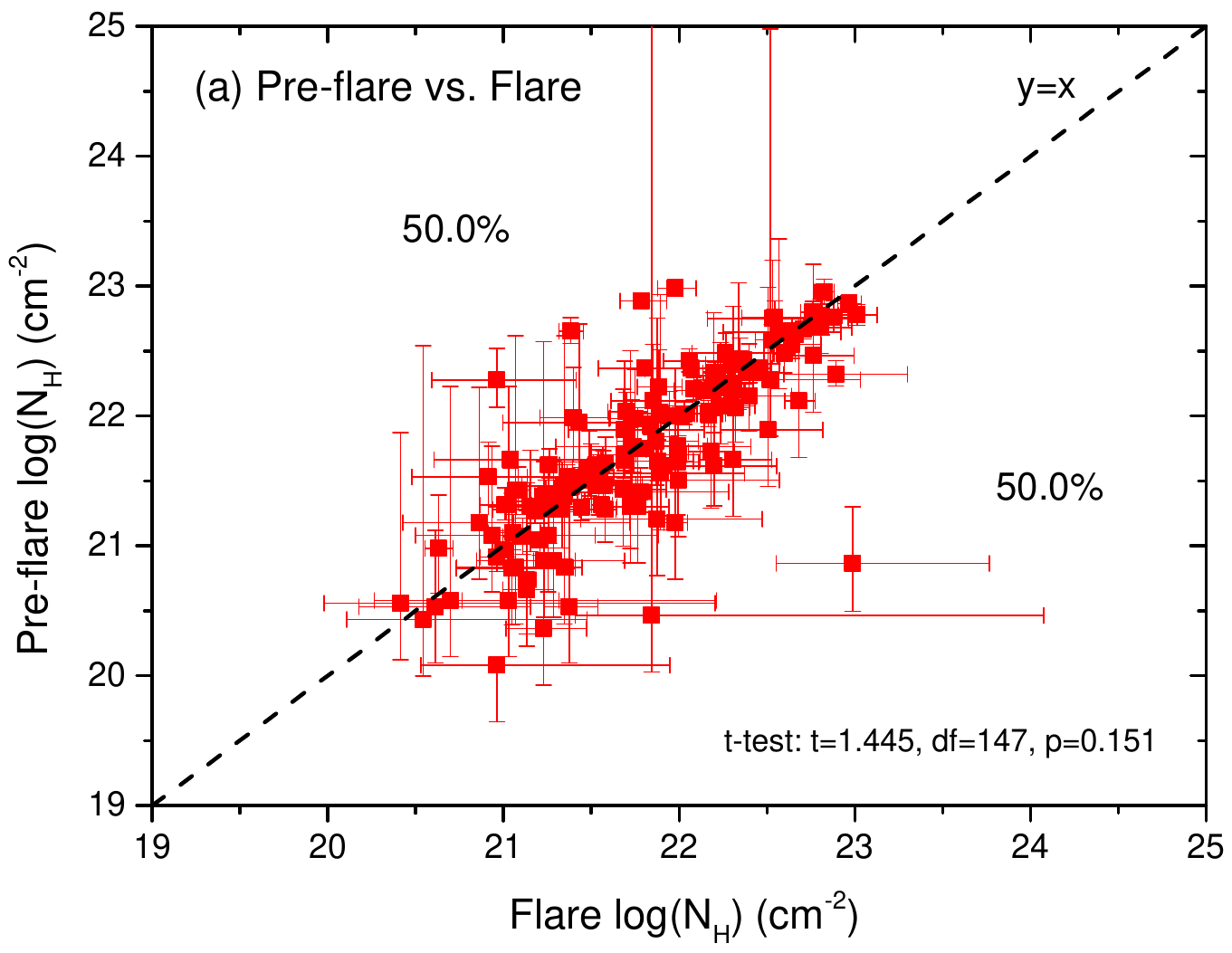}
  %\hspace{-1.5ex}
  \includegraphics[width=0.5\linewidth]{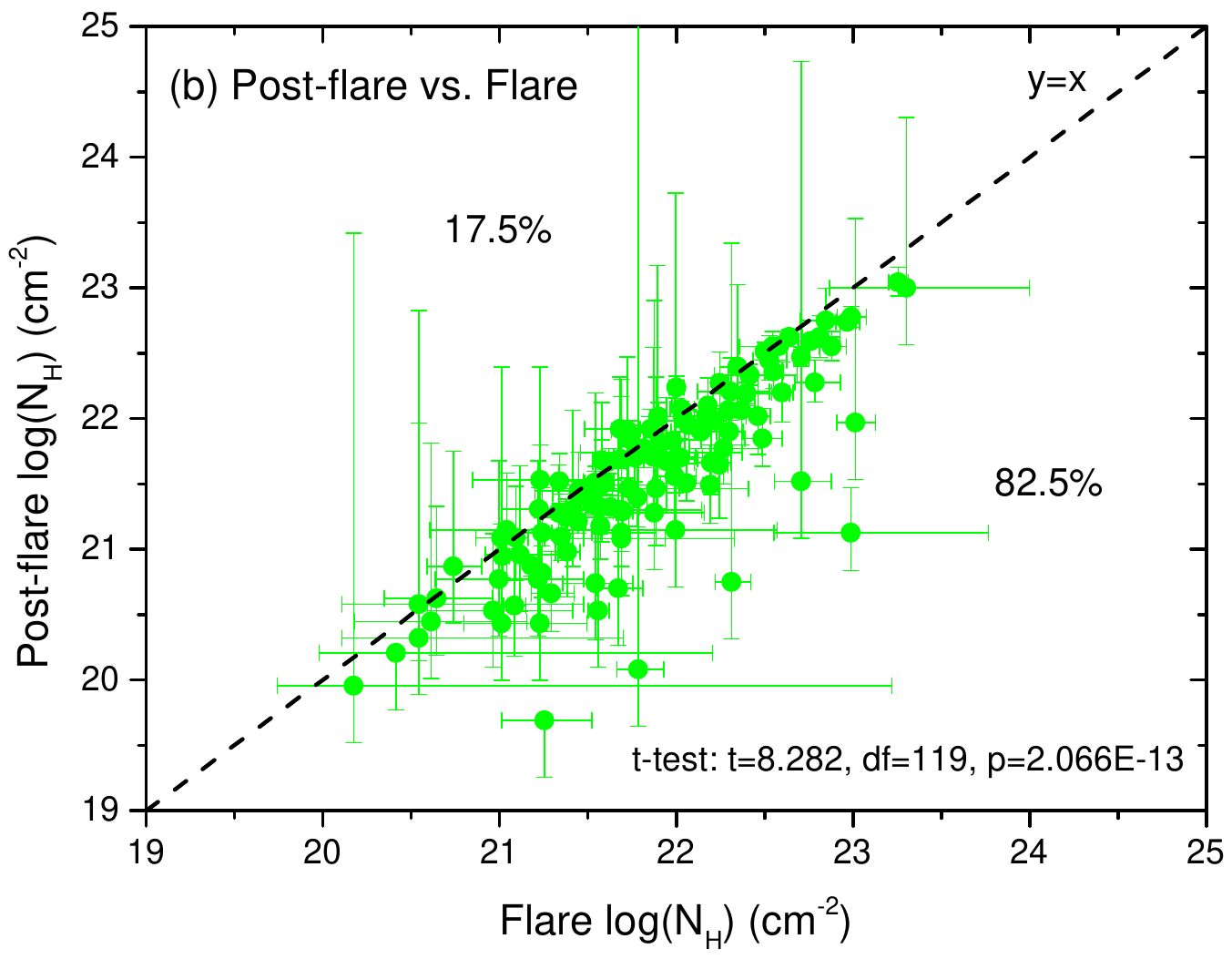}
  \includegraphics[width=0.5\linewidth]{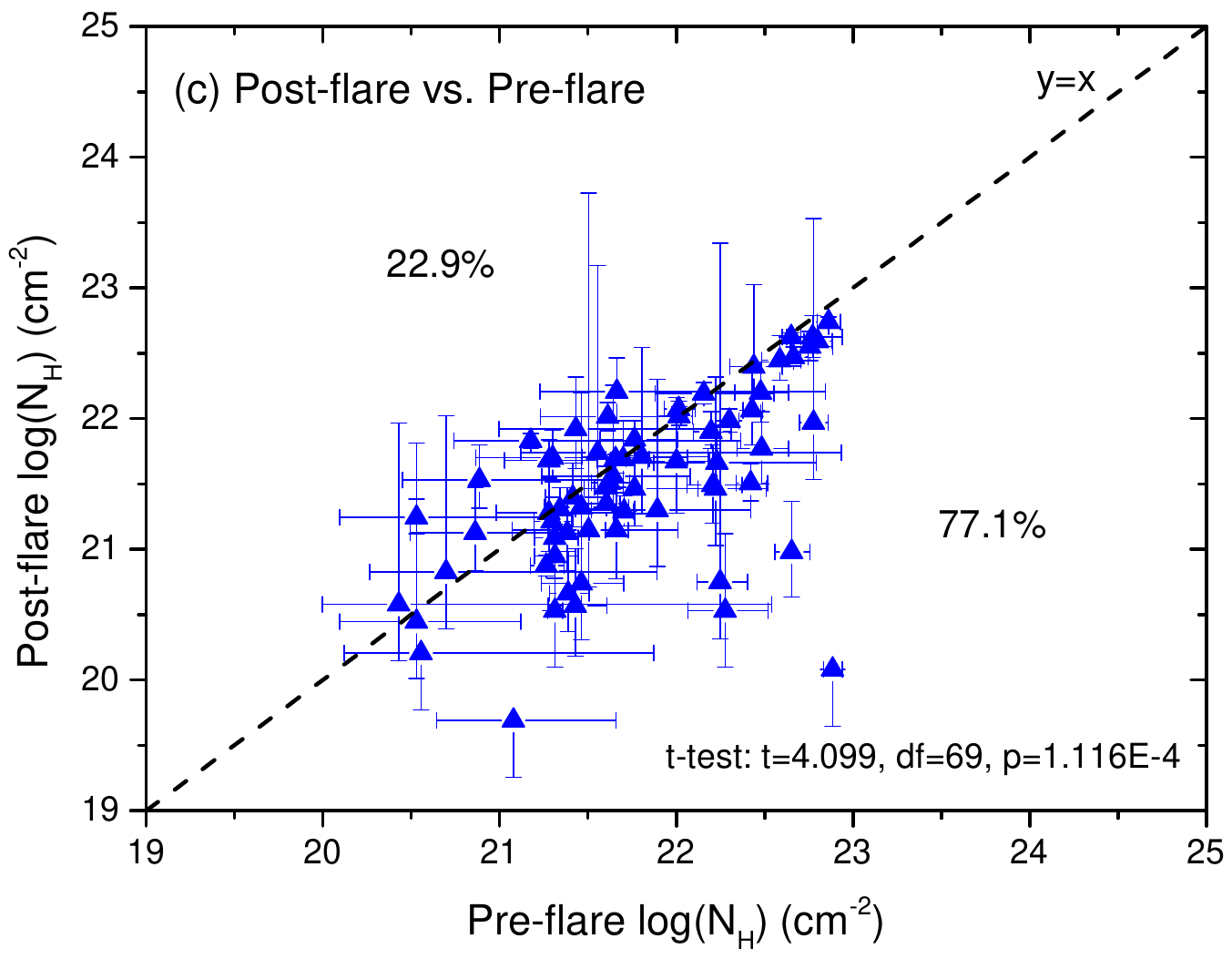}
  \caption{Comparison of $N_{\rm H}$ in the preflare, flare, and postflare phases (total-flare sample). The dashed line represents the identity line ($y = x$). (a) Flare versus preflare, (b) flare versus postflare, and (c) preflare versus postflare.}
  \label{fig14:NH_paired_total_flare}
\end{figure*}

We performed paired comparisons of the hydrogen column density \(\log(N_{\rm H})\) in three phases (preflare, flare, and postflare phases), and the results are shown in Figure 14. In the comparison between preflare phase and flare phase (Figure~14(a)), the paired t-test reveals no significant difference between the two phases on $N_{\rm H}$ ($t = 1.445$, $df = 147$, $p = 0.151$). In the comparison between flare phase and postflare phase (Figure~14(b)), the paired t-test indicates an extremely significant difference between the two phases on $N_{\rm H}$ ($t = 8.282$, df $= 119$, $p = 2.066 \times 10^{-13}$). In the comparison between preflare phase and postflare phase (Figure~14(c)), the paired t-test shows a significant difference between the two phases on $N_{\rm H}$ ($t = 4.099$, $df = 69$, $p = 1.116 \times 10^{-4}$).

\begin{figure*}[!htb]
 \centering
  \includegraphics[width=0.7\linewidth]{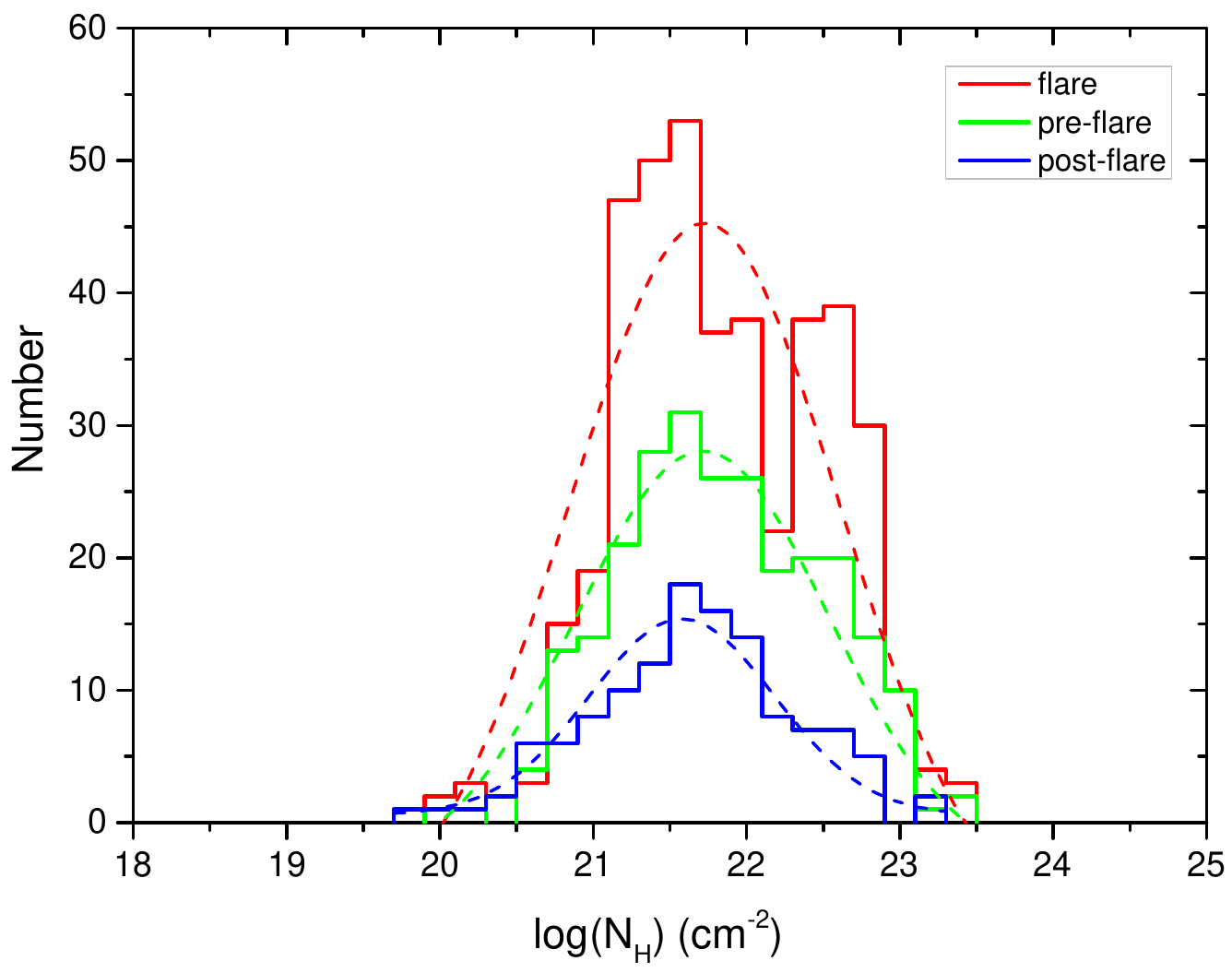}
  \caption{Histograms of $N_{\rm H}$ distributions in the preflare, flare, and postflare phase (all-flare sample). The red, green, and blue solid lines represent the data for the flare, preflare, and postflare phases, respectively, and the dashed lines are the Gaussian fits.}\label{fig15:NH_hist_all_flare}
 \end{figure*}

Figure 15 presents the histograms and Gaussian fits of the hydrogen column density for all-flare sample. The central peaks of the Gaussian fits are located at log($N_{\rm H}$) = $21.72 \pm 0.07$ for the flare, log($N_{\rm H}$) = $21.71 \pm 0.04$ for the preflare, and log($N_{\rm H}$) = $21.58 \pm 0.04$ for the postflare. The K-S test results are as follows: $p = 0.970$ for flare versus preflare, $p = 0.032$ for flare versus postflare, $p = 0.108$ for preflare verus postflare.

\begin{figure*}[!htb]
  \centering
  \includegraphics[width=0.5\linewidth]{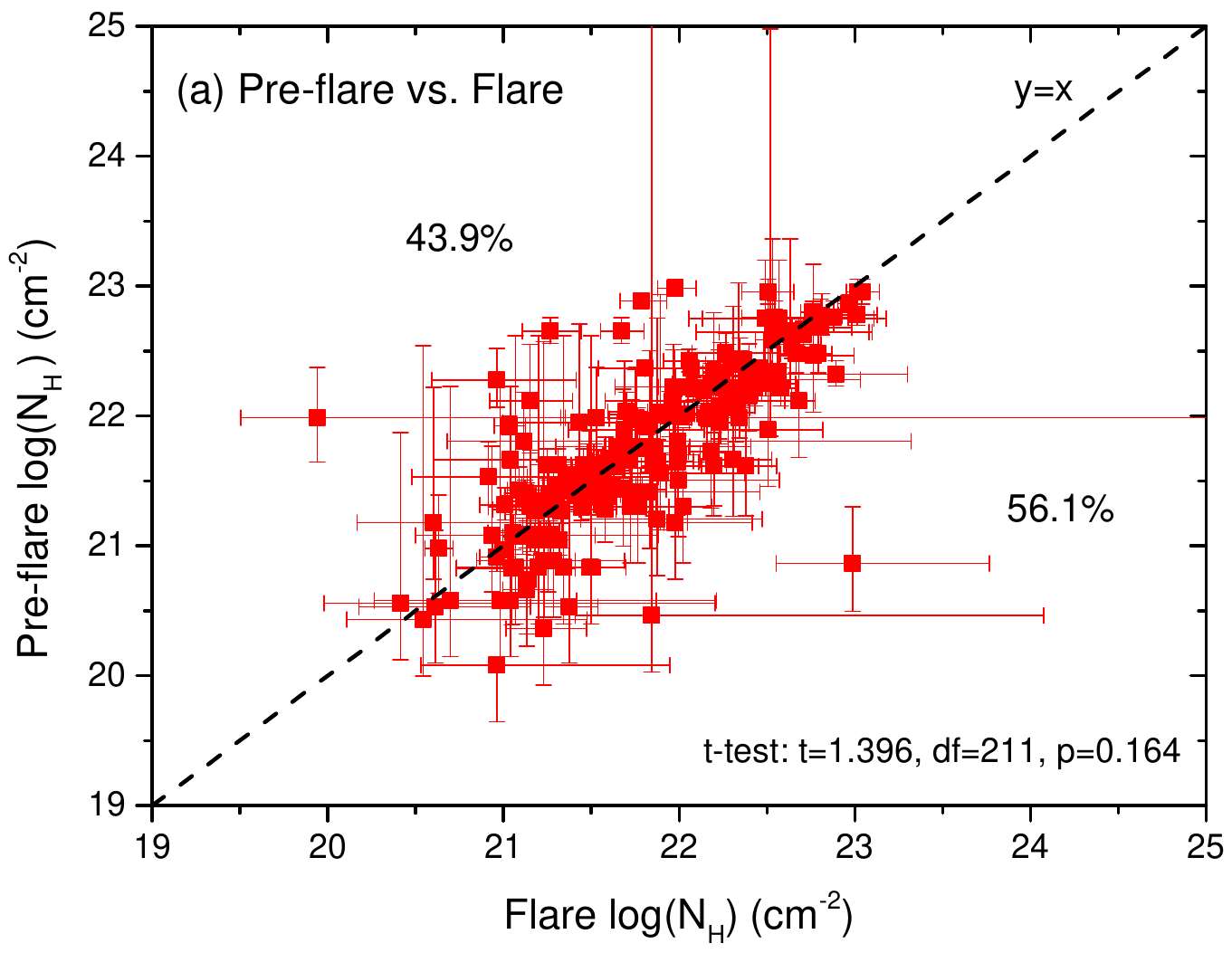}
  \includegraphics[width=0.5\linewidth]{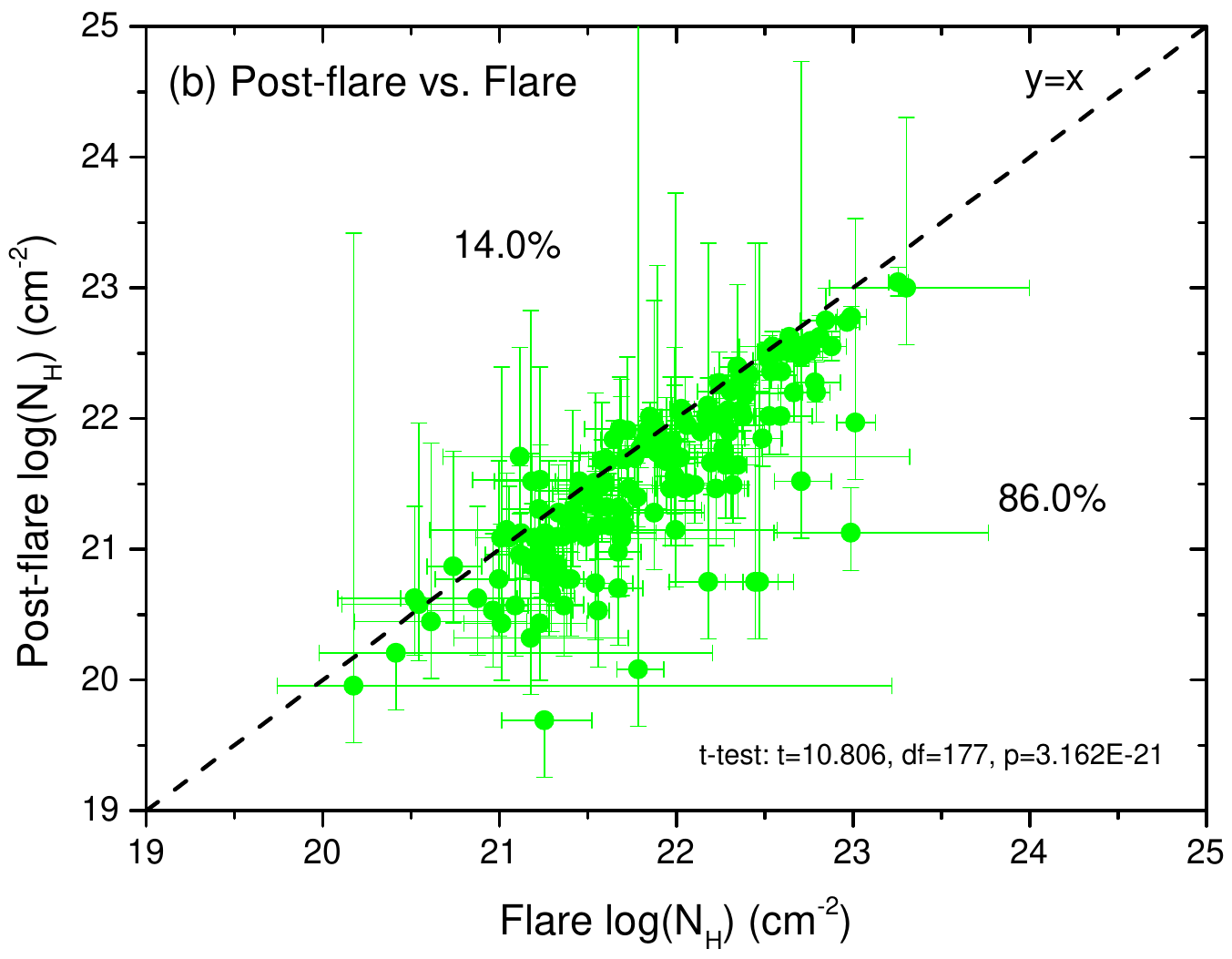}
  \includegraphics[width=0.5\linewidth]{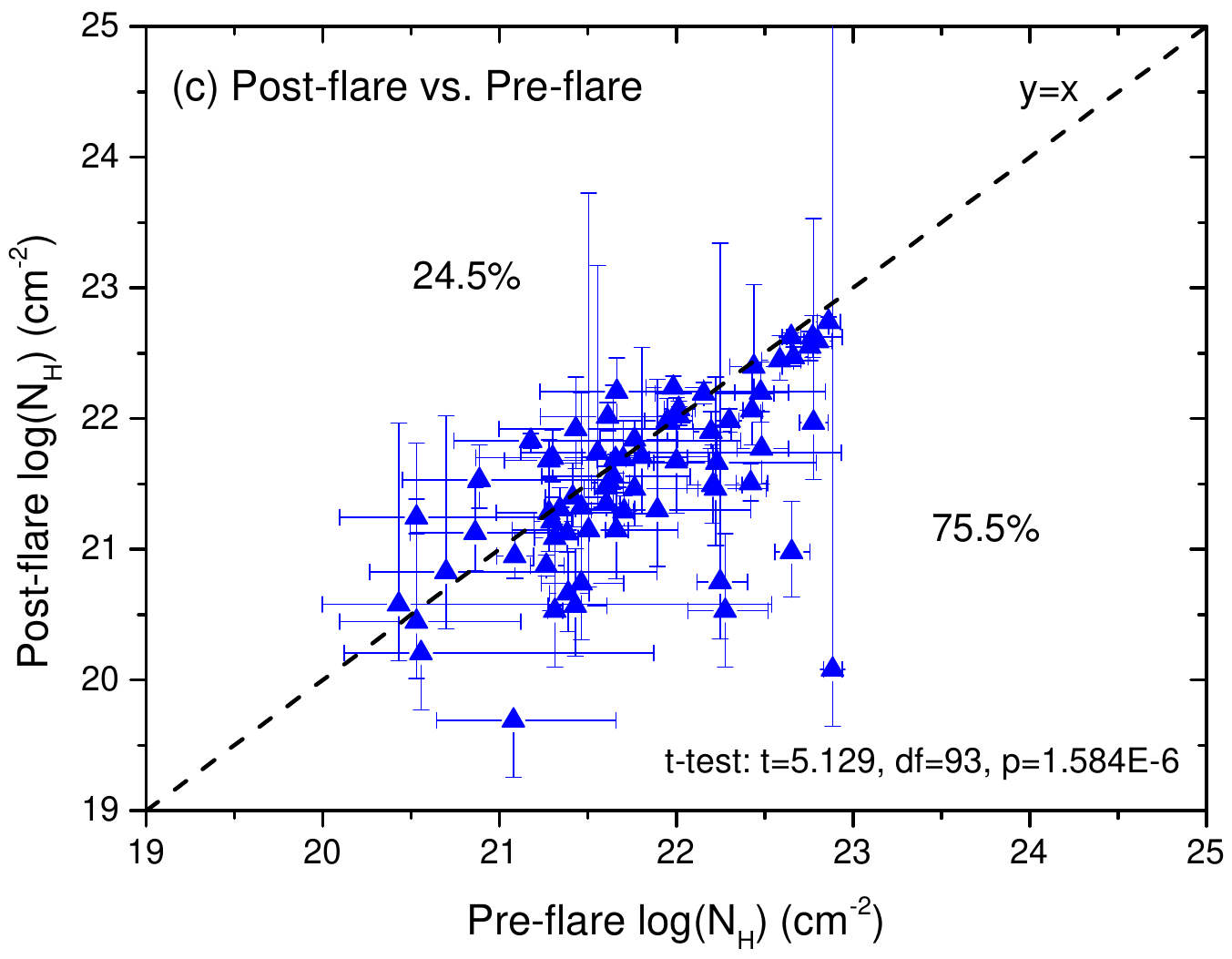}
  \caption{Comparison of $N_{\rm H}$ in the preflare, flare, and postflare phases (all-flare sample). The dashed line represents the identity line ($y = x$). (a) Flare versus preflare, (b) flare versus postflare, and (c) preflare versus postflare.}
  \label{fig16:NH_paired_all_flare}
\end{figure*}

Figure 16 presents a paired comparison of the hydrogen column density ($N_{\rm H}$) in the preflare and the postflare phases for the all-flare sample. A statistical analysis was conducted on $N_{\rm H}$ measured in three phases: preflare, flare, and postflare. In the comparison between preflare phase and flare phase (Figure~16(a)), the paired t-test reveals no significant difference between the two phases on $N_{\rm H}$ ($t = 1.396$, $df = 211$, $p = 0.164$). In the comparison between flare phase and postflare phase (Figure~16(b)), the paired t-test indicates a highly significant difference between the two phases on $N_{\rm H}$ ($t = 10.806$, $df = 177$, $p = 3.162 \times 10^{-21}$). In the comparison between preflare phase and postflare phase (Figure~16(c)), the paired t-test shows a highly significant difference between the two phases on $N_{\rm H}$ ($t = 5.129$, $df = 93$, $p = 1.584 \times 10^{-6}$).

From the K-S tests of the $N_{\rm H}$ histograms shown in Figures 13 and 15, we obtained the following results. For the total-flare sample, the $p$-values for the three comparisons are: $p=0.948$ (flare versus preflare), $p=0.108$ (flare versus postflare), and $p=0.177$ (preflare versus postflare). For the all-flare sample, the corresponding $p$-values are $0.970$ (flare versus preflare), $0.032$ (flare versus postflare), and $0.108$ (preflare versus postflare). In the all-flare sample, the comparison between the flare and postflare phases yields $p=0.032$, which is below the $0.05$ significance level, indicating a statistically significant difference between the two distributions.

The paired-sample t-tests shown in Figures 14 and 16 further confirm a significant difference between the $N_{\rm H}$ values derived for the postflare phase and those for the preflare and flare phases. For the flare versus postflare comparison (Figures 14(b) and 16(b)), the total-flare sample gives $t=8.282$, $df = 119$, $p=2.066\times10^{-13}$, and the all-flare sample gives $t=10.806$, $df = 177$, $p=3.162\times10^{-21}$. These results demonstrate that the $N_{\rm H}$ in the postflare phase is significantly lower than that in the flare phase. For the preflare versus postflare comparison (Figures 14(c) and 16(c)), the total-flare sample yields $t=4.099$, $df = 69$, $p=1.116\times10^{-4}$, and the all-flare sample yields $t=5.129$, $df = 93$, $p=1.584\times10^{-6}$, further confirming that the $N_{\rm H}$ in the postflare phase is also significantly lower than that in the preflare phase. We also note that most individual $N_{\rm H}$ data points in Figures 14 and 16 exhibit relatively large statistical uncertainties. Thus, the statistical trend reported in this work requires further verification with larger GRB samples in future studies.

We emphasize again that a fraction of our spectral fits do not provide strong constraints on $N_{\rm H}$, yielding either only upper limits or formally constrained values with large uncertainties. We have quantified the fraction of such cases across the three phases: 5.0\% ($26/521$) of flare spectral fits, 11.7\% ($23/197$) of preflare spectral fits, and 24.0\% ($47/196$) of postflare spectral fits yield only upper limits on $N_{\rm H}$. All upper-limit values are listed in Table~1, but we have excluded all such data points from our statistical analysis of $N_{\rm H}$ to ensure the reliability of our conclusions.

To further investigate the impact of photon counts on the constraint of $N_{\rm H}$, we have selected three representative examples of $N_{\rm H}$ constraints from our spectral fitting results. The three examples correspond to: (1) well-constrained $N_{\rm H}$ with small uncertainties; (2) constrained $N_{\rm H}$ with relatively large uncertainties; and (3) only upper limits on $N_{\rm H}$. The spectral-fit plots and residual distributions for these three examples are shown in Figures A1, A2, and A3, respectively (see the Appendix), and the corresponding best-fit parameters are summarized in Table A1. As demonstrated by these three examples, the quality of the $N_{\rm H}$ constraint depends critically on the photon counts, particularly in the soft 0.3--1.0\,keV band.

Based on the above statistical results, we conclude that the hydrogen column density in the flare phase shows no significant difference from that in the preflare phase, implying that the flare activity does not significantly modify the absorption environment. However, $N_{\rm H}$ in the postflare phase is significantly lower than that in both the preflare and flare phases, which may be related to the evolution of the radiative environment following the event.

Based on the above comparative statistical analysis of two key parameters---the photon index and $N_{\rm H}$ obtained from spectral fitting of X-ray flares and their preflare and postflare phases---the statistical results indicate the following: (1) the photon index during X-ray flares differs significantly from that in the preflare and postflare phases, exhibiting a clear trend toward hardening (i.e., a harder spectrum) during the flare, while no significant difference is observed between the preflare and postflare photon indices; (2) the hydrogen column density shows no significant change between the preflare and flare phases, but significant differences are found between the postflare and flare phases, as well as between the postflare and preflare phases, manifested as a significantly lower $N_{\rm H}$ after the flare than during both the preflare and flare periods.

\subsection{The Photon Index Distributions of X-ray Flares for XRFs, XRRs, and C-GRBs}
\begin{figure*}[!htb]
  \centering
  \includegraphics[width=0.7\linewidth]{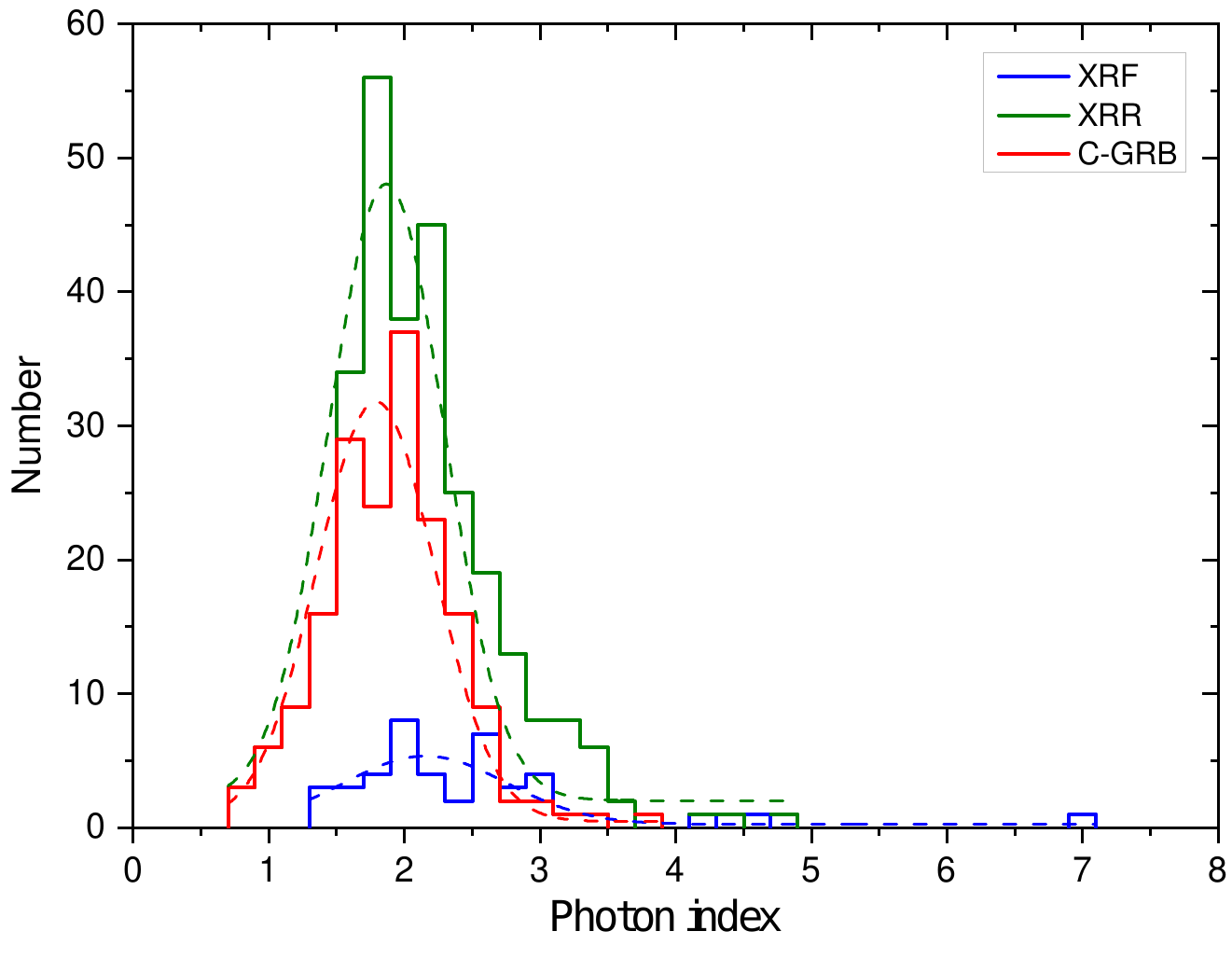}
  \caption{Histogram of X-ray flare photon index distributions for three GRB subclasses: XRF (solid blue line), XRR (solid green line), and C-GRB (solid red line). The dashed lines are Gaussian fits.} \label{fig17:PI_XRR_XRF_CGRB}
 \end{figure*}

We performed a statistical analysis of the X-ray flare photon indices for three GRB subclasses (XRF, XRR, and C-GRB), with results presented in Figure 17. Gaussian fitting was applied to the photon index distributions of each subclass, and K-S tests were conducted to compare these distributions. The central values of the Gaussian fits are as follows: $x_c = 2.15 \pm 0.11$ for XRF, $x_c = 1.87 \pm 0.03$ for XRR, and $x_c = 1.79 \pm 0.03$ for C-GRB. The K-S test results show: $p = 0.007$ for XRF versus XRR, $p = 3.35 \times 10^{-5}$ for XRF versus C-GRB, and $p = 0.007$ for XRR versus C-GRB, indicating significant differences in the photon index distributions of X-ray flares among the three subclasses. In particular, the photon indices of X-ray flares in XRF and C-GRB exhibit a marked difference, with those in XRF being generally the softer.

\subsection{Statistical Analysis of the Photon Indices of X-ray Flares and Prompt Emission }
\cite{Chincarini2010} suggested that the photon indices of X-ray flares are generally softer than those of prompt emission. To further investigate the difference in photon indices between prompt emission and X-ray flares, we conducted a comparative analysis of the photon index relationship between X-ray flares and prompt emission in our sample.

\subsubsection{Relationship between the Photon Index of X-ray Flares and the Power-law Photon Index of Prompt Emission}
Figure~18 presents the histograms of the photon index distributions for both the X-ray flares and the prompt emission (fitted with a power-law model) for the full GRB sample as well as for the three subclasses (XRF, XRR, and C-GRB). It can be seen from the figure that the spectral indices of X-ray flares and prompt emission show no significant overlap in distribution. Gaussian fits to the histograms for the full sample yield a peak value of \(x_c = 1.62 \pm 0.03\) for the prompt emission photon index and \(x_c = 1.85 \pm 0.02\) for the X-ray flare photon index. A K-S test reveals a significant difference between the two distributions, with \(p \approx 4.89 \times 10^{-22}\), indicating that the photon indices of X-ray flares are systematically larger than those of the prompt emission.
\begin{figure*}[!htb]
  \centering
  \includegraphics[height=5cm,width=7cm]{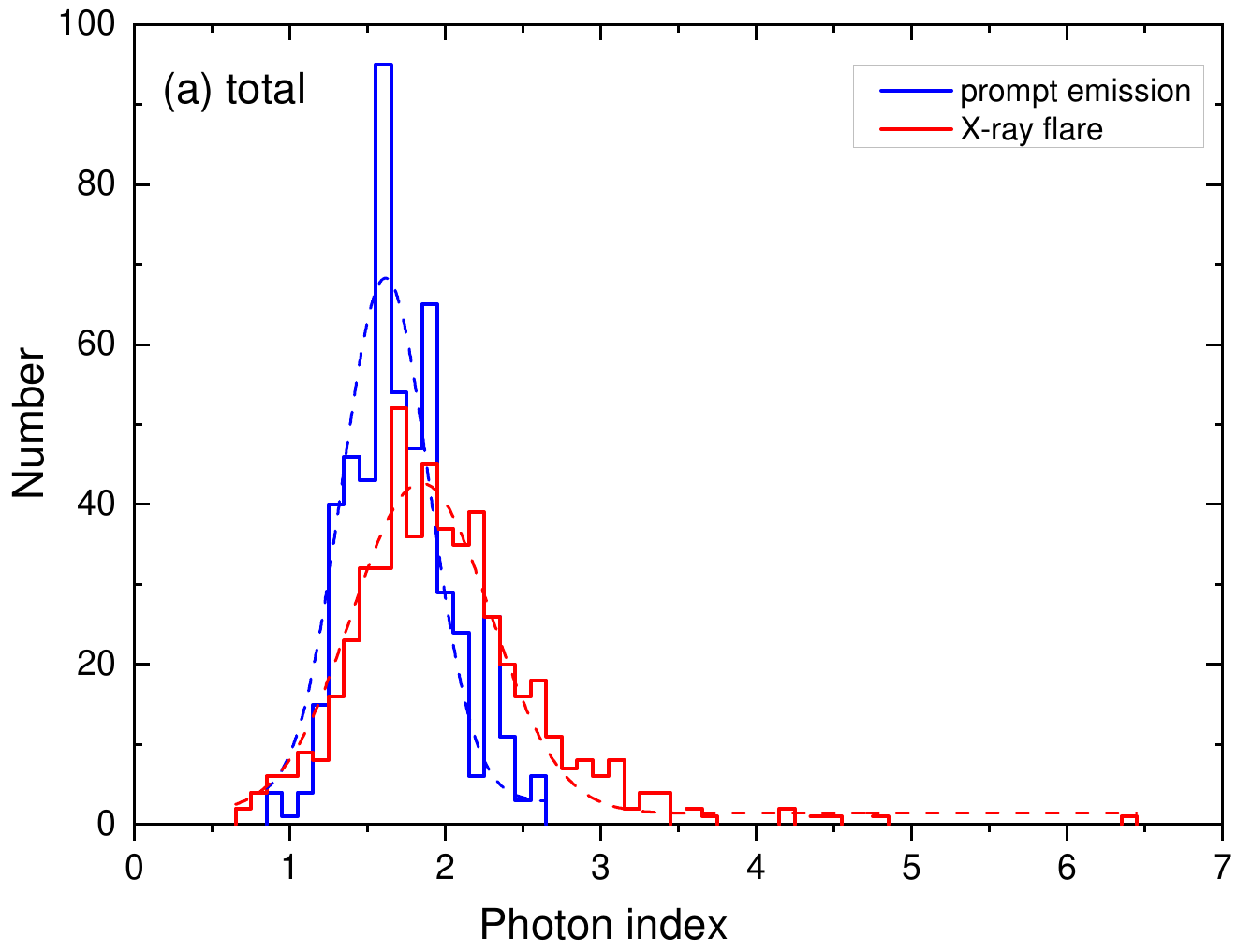}
  \includegraphics[height=5cm,width=7cm]{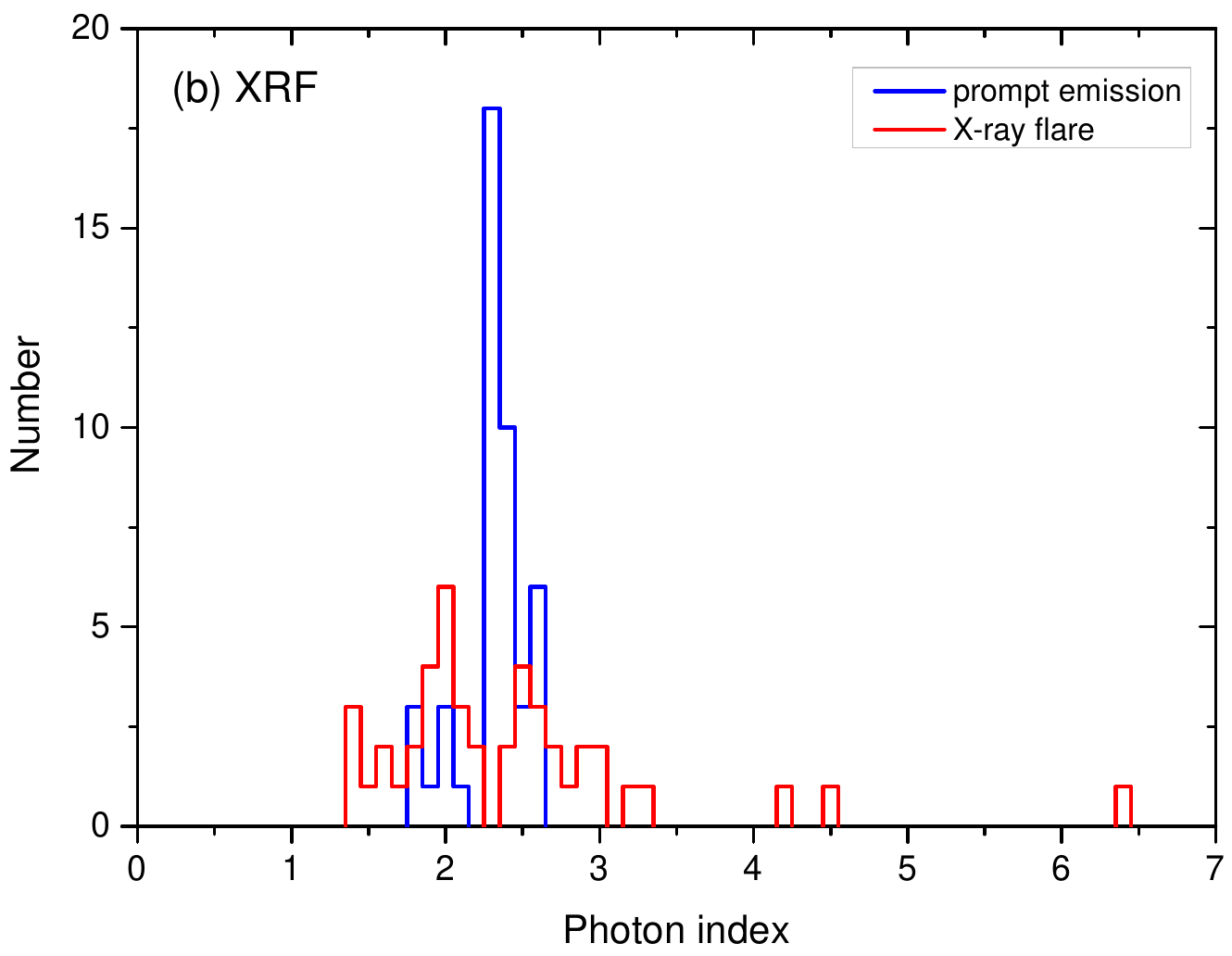}
  \includegraphics[height=5cm,width=7cm]{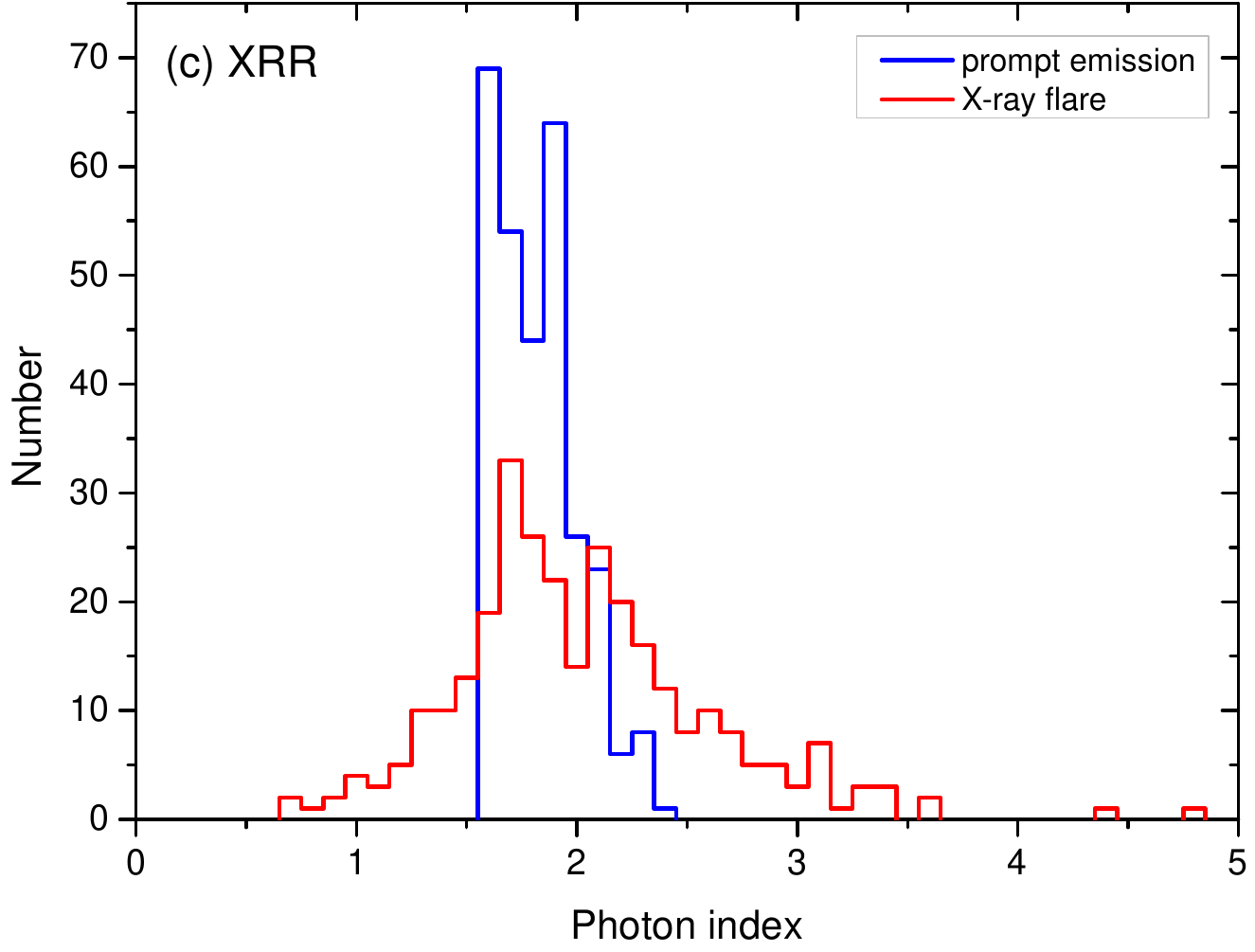}
  \includegraphics[height=5cm,width=7cm]{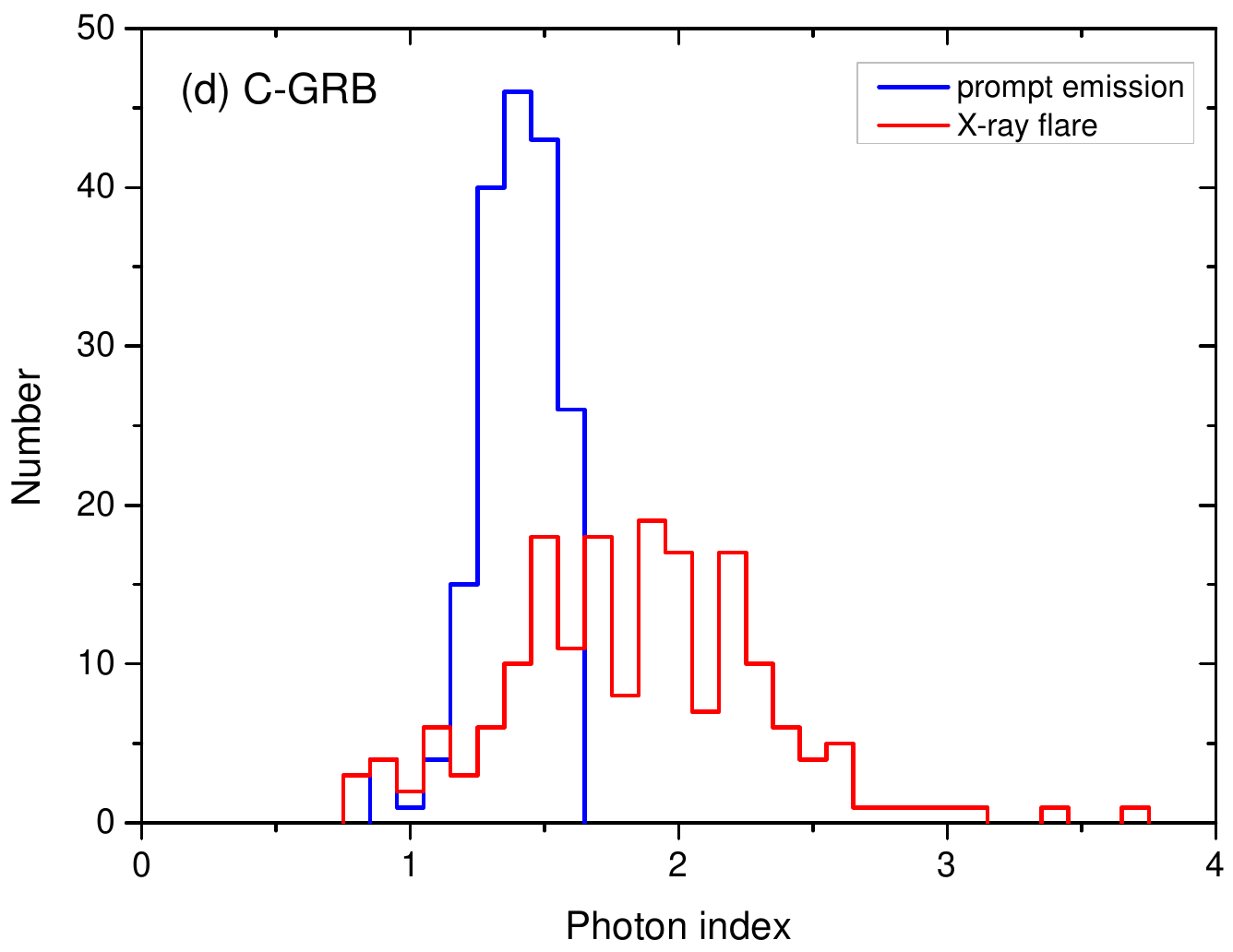}
  \caption{Histograms of the photon index distribution for X-ray flares (red solid line) and prompt emission (power-law model, blue solid line). (a) Distribution for the full GRB sample; (b) XRF subclass, (c) XRR subclass, (d) C-GRB subclass.}\label{fig18:powerlaw_hist}
  \end{figure*}

\begin{figure*}[!htb]
  \centering
  \includegraphics[width = 0.5\linewidth]{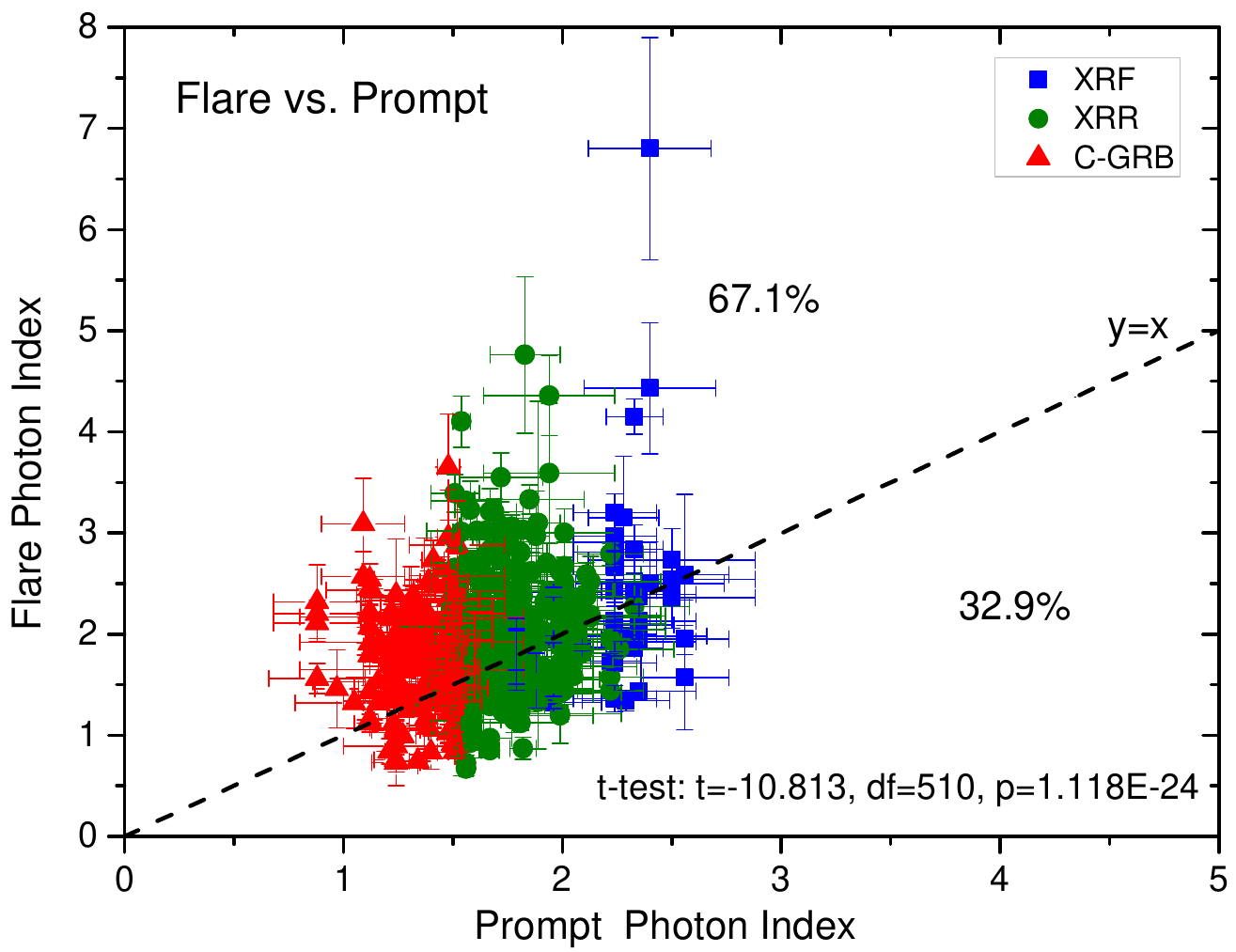}
  \caption{Comparison of the photon indices between X-ray flares and the prompt emission (power-law model). The black dashed line denotes $y=x$.}\label{fig19}
 \end{figure*}

In Figure~19, we present a comparative analysis of the photon indices between X-ray flares and prompt emission for the GRB sample (including XRF, XRR, and C-GRB). The dashed line in the figure represents the equality of the two quantities. Overall, the photon index of X-ray flares is larger than that of the prompt emission in 67.1\% of the sources. The spectrum of X-ray flares is softer than that of the prompt emission. A two-sample t-test reveals a significant difference between the two distributions ($t = -10.813$, $df = 510$, $p = 1.118 \times 10^{-24}$), indicating that the spectra of X-ray flares are systematically softer than those of the prompt emission.

\subsubsection{Relationship between the Photon Index of X-ray Flares and the Cutoff Power-law Photon Index of Prompt Emission}

\begin{figure*}[!htb]
  \centering
  \includegraphics[height=5cm,width=7cm]{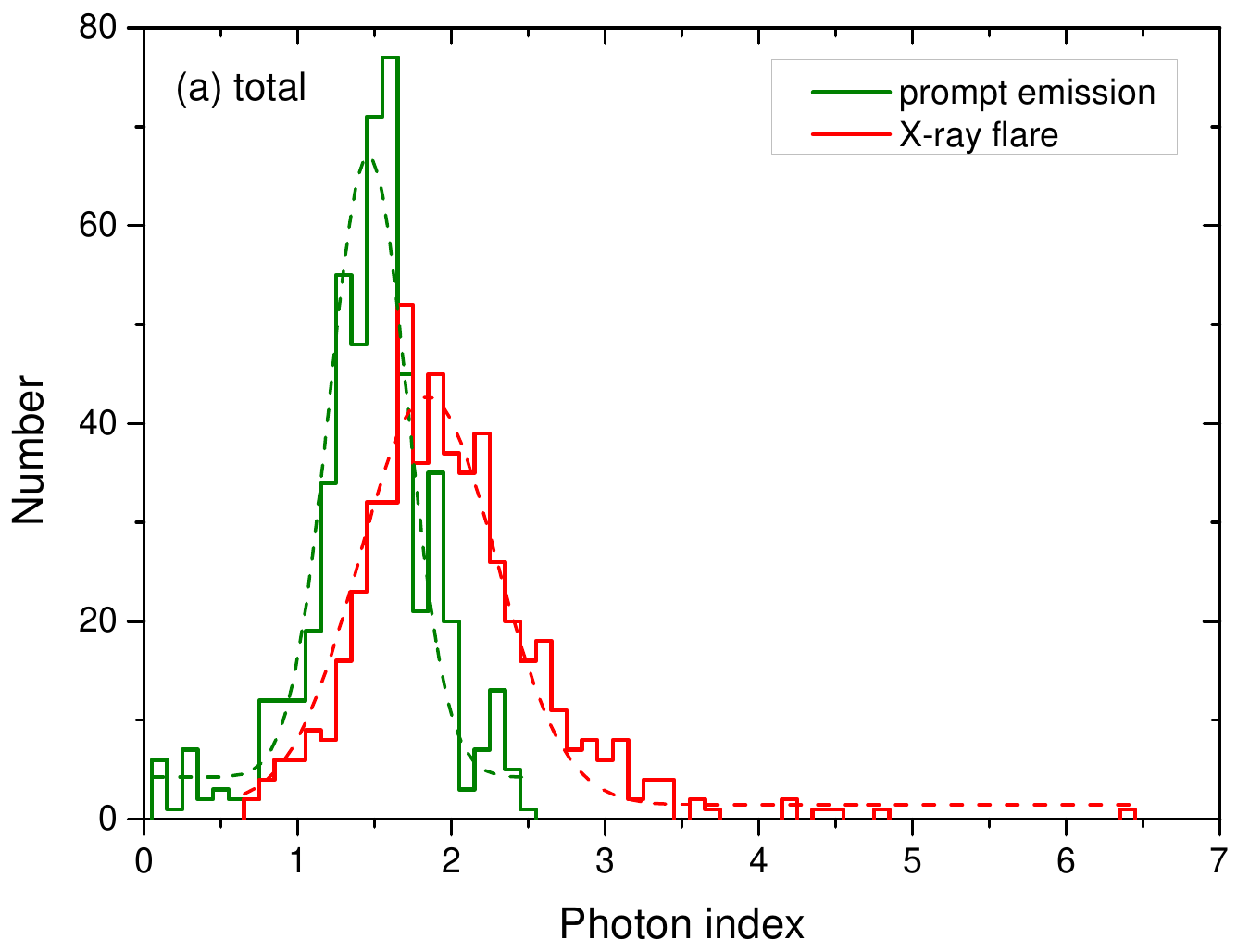}
  \includegraphics[height=5cm,width=7cm]{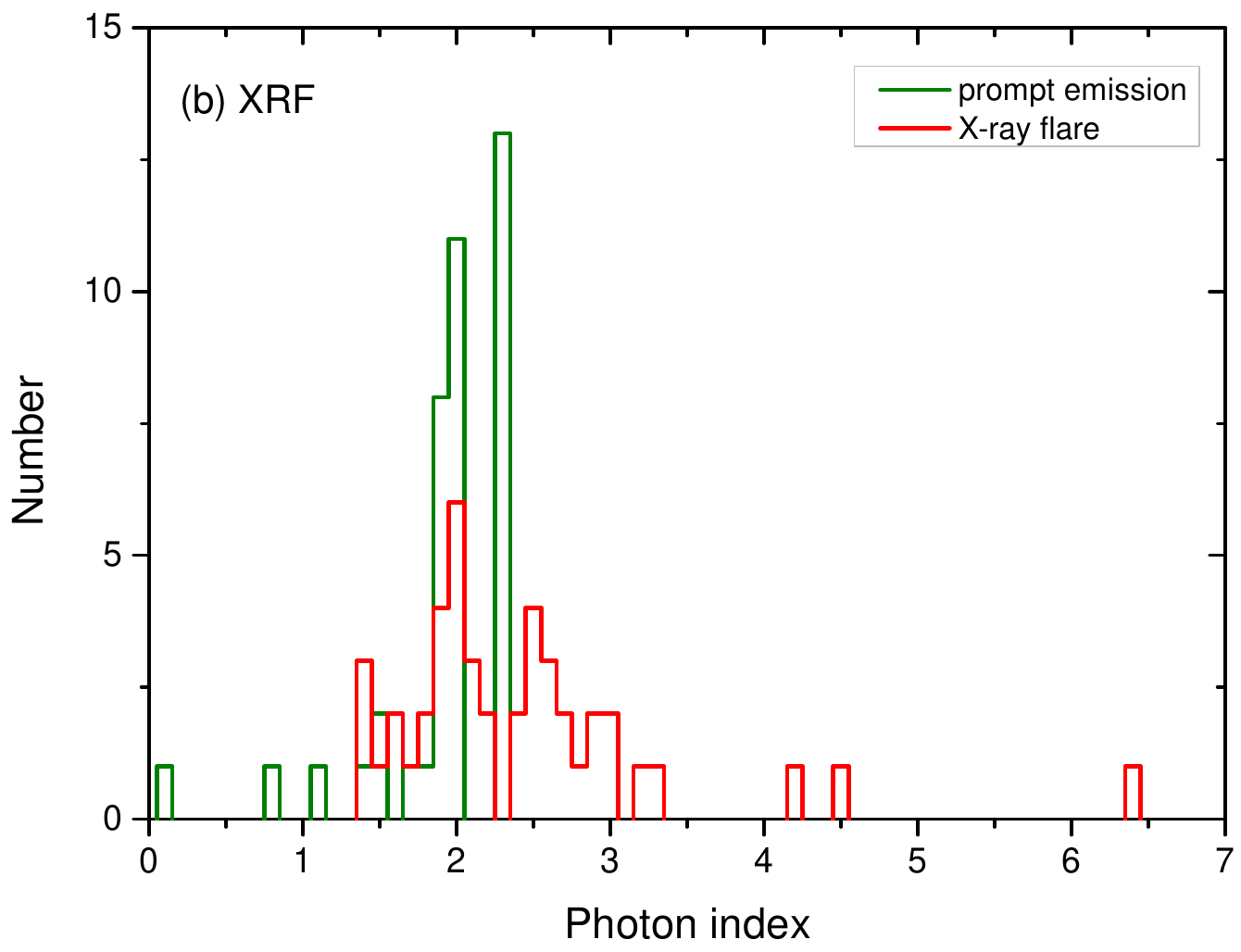}
  \includegraphics[height=5cm,width=7cm]{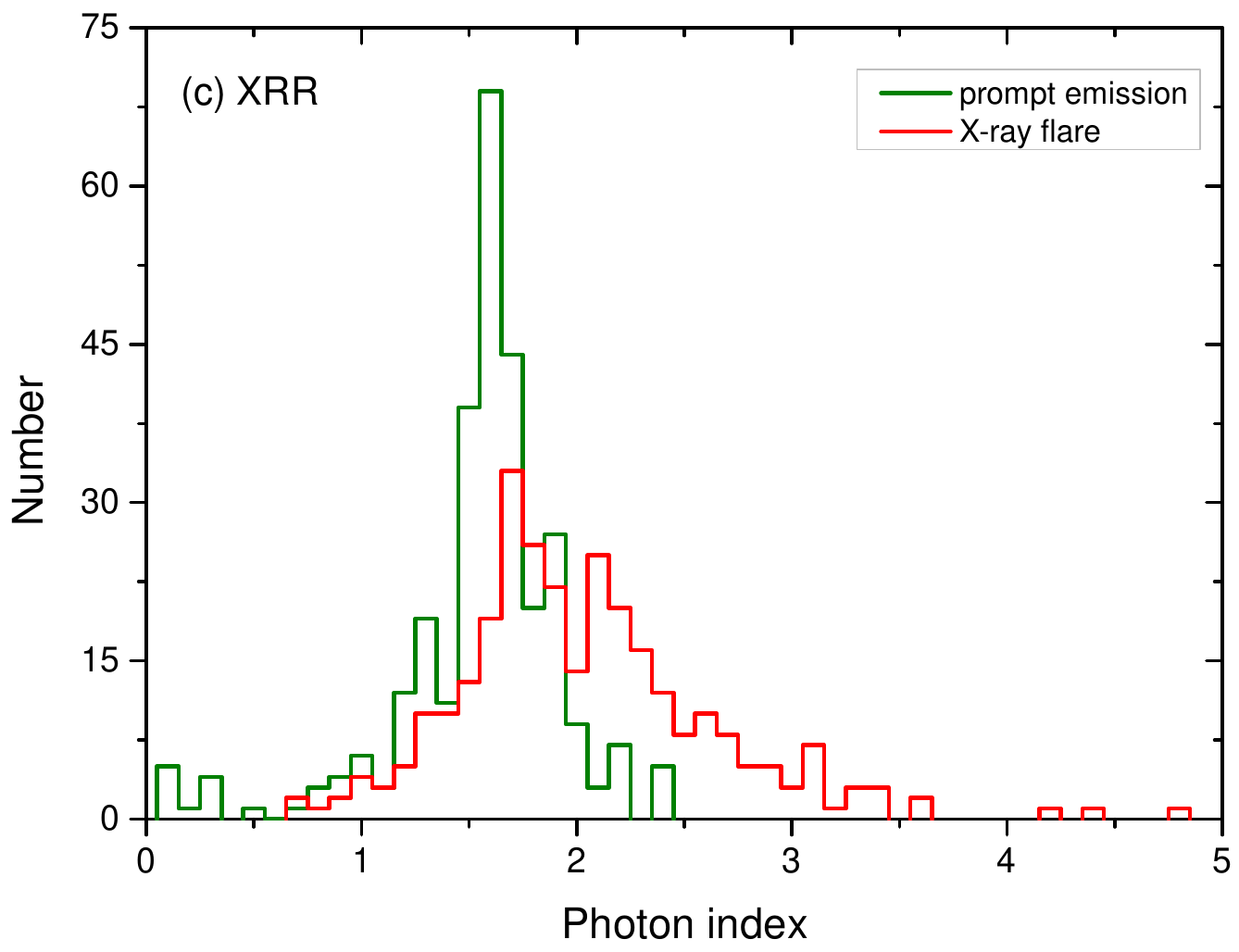}
  \includegraphics[height=5cm,width=7cm]{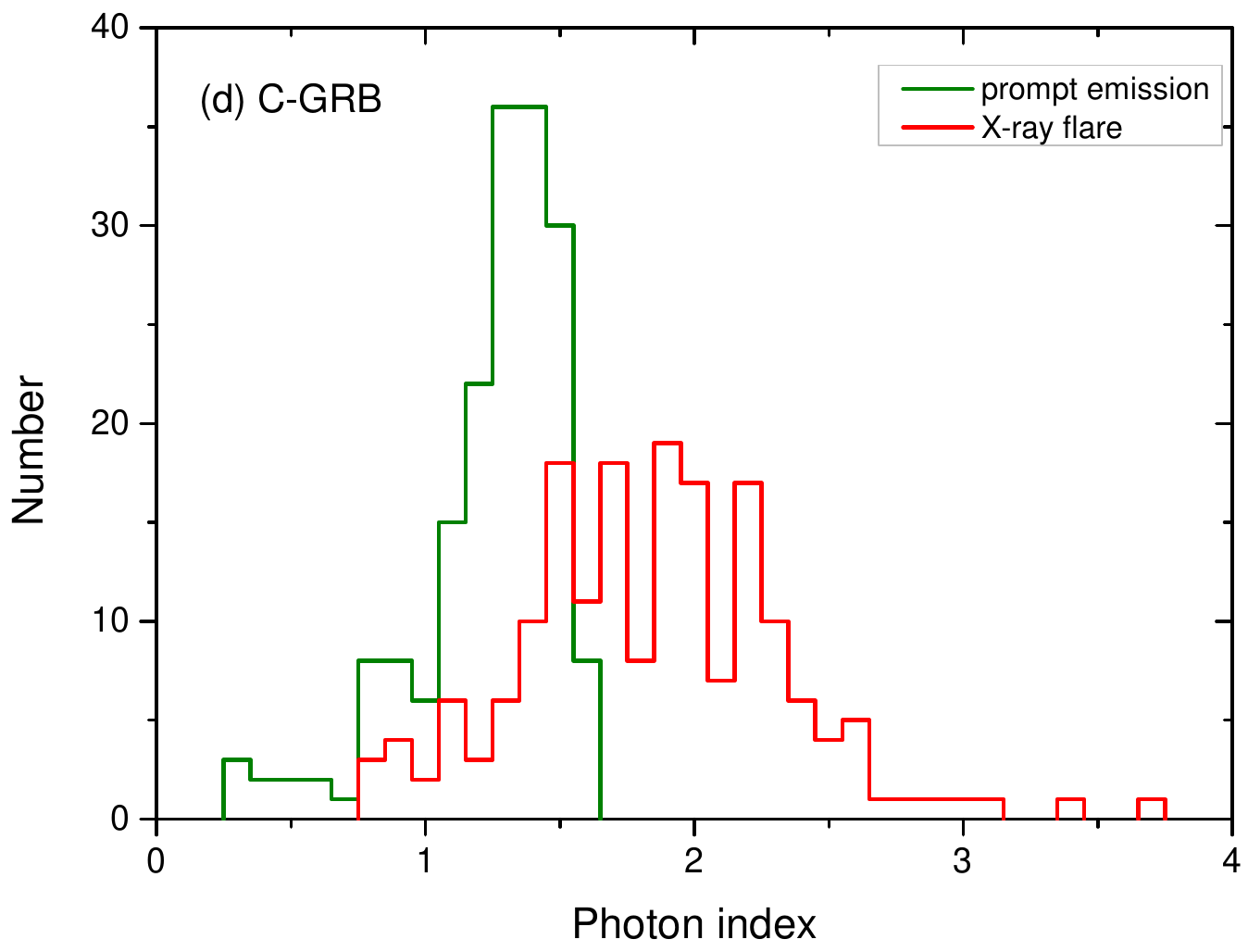}
  \caption{Histograms of the photon index distributions for X-ray flares (red solid line) and prompt emission (cutoff power-law model, green solid line). (a) Distribution for the full GRB sample; (b) XRF subclass, (c) XRR subclass, (d) C-GRB subclass.\label{fig20:cutoff_powerlaw_hist}}
  \end{figure*}

We further analyzed the relationship between the cutoff power-law photon indices of GRB prompt emission and those of X-ray flares. Figure~20 presents the histograms of the photon index distributions for X-ray flares and prompt emission (cutoff power-law model) for the full GRB sample as well as for the XRF, XRR, and C-GRB subclasses. Gaussian fits to the histogram of the full sample in Figure~20(a) yield a peak value of \(x_c = 1.46 \pm 0.02\) for the prompt emission photon index and \(x_c = 1.85 \pm 0.02\) for the X-ray flare photon index. A K-S test performed on the two spectral distributions yields \(p \approx 7.36 \times 10^{-48}\) (\(p \ll 0.05\)), further confirming a significant difference between the photon index distributions of prompt emission and X-ray flares.\footnote{In our sample, four GRBs have negative cutoff power-law (CPL) spectral indices for prompt emission: GRB 050406 ($\alpha_{\rm CPL} = -0.24 \pm 2.38$), GRB 070330 ($\alpha_{\rm CPL} = -0.67 \pm 1.70$), GRB 080319D ($\alpha_{\rm CPL} = -0.39 \pm 3.62$), and GRB 131117A ($\alpha_{\rm CPL} = -0.17 \pm 1.35$). The uncertainties of the CPL indices for these four GRBs are relatively large. While these four bursts are listed in Table~1 with their $\alpha_{\rm CPL}$ values, they are excluded from all figures involving $\alpha_{\rm CPL}$ (e.g., Figures~20, 21).}

\begin{figure*}[!htb]
  \centering
  \includegraphics[width=0.5\linewidth]{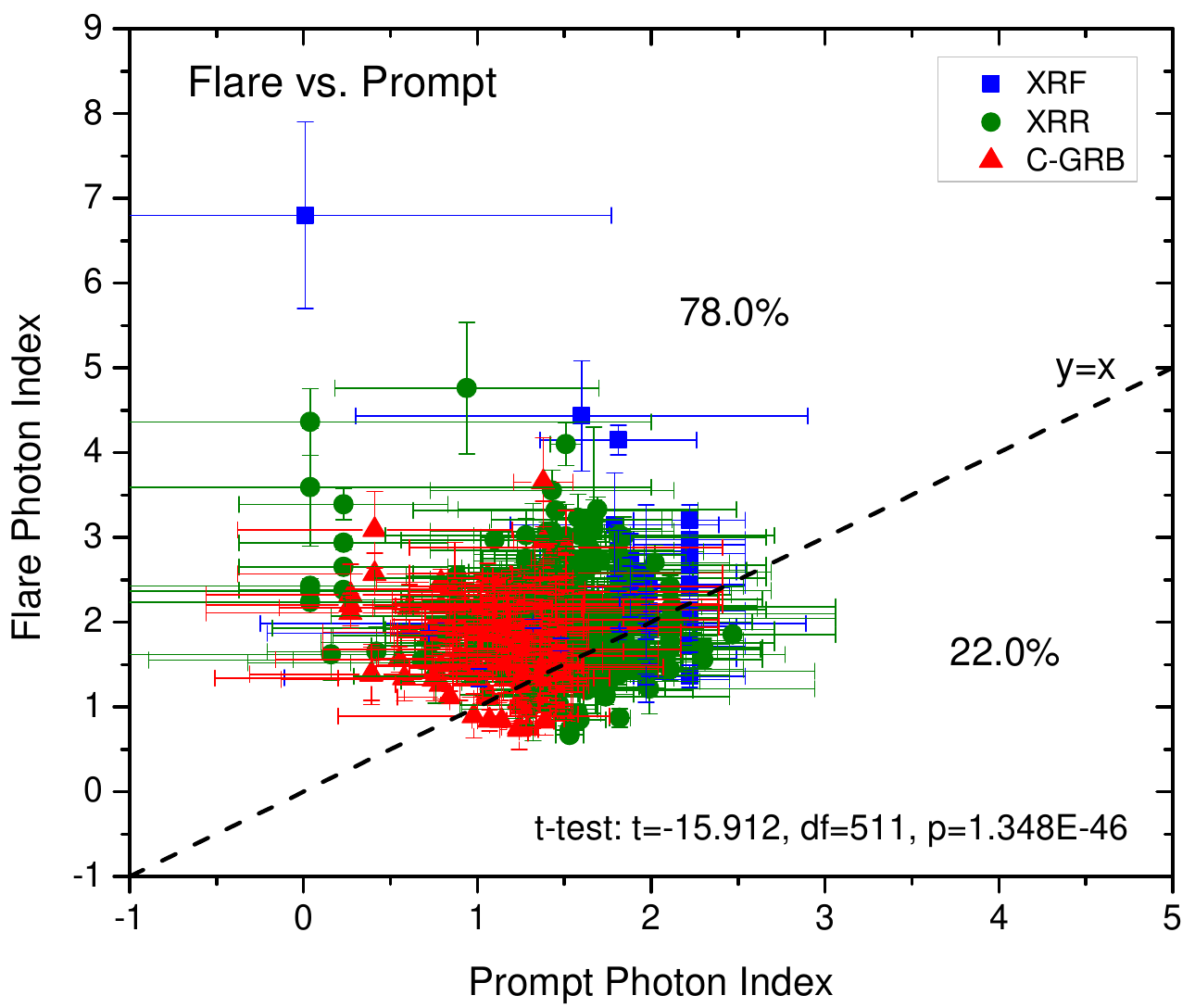}
  \caption{Comparison of the photon indices between X-ray flares and the prompt emission (cutoff power law model). The black dashed line denotes $y=x$.}\label{fig21}
 \end{figure*}

 In Figure~21, we present a comparative analysis of the photon indices between X-ray flares and the prompt emission (cutoff power-law model) for the full GRB sample, including XRF, XRR, and C-GRB. The results show that for 78.0\% of the bursts, the photon index of the X-ray flare is larger than that of the prompt emission, indicating a softer spectrum during the flare phase. In contrast, only 22.0\% of the bursts exhibit a smaller photon index (harder spectrum) during the flare. A two-sample t-test reveals a statistically significant difference between the two distributions ($t = -15.912$, $df = 511$, $p = 1.348 \times 10^{-46}$), demonstrating that the spectra of X-ray flares are systematically softer than those of the prompt emission.

 Based on the comparative analysis of the photon indices of GRB prompt emission (both the power-law and cutoff power-law models) and those of X-ray flares, we find that the spectra of X-ray flares are systematically softer than the prompt emission spectra, with a high statistical significance.

\section{Discussion}
In this paper, we adopt the simple power-law model to fit the X-ray spectra of GRB X-ray flares, as well as to fit the afterglow emission before and after the flares. This model is a classical one for the spectral analysis of X-ray emission, which can effectively characterize the power-law nature of the radiation and the effect of absorption due to the interstellar medium. It has been widely used in the analysis of X-ray observational data from the Swift satellite \citep{ButlerKocevski2007a,Falcone2007,Chincarini2010}, ensuring that our results are comparable with previous studies. However, this model has limitations: it does not take into account effects of spectral curvature (e.g., spectral breaks, high-energy cutoffs) and line emission/absorption features. These factors may introduce systematic biases in the fitted photon index and column density \citep{O'Brien2006,Falcone2007,ButlerKocevski2007a,Chang2021,Li2022}.

In addition, in time-sliced spectral fitting, the temporal boundaries between the flare interval and the preceding/following afterglow intervals rely on the morphological identification of the light curve. For GRBs with multiple overlapping flares, such classification is somewhat subjective, which may lead to some afterglow emission being incorrectly included in the flare sample, or flare data being assigned to the afterglow phase, thereby affecting the statistical results of the spectral parameters.

In this work, we perform independent spectral fittings for the preflare, flare, and postflare phases of each X-ray flare. The underlying afterglow persists throughout the flare episode and its emission may contaminate the flare spectra. We adopt a single-component spectral fitting approach without subtracting the underlying afterglow. In the following, We will discuss the potential impact of this underlying afterglow on the spectral fitting of flares.

To avoid interference from overlapping multiple flares and complex light-curve variations, we select GRB samples that contain only one isolated X-ray flare and exhibit steady continuous afterglow emission both before and after the flare. This yields a final sample of 10 GRBs. We divide these sources into bright-flare and faint-flare groups according to the integrated fluence $S_{\rm F}$ of the X-ray flare. Sources with $S_{\rm F}\gtrsim 100\times10^{-8}\,\mathrm{erg\,cm^{-2}}$ are classified as bright flares (four GRBs): GRB 060904B, GRB 070704, GRB 100902A, and GRB 121123A. Sources with $S_{\rm F}<10\times10^{-8}\,\mathrm{erg\,cm^{-2}}$ are classified as faint flares (six GRBs): GRB 060512, GRB 091029, GRB 091104, GRB 110520A, GRB 121125A, and GRB 130427B. The fluence values are compiled from \citet{Chincarini2010} and \citet{Yi2016}. The X-ray light curves of the bright-flare and faint-flare samples are shown in Figure 22 and Figure 23, respectively. Table 2 summarizes the spectral fitting parameters of the two groups in the preflare, flare, and postflare phases. These parameters include the flare fluence, photon index, and hydrogen column density $N_{\rm H}$.

To investigate the potential impact of the underlying afterglow on the flare spectral fitting, we compare the temporal evolution of key spectral parameters (photon index $\Gamma$, intrinsic hydrogen column density $N_{\rm H}$) between the bright-flare and faint-flare groups across the three phases. The temporal evolution of $\Gamma$ and $N_{\rm H}$ is presented in Figures 24 and 25, respectively.

As shown in Figure 24, the temporal evolution of the photon index shows that the bright-flare and faint-flare samples exhibit consistent trends, with no significant differences between the two groups. For example, the photon index $\Gamma$ of the bright-flare source GRB 060904B rises from $1.57 \pm 0.24$ before the flare to $1.82 \pm 0.06$ during the flare, and further to $3.25 \pm 0.10$ afterward. The faint-flare source GRB 091104 shows a similar trend, with $\Gamma$ rising from $2.02 \pm 0.14$ before the flare to $ 2.26 \pm 0.14$ during the flare, and continuing to $2.31 \pm 0.18$ afterward.

As shown in Figure 25, the $N_{\rm H}$ evolution of the bright-flare and faint-flare samples exhibits consistent trends, with no significant differences between the two groups. For example, the bright-flare source GRB 060904B shows $N_{\rm H}$ rising from $(15.0^{+41.0}_{-15.0})\times10^{20}\,\mathrm{cm^{-2}}$ before the flare to $(95.0^{+15.0}_{-14.0})\times10^{20}\,\mathrm{cm^{-2}}$ during the flare, and decreasing to $(67.3^{+8.7}_{-8.4})\times10^{20}\,\mathrm{cm^{-2}}$ afterward. The faint-flare source GRB 130427B exhibits a similar evolution, with $N_{\rm H}$ increasing from $(36.0^{+114.0}_{-36.0})\times10^{20}\,\mathrm{cm^{-2}}$ before the flare to $(78.0^{+86.0}_{-78.0})\times10^{20}\,\mathrm{cm^{-2}}$ during the flare, and falling back to $(55.0^{+181.0}_{-55.0})\times10^{20}\,\mathrm{cm^{-2}}$ afterward.

As shown in Figures 24 and 25, we find that the underlying afterglow does not introduce systematic bias in the flare spectral fits. Nevertheless, these conclusions are based on the single-component fitting scheme adopted in our analysis. While no systematic deviations are detected, this finding does not exclude genuine physical contributions from the afterglow component to the observed X-ray flux. In the meanwhile, we also speculate that the X-ray flare itself may have strong spectral evolution.

To remove contamination from the underlying afterglow and obtain more reliable intrinsic spectral parameters of flares, we recommend a two-component joint fitting approach. The implementation is as follows. We first fit the pure afterglow spectrum extracted from the preflare interval and fix the afterglow photon index $\Gamma_{\rm pre-flare}$, allowing only the normalization factor $C_{\rm pre-flare}$ to decay following the afterglow light curve. Spectra taken during the flare are fitted with a two-component model consisting of the afterglow component plus the flare emission, expressed as
\[
F(E)=C_{\rm pre-flare}\,E^{-\Gamma_{\rm pre-flare}}+C_{\rm flare}\,E^{-\Gamma_{\rm flare}},
\]
where $C_{\rm flare}$ and $\Gamma_{\rm flare}$ are the normalization and photon index of the flare component. This approach enables a reliable decomposition of the two components, removes background contamination, and provides accurate spectral fitting for X-ray flares.

\begin{figure*}[!htb]
  \centering
  \includegraphics[height=5cm,width=7cm]{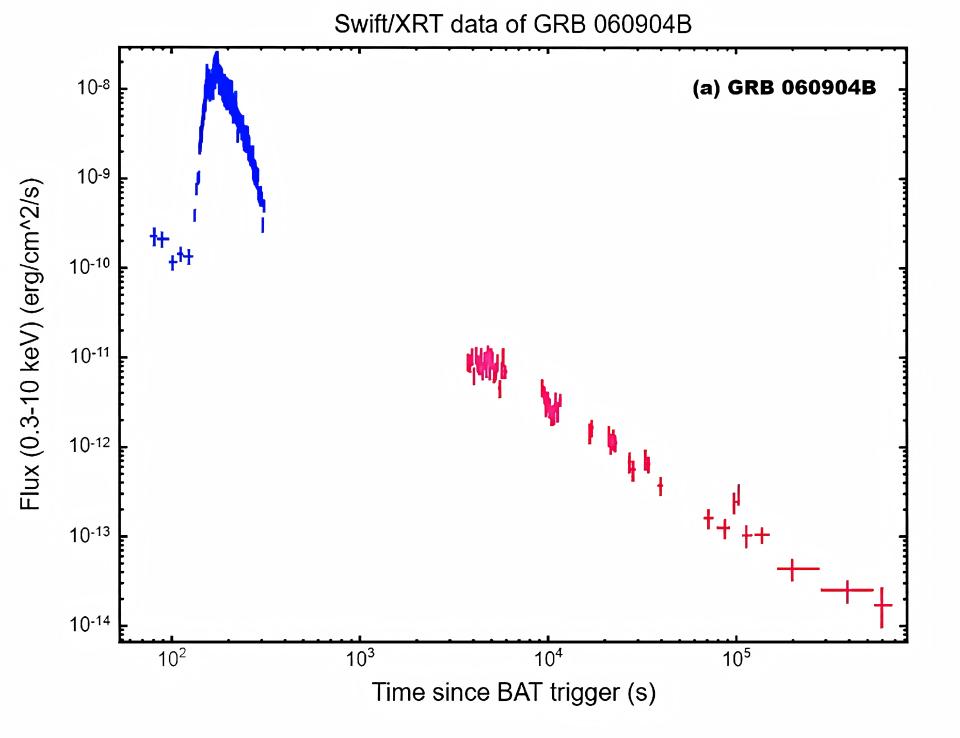}
  \includegraphics[height=5cm,width=7cm]{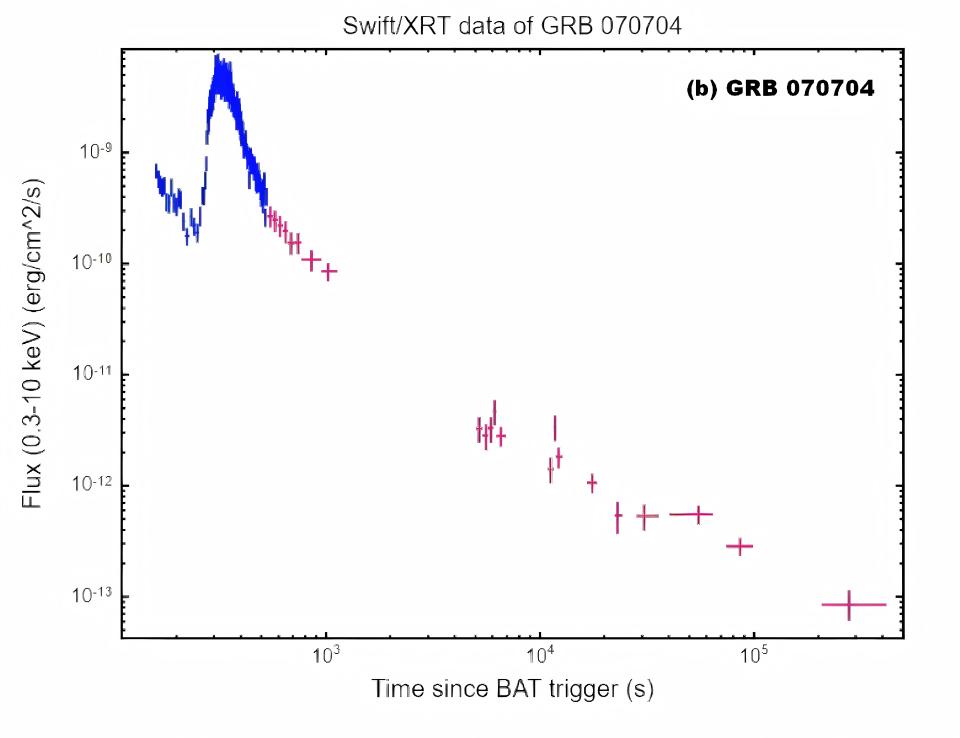}
  \includegraphics[height=5cm,width=7cm]{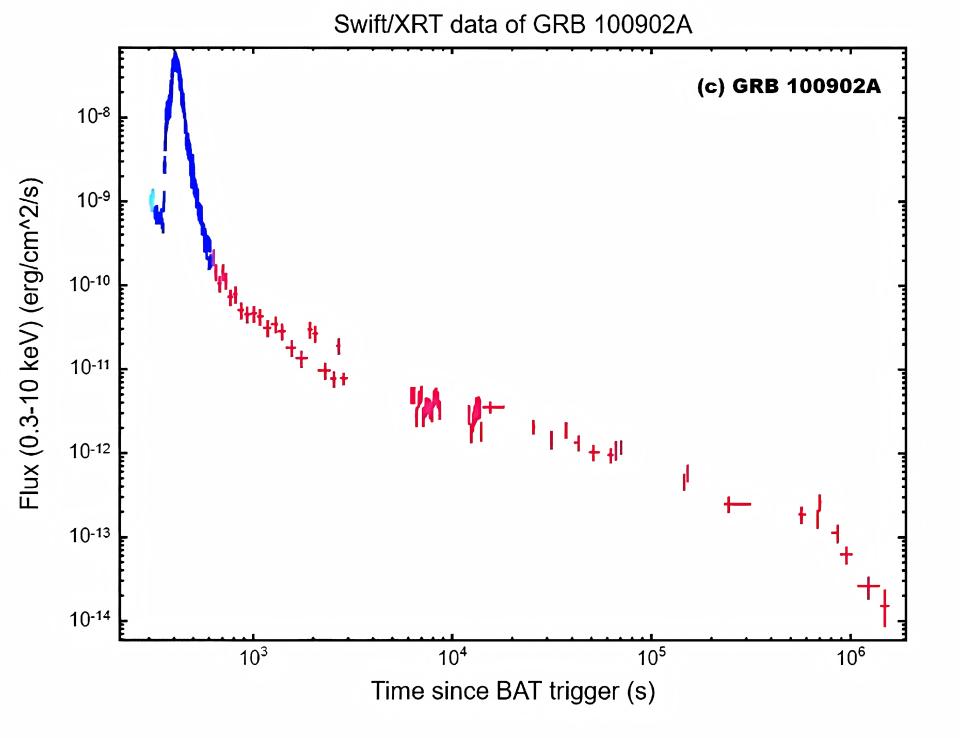}
  \includegraphics[height=5cm,width=7cm]{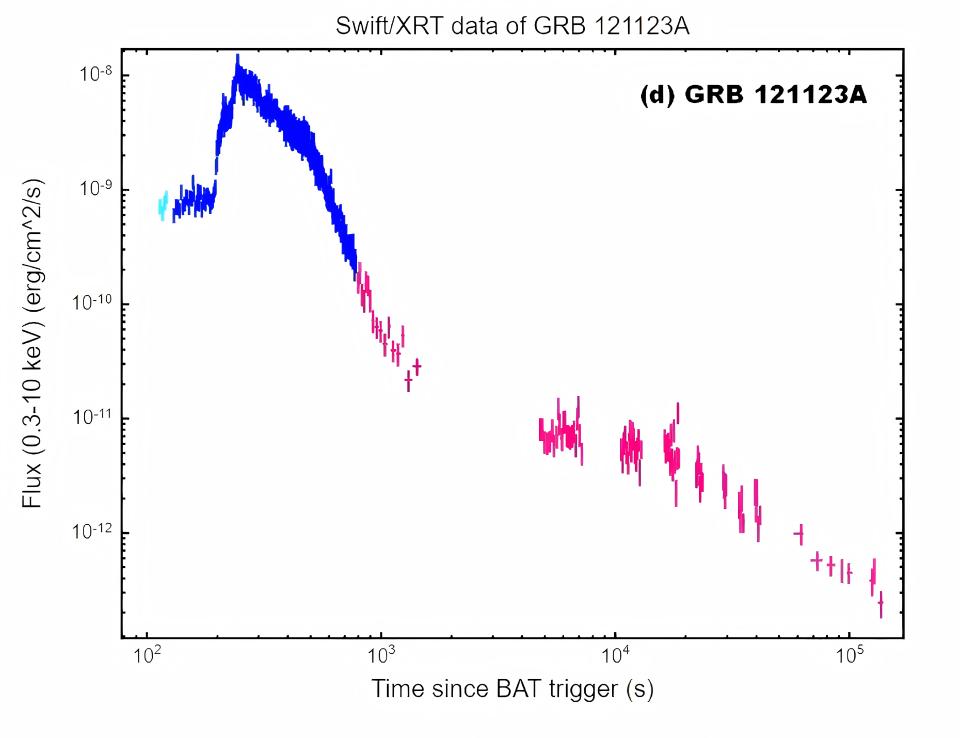}
  \caption{X-ray light curves exhibiting prominent bright flares for four GRBs. (a) GRB 060904B, (b) GRB 070704, (c) GRB 100902A, (d) GRB 121123A.}\label{fig22:Bright flare}
  \end{figure*}

\begin{figure*}[!htb]
  \centering
  \includegraphics[width=0.3\linewidth]{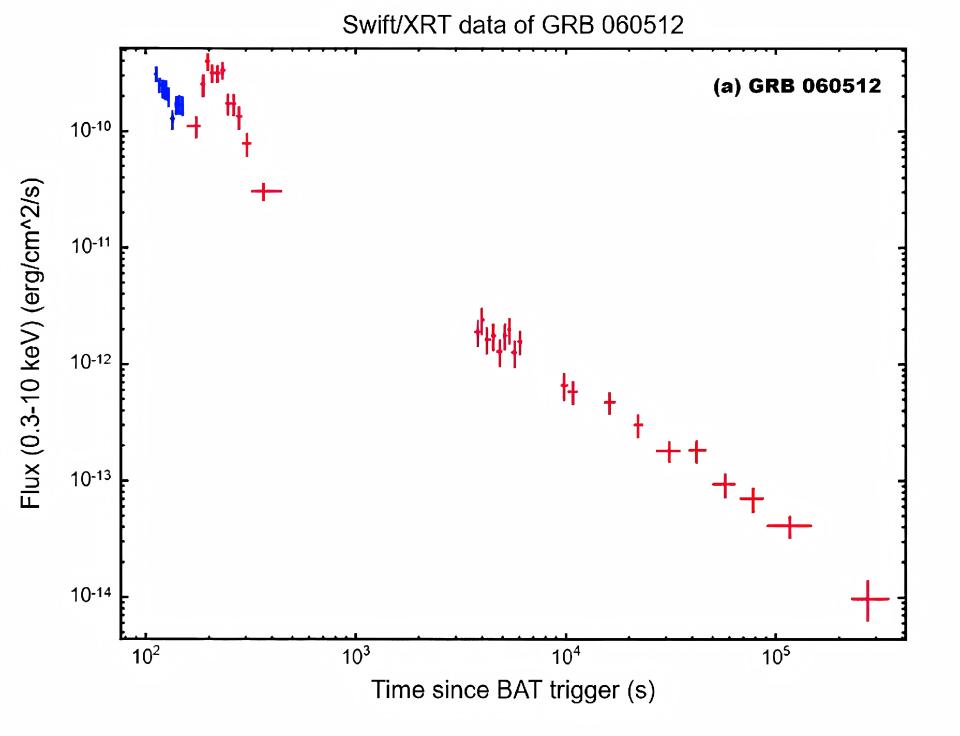}
  \includegraphics[width=0.3\linewidth]{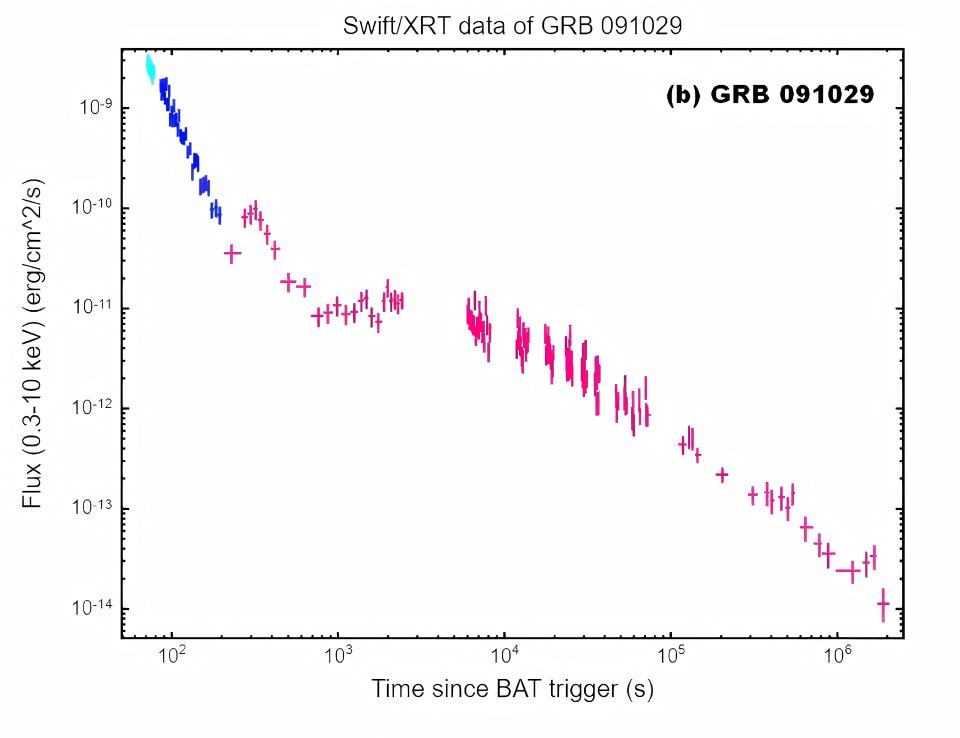}
  \includegraphics[width=0.3\linewidth]{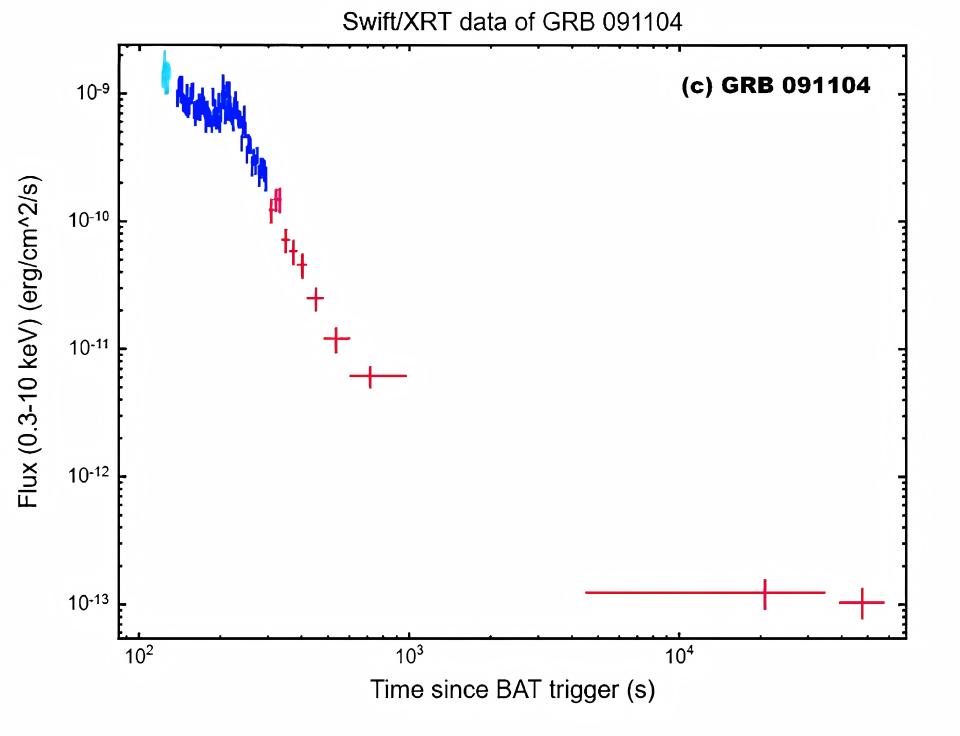}
  \includegraphics[width=0.3\linewidth]{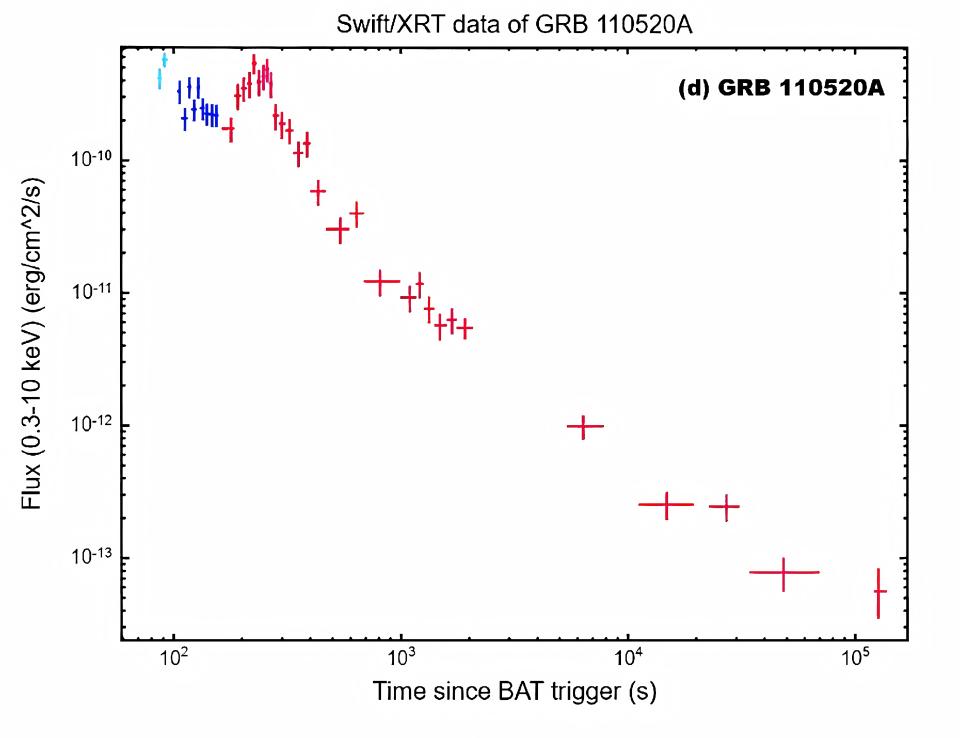}
  \includegraphics[width=0.3\linewidth]{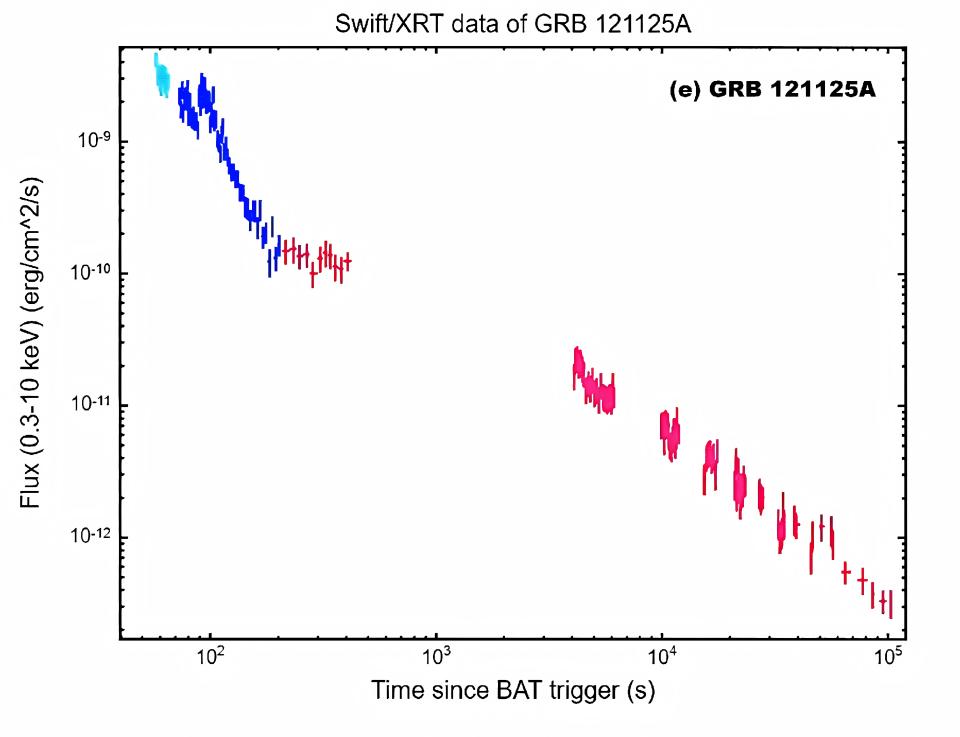}
  \includegraphics[width=0.3\linewidth]{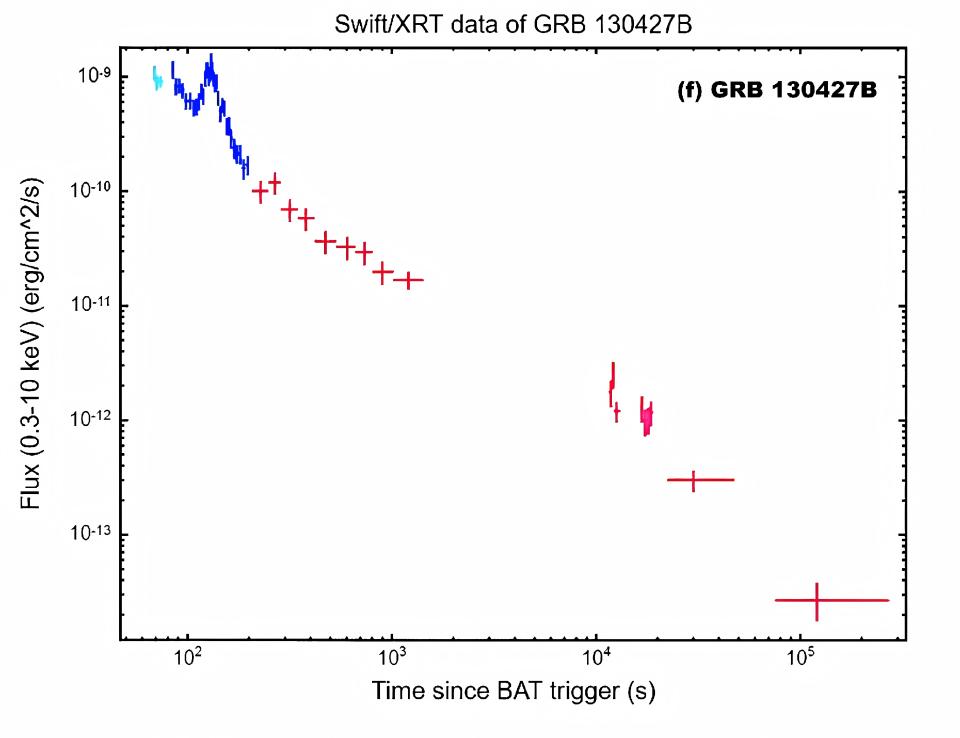}
  \caption{X-ray light curves featuring faint flares for six GRBs. (a) GRB 060512, (b) GRB 091029, (c) GRB 091104, (d) GRB 110520A, (e) GRB 121125A, (f) GRB 130427B.}\label{fig23: Faint flare}
  \end{figure*}

\begin{figure*}[!htb]
  \centering
  \includegraphics[width=0.45\linewidth]{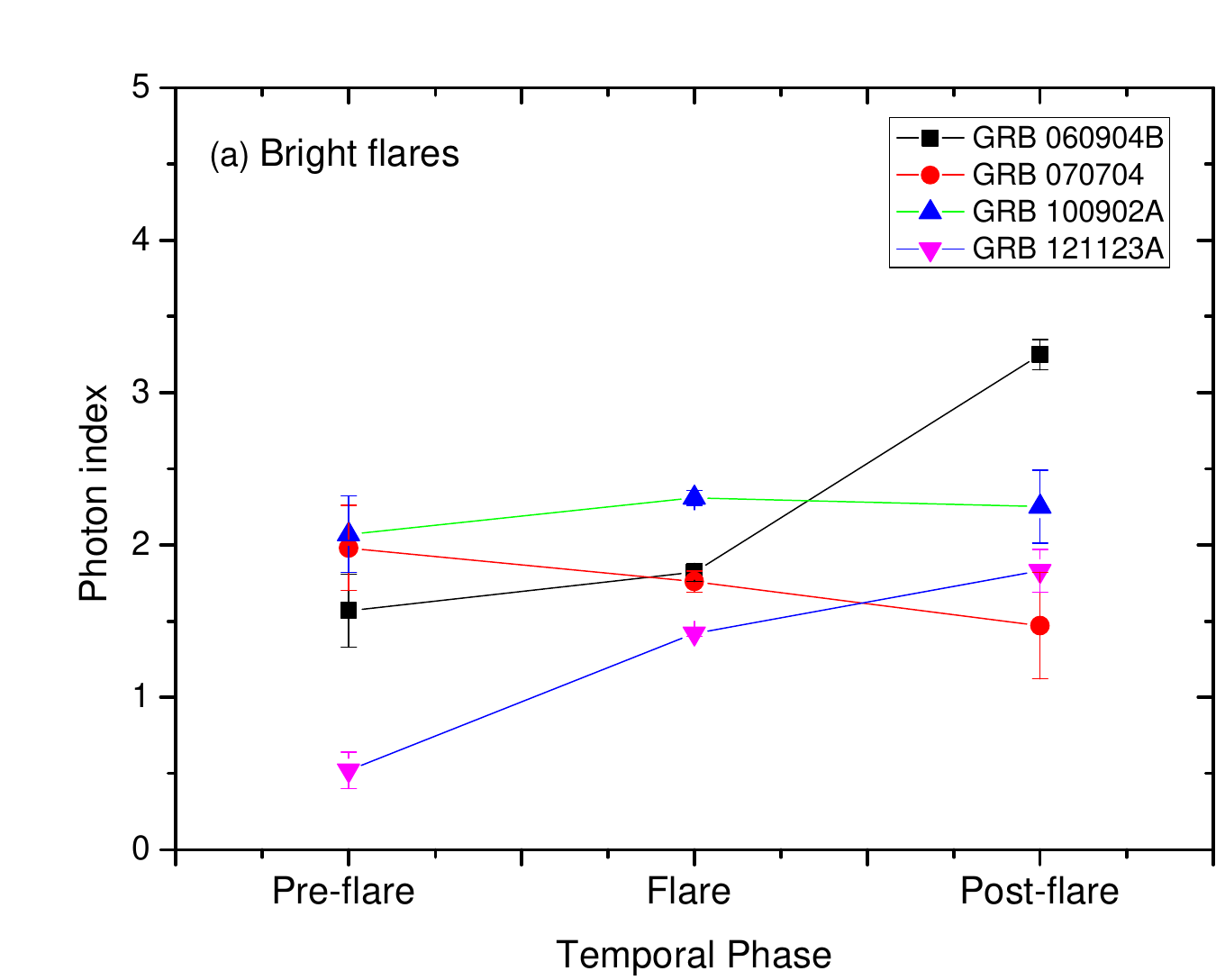}
  \includegraphics[width=0.45\linewidth]{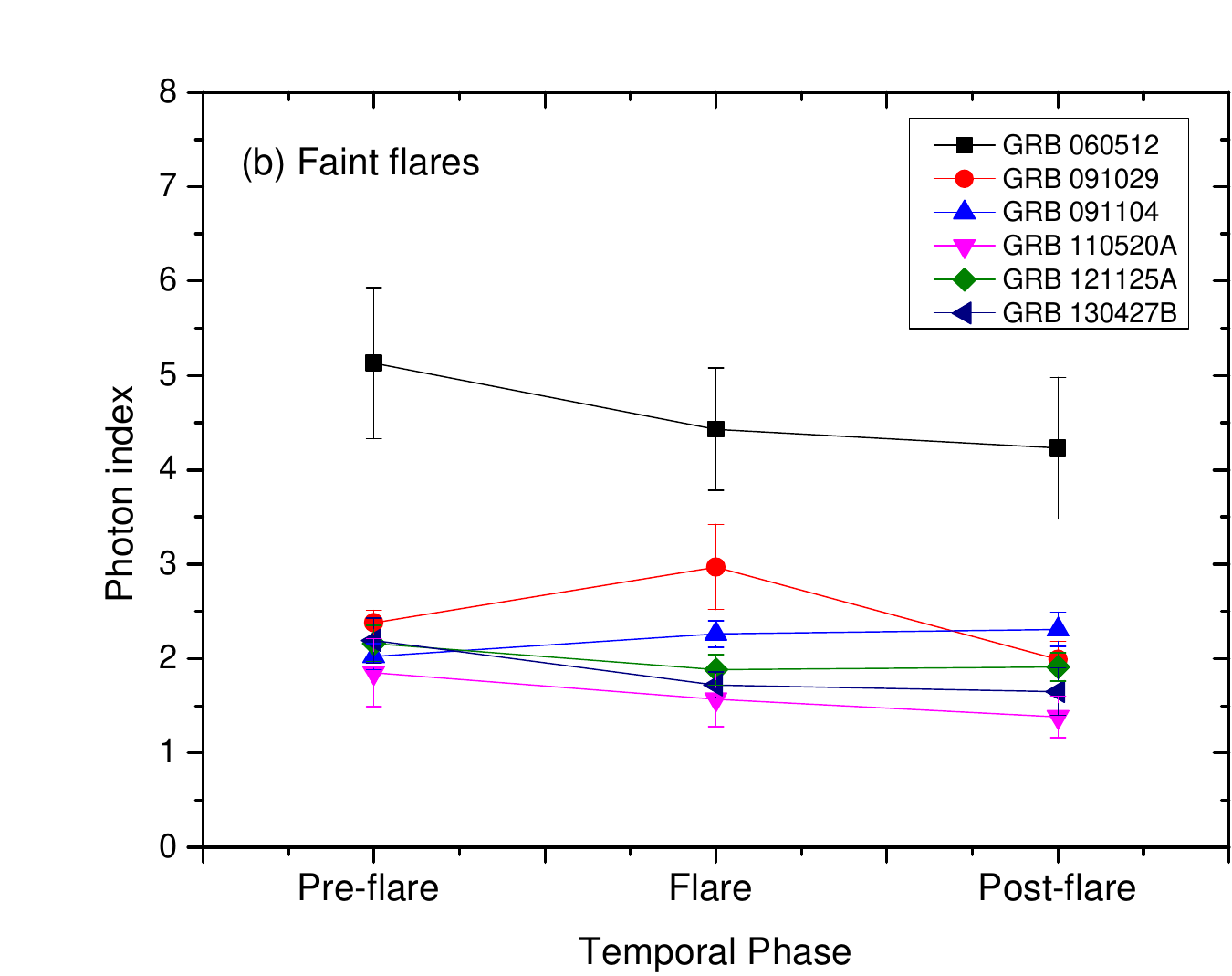}
   \caption{Temporal evolution of photon indices during preflare, flare, and postflare phases of GRB X-ray flares. (a) Bright-flare sample; (b) faint-flare sample.}\label{fig24:Bright flare}
  \end{figure*}

\begin{figure*}[!htb]
  \centering
  \includegraphics[width=0.45\linewidth]{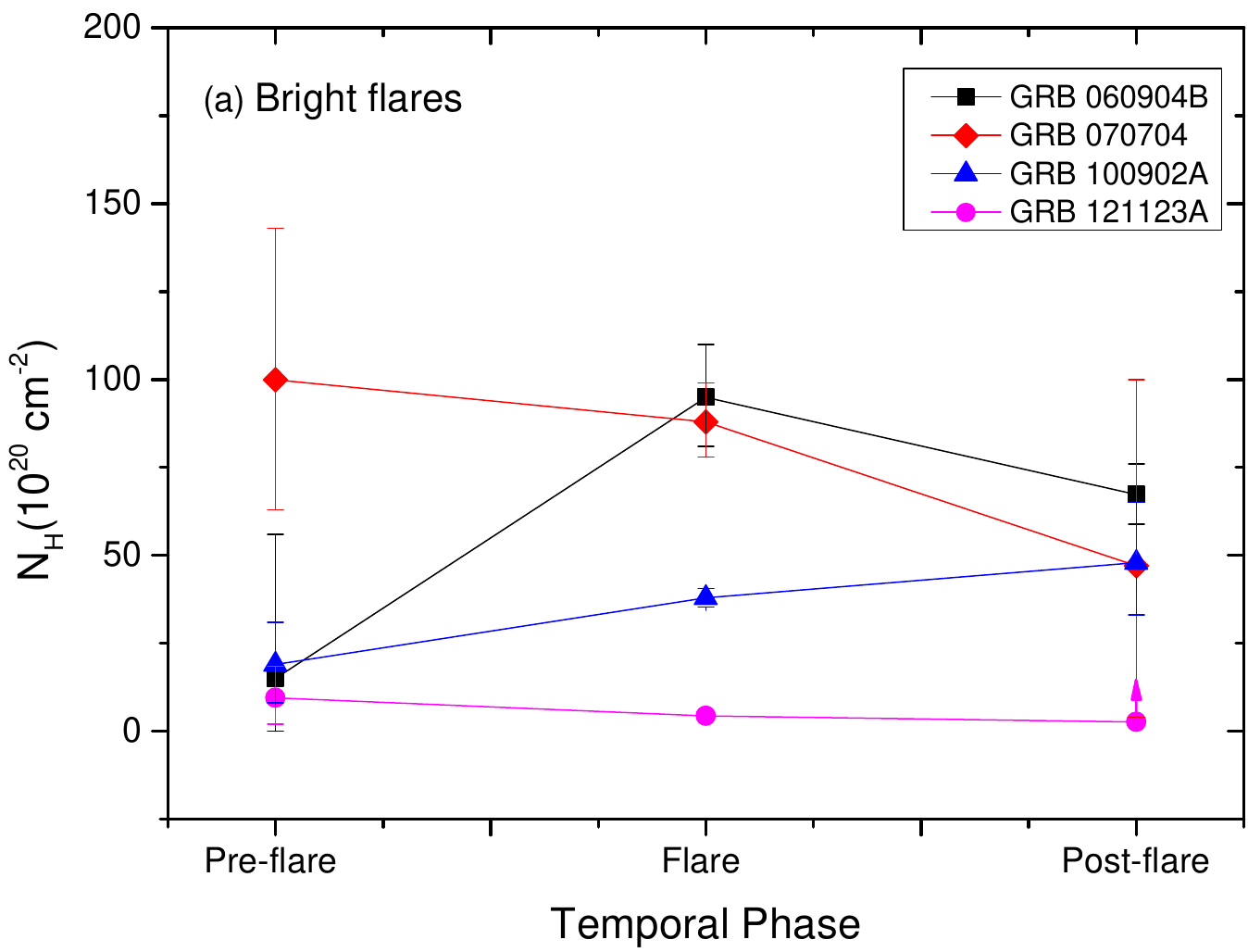}
  \includegraphics[width=0.45\linewidth]{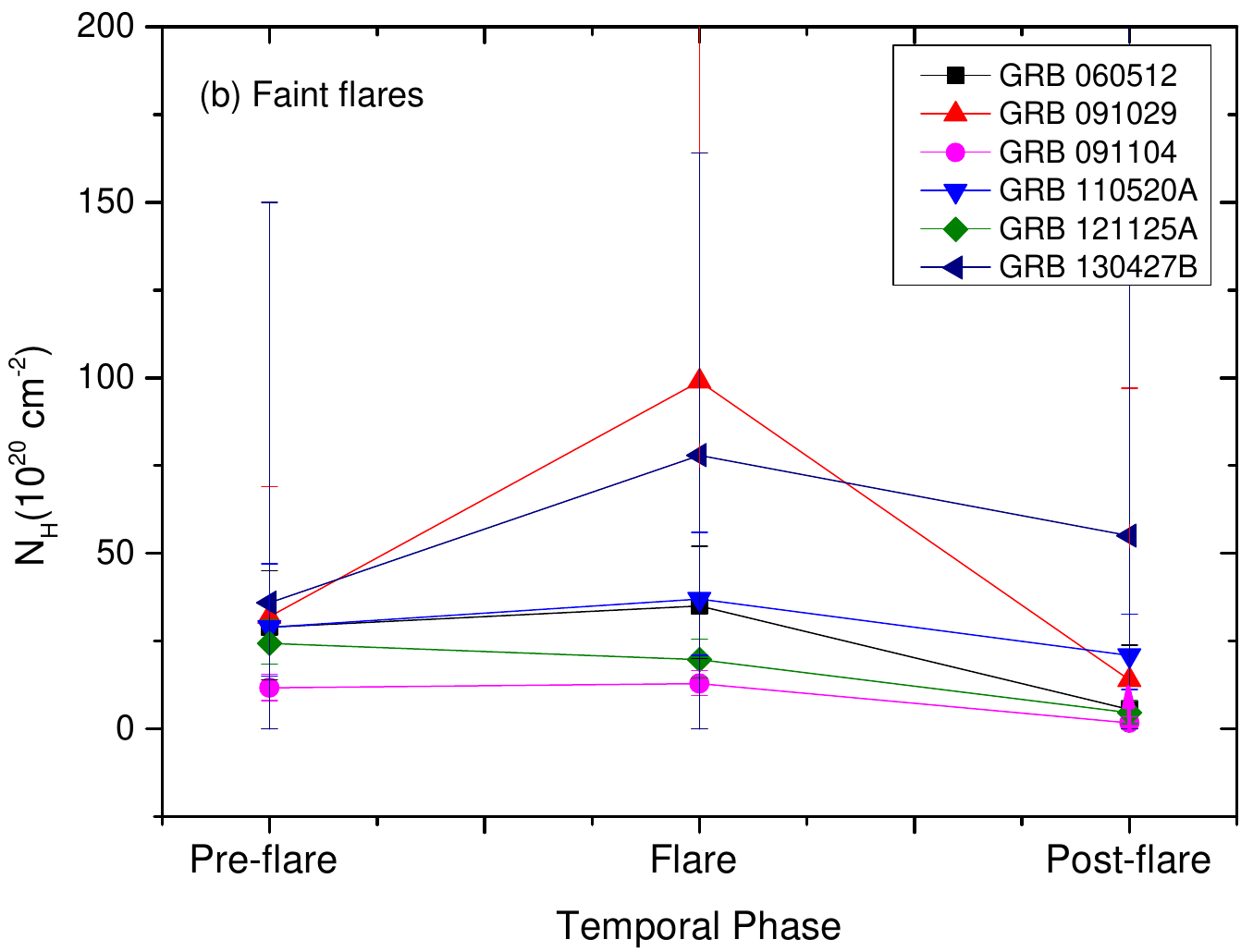}
   \caption{Temporal evolution of $N_{\rm H}$ during preflare, flare, and postflare phases of GRB X-ray flares. (a) Bright-flare sample; (b): faint-flare sample.}\label{fig25:Faint flare}
  \end{figure*}

Although this study has constructed one of the largest X-ray flare samples, sample selection effects may still affect the statistical results. Our sample consists of 521 X-ray flares from 206 GRBs, selected based on \cite{Chincarini2010} and \cite{Yi2016}, including four short GRBs, 202 long GRBs, and 99 GRBs with measured redshifts. First, the incompleteness of sample data can affect statistical results. In our sample, some flare events lack relatively complete and continuous afterglow radiation before and after the flare, making it impossible to obtain the corresponding spectral data. Consequently, these flares are excluded from the comparative analysis. This omission introduces bias into the statistical results, potentially failing to accurately reflect the true characteristics of X-ray flares. Second, long GRBs account for as much as 98.06\%, making it impossible for us to conduct effective statistical analysis on short GRBs, and there is an essential difference in the physical origins of long GRBs and short GRBs \citep{Margutti2011b,Berger2014}. In addition, only 99 GRBs have redshift values, which may limit the research on the cosmological evolution of spectral characteristics, and the results tend to be biased toward bright bursts and low-redshift bursts. Finally, the sample only includes significant flares, which may miss weak flare samples that reflect weak activities of the central engine \citep{Margutti2010,Willingale2007}, thereby affecting the understanding of the overall distribution of central engine activities.

\citet{Falcone2007} conducted spectral analysis of X-ray flares from early GRBs observed by the Swift satellite. In our sample, some X-ray flares overlap with the sample used by \citet{Falcone2007}. To verify the reliability of our derived X-ray flare spectral indices, we conducted a comparative analysis with the result of \citet{Falcone2007}, and the results are shown in Figure~26.

\begin{figure*}[!htb]
  \centering
  \includegraphics[width = 0.45\linewidth]{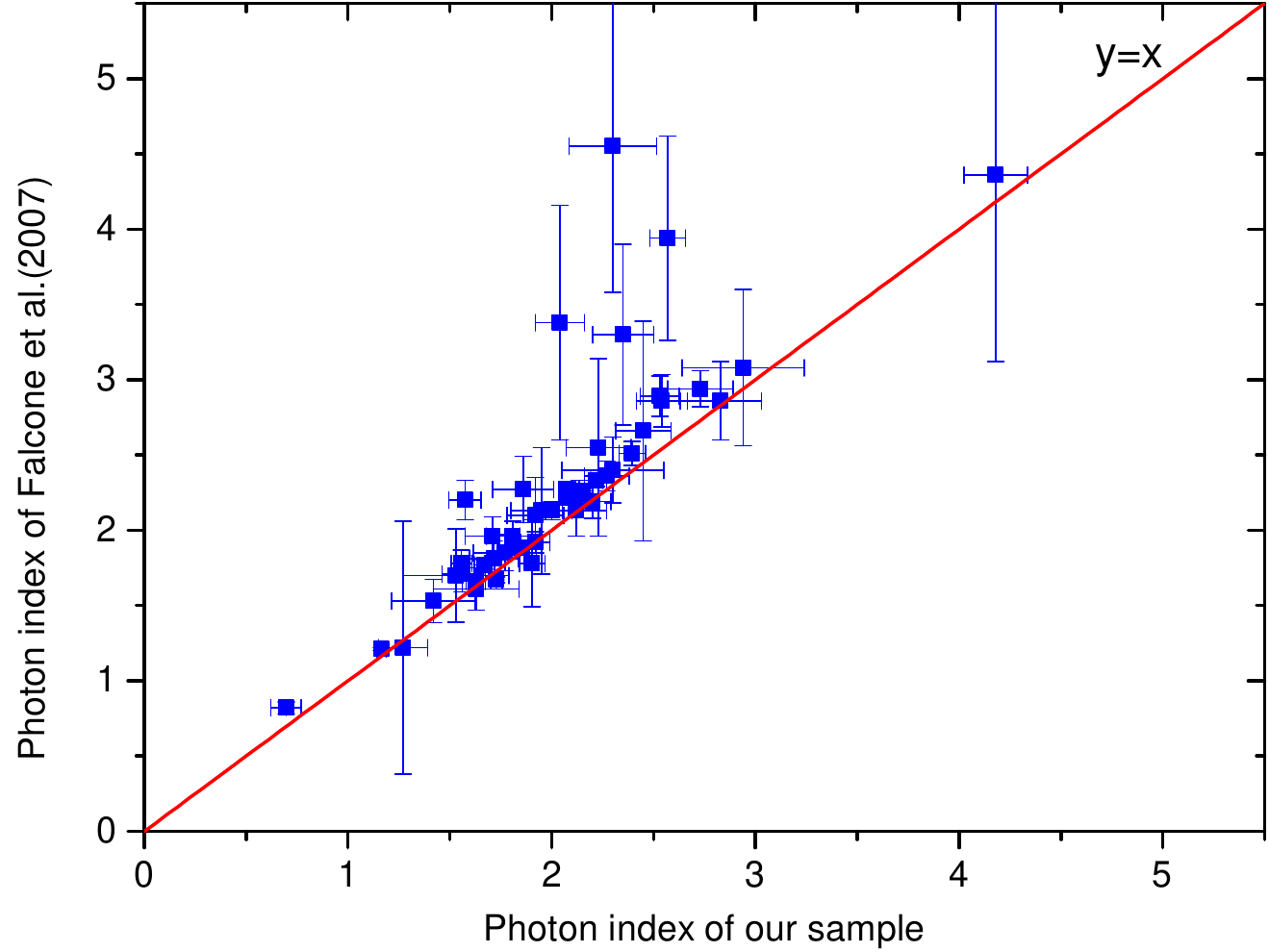}
  \caption{Comparison of photon indices between our sample and the \cite{Falcone2007} sample. The solid red line represents $y=x$.\label{fig26}}
 \end{figure*}

Based on the results of our statistical analysis of the spectral characteristics of X-ray flares, we found that the photon index of the flare phase showed a significant decrease compared to that of the preflare and postflare afterglow radiation, indicating a spectral hardening during the flare phase. This further suggests that X-ray flares and afterglow radiation may have essential differences. It is generally believed that X-ray flares may originate from the reactivation of the central engine, resulting in internal shock radiation, while the preflare and postflare afterglow radiation may mainly come from external shock radiation \citep{ ZhangMeszaros2004,Fan2005,Falcone2006,Chincarini2010,Yi2015}. A giant X-ray flare observed in GRB 050502B serves as one of the best examples supporting the idea that X-ray flares may originate from sustained activity of the central engine \citep{Burrows2005}.

Our statistical results indicate that for GRBs with multiple consecutive X-ray flares, the photon index of X-ray flares increases with increasing peak time, showing a trend in which the spectrum gradually softens with the peak time of the flare. This characteristic of spectral evolution with time may be caused by multiple intermittent activities of the GRB central engine, and the central engine activity has a gradually decaying feature \citep{Godet2006,ButlerKocevski2007a,Chincarini2010}. In the early X-ray flares of GRB 050421, the spectrum showed a gradually softening feature, further confirming that the spectrum of X-ray flares has a trend of gradually softening over time \citep{Godet2006}.

Our statistical results show that the overall spectrum of X-ray flares is softer than that of the prompt emission. The main reason may be that the prompt emission originates from early internal shock radiation, while X-ray flares originate from late internal shock radiation \citep{Burrows2005,Chincarini2010,Yi2016}. Based on the above statistical analysis of the spectral characteristics of X-ray flares, our statistical results further support the theoretical model in which X-ray flares may originate from internal shock radiation caused by the continuous activity of the GRB central engine \citep{ZhangMeszaros2004,Burrows2005,Falcone2006,ButlerKocevski2007b,Chincarini2010,Yi2016}.

\section{Summary}

We present a spectral analysis of 521 X-ray flares from 206 GRBs observed by the Swift satellite. Using a simple power-law model, we fit the X-ray flare spectra as well as the preflare and postflare afterglow emission, thereby constraining key physical parameters including the photon index and $N_{\rm H}$. Through systematic statistical analyses, we compare the spectral properties of X-ray flares to those of prompt emission. We investigate the temporal evolution of flare spectra, the spectral differences among three GRB subclasses (XRF, XRR, and C-GRB), and the variation of $N_{\rm H}$ during different flare stages. Our main findings are summarized as follows.
\begin{enumerate}

    \item The distribution of spectral indices for X-ray flares shows no significant difference between the WT and PC observation modes.

    \item The spectral index during the flare phase differs significantly from that in the preflare and postflare phases. Specifically, the photon index during the flare phase is significantly smaller, indicating a distinct spectral hardening during the flare. In contrast, there is no significant difference in the photon index between the preflare and postflare phases.

    \item The hydrogen column density $N_{\rm H}$ shows no significant variation between the preflare and flare phases. This implies that the flare activity does not modify the initial absorbing environment. However, the postflare $N_{\rm H}$ is significantly lower than that in the preflare and flare phases.

    \item The spectral indices of X-ray flares are systematically larger than those of prompt emission.

    \item For GRBs with multiple consecutive X-ray flares, the spectral index increases with the peak time of each flare, i.e., the spectrum evolves toward softness over time.

    \item Significant differences in X-ray flare spectra exist among the three subclasses (XRF, XRR, and C-GRB). XRF flares exhibit softer spectra than C-GRB flares, while XRR spectra lie intermediate between the two.

\end{enumerate}

This work presents a systematic characterization of the spectral properties and evolutionary patterns of X-ray flares in GRBs through a large-sample statistical analysis. The findings provide new observational constraints on the prolonged activity of GRB central engines. Future studies with larger samples and more comprehensive data analyses will be required to further reveal the physical origin of X-ray flares. This work is also helpful for the study of X-ray transients from the Einstein Probe (EP) satellite.

\appendix
\setcounter{figure}{0}
\setcounter{table}{0}
\renewcommand{\thefigure}{A\arabic{figure}}
\renewcommand{\thetable}{A\arabic{table}}

\section{Supplementary spectral fitting results}

In this appendix, we present the spectral fit plots and residual distributions for three representative examples of $N_{\rm H}$ constraints from our spectral fitting results. The three examples correspond to: (1) well-constrained $N_{\rm H}$ with small uncertainties; (2) constrained $N_{\rm H}$ with relatively large uncertainties; and (3) only upper limits on $N_{\rm H}$. The spectral fit plots for Examples 1, 2, and 3 are shown in Figures A1, A2, and A3, respectively. The corresponding best-fit parameters for each example are listed in Table A1. As discussed in Section 3.4, the quality of the \(N_{\rm H}\) constraint depends critically on the photon counts, particularly in the soft 0.3--1.0 keV band.

\begin{figure*}[!htb]
  \centering
  \includegraphics[width=0.3\linewidth]{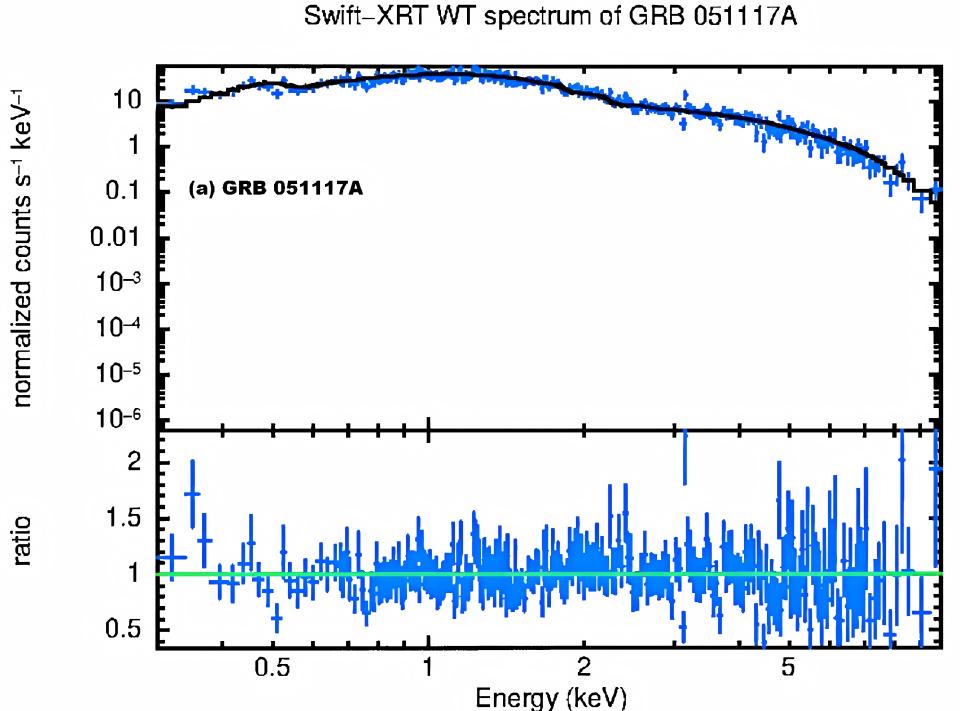}
  \includegraphics[width=0.3\linewidth]{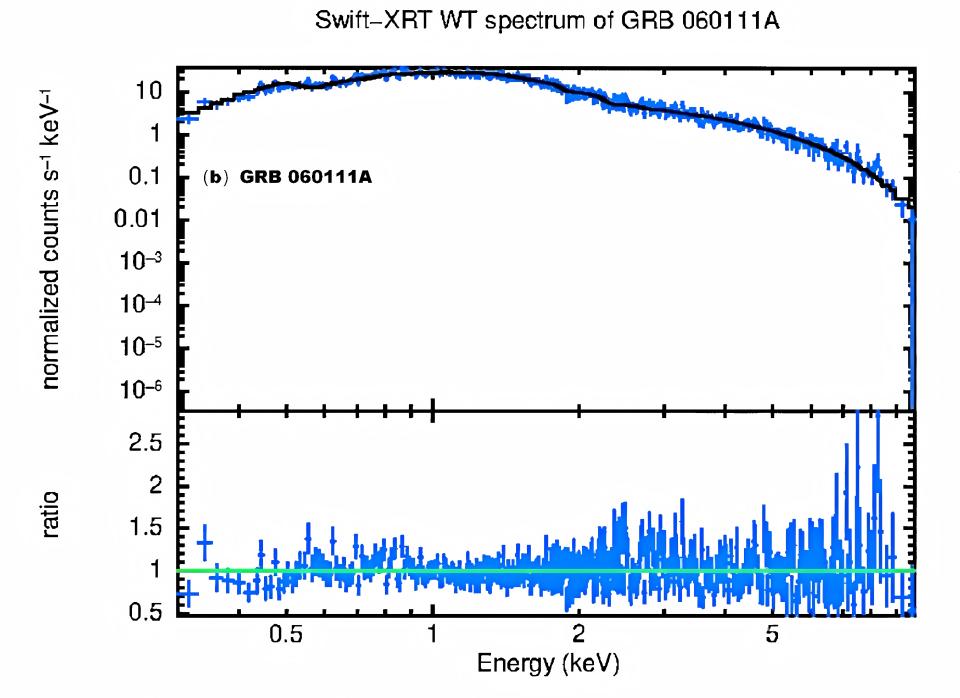}
  \includegraphics[width=0.3\linewidth]{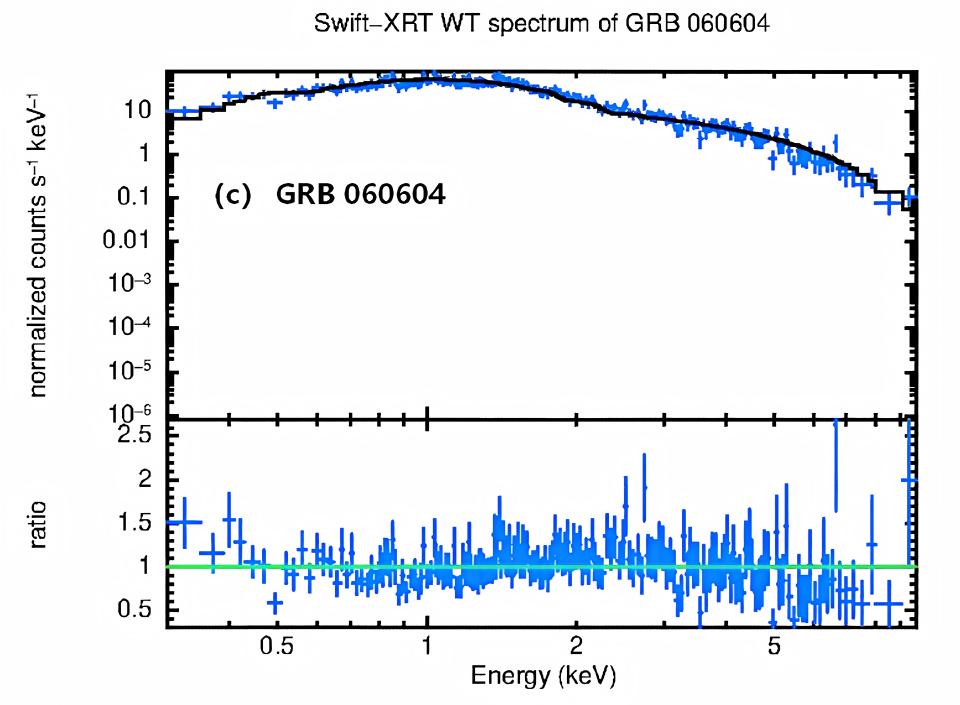}
  \includegraphics[width=0.3\linewidth]{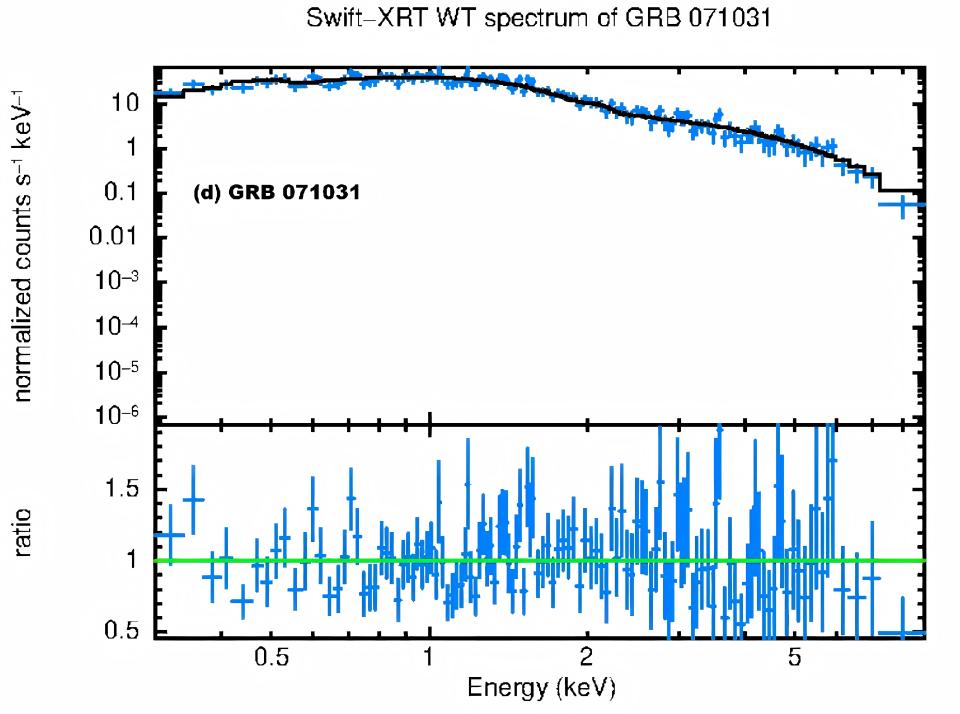}
  \includegraphics[width=0.3\linewidth]{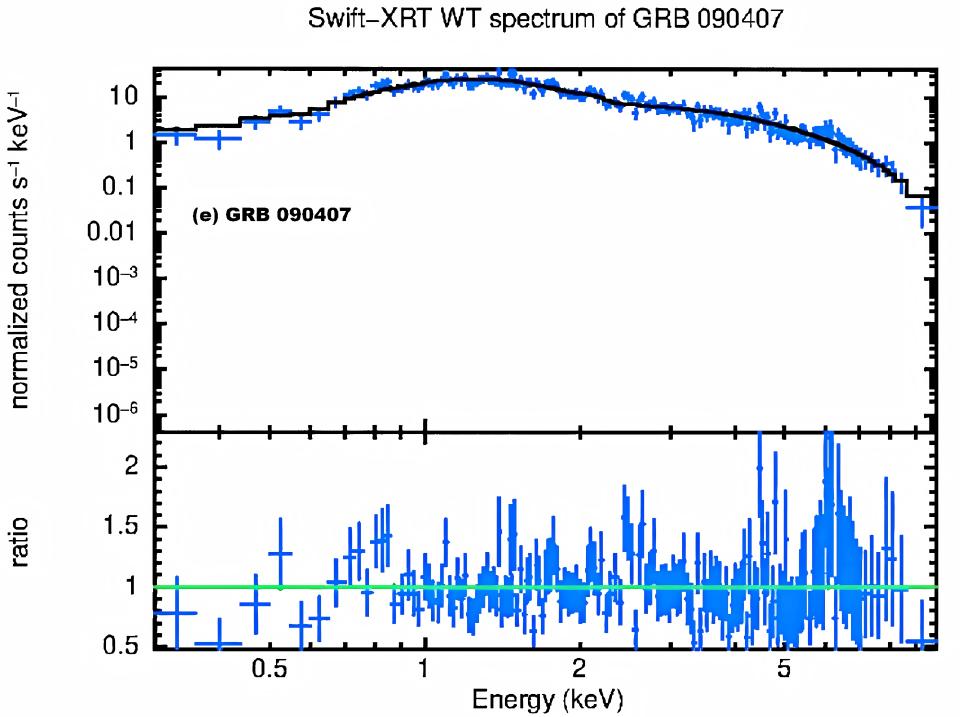}  \includegraphics[width=0.3\linewidth]{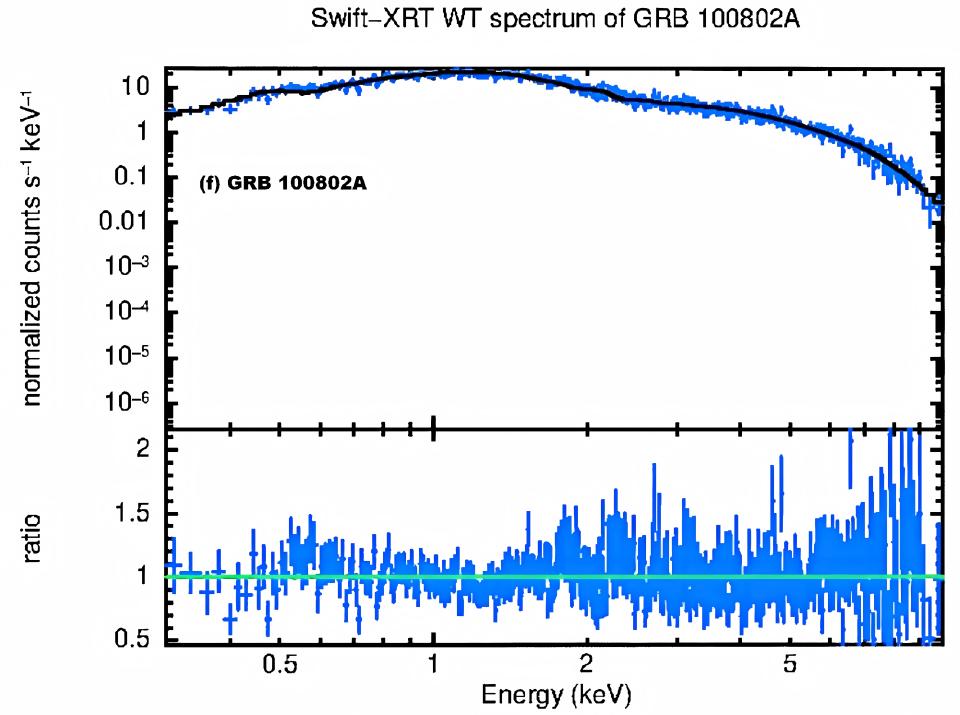}
  \includegraphics[width=0.3\linewidth]{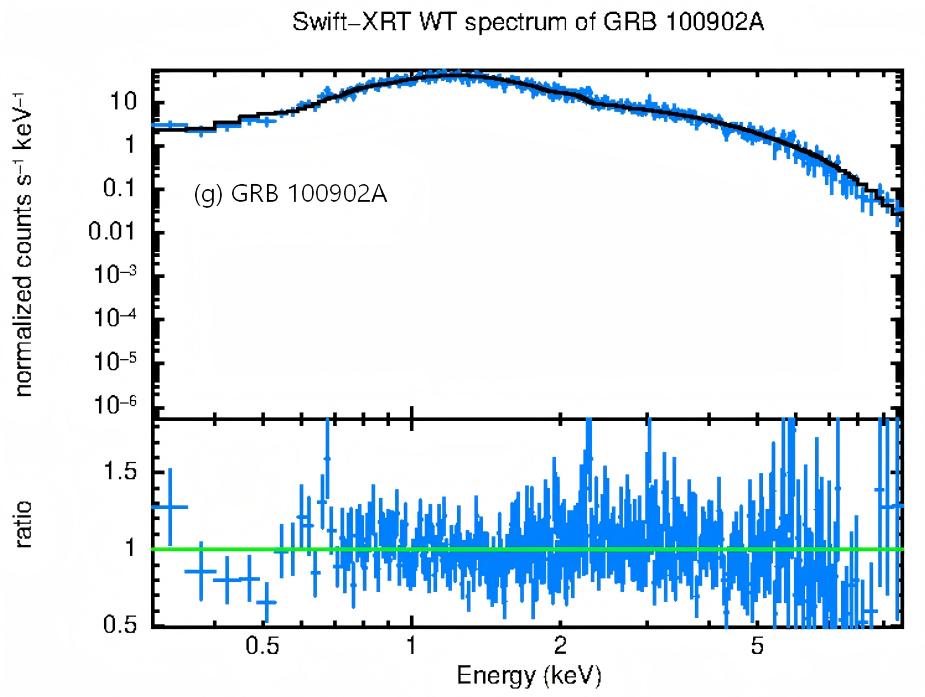}
  \includegraphics[width=0.3\linewidth]{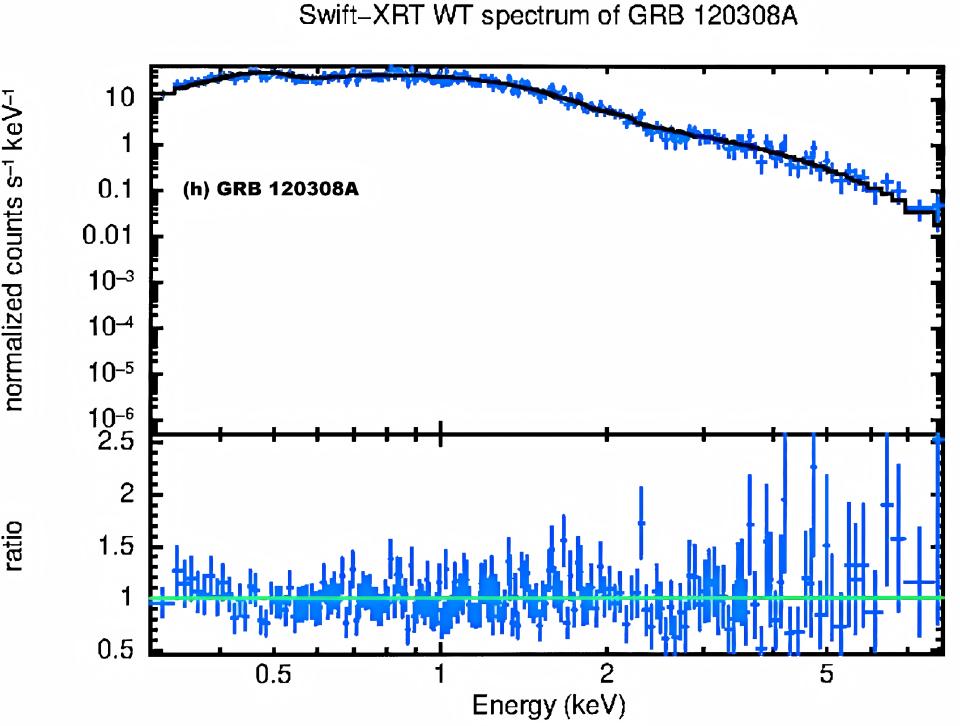}
  \includegraphics[width=0.3\linewidth]{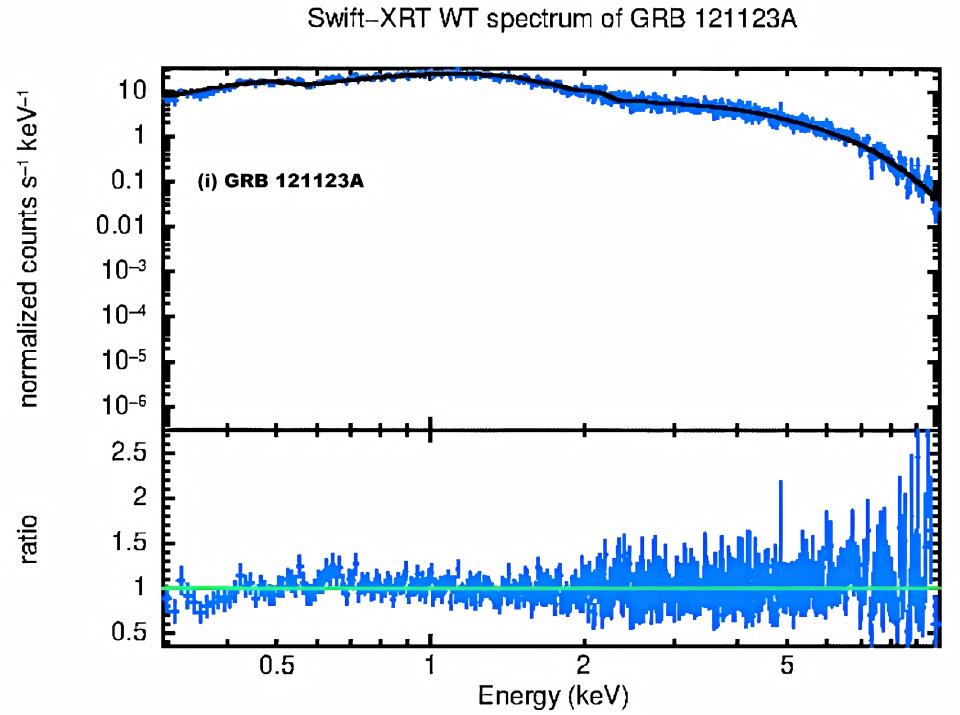}
  \caption{Spectral fit and residual plots for Example 1: well-constrained $N_{\rm H}$ with small uncertainties. Panels (a)--(i) correspond to GRB 051117A, GRB 060111A, GRB 060604, GRB 071031, GRB 090407, GRB 100802A, GRB 100902A, GRB 120308A, and GRB 121123A, respectively. The best-fit parameters are listed in Table A1.}
  \label{fig:A1}
\end{figure*}

\begin{figure*}[!htb]
  \centering
  \includegraphics[width=0.3\linewidth]{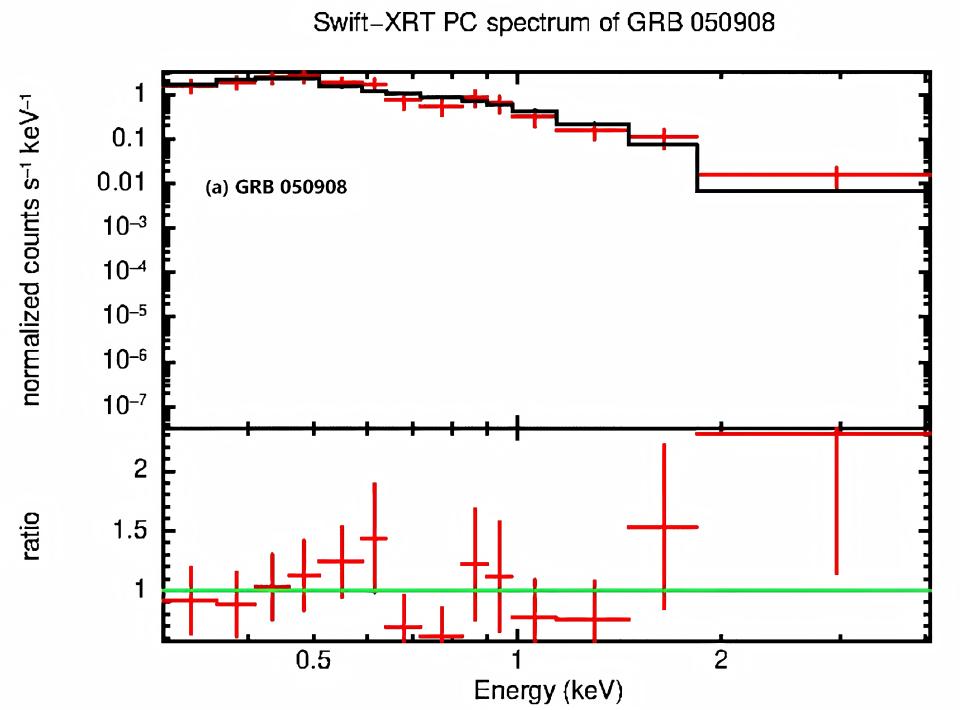}
  \includegraphics[width=0.3\linewidth]{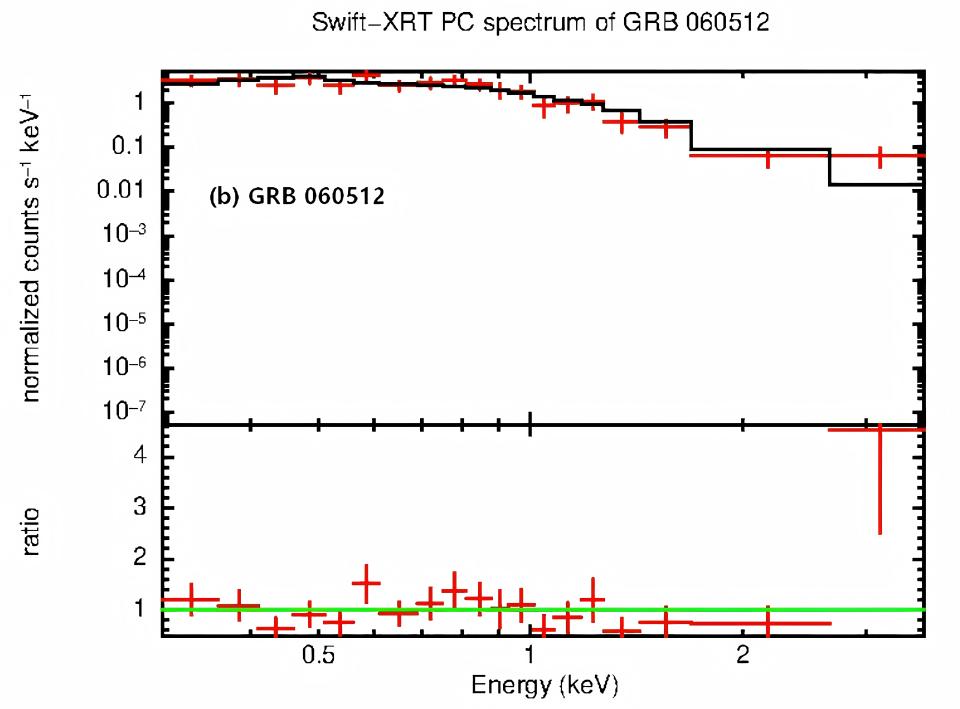}
  \includegraphics[width=0.3\linewidth]{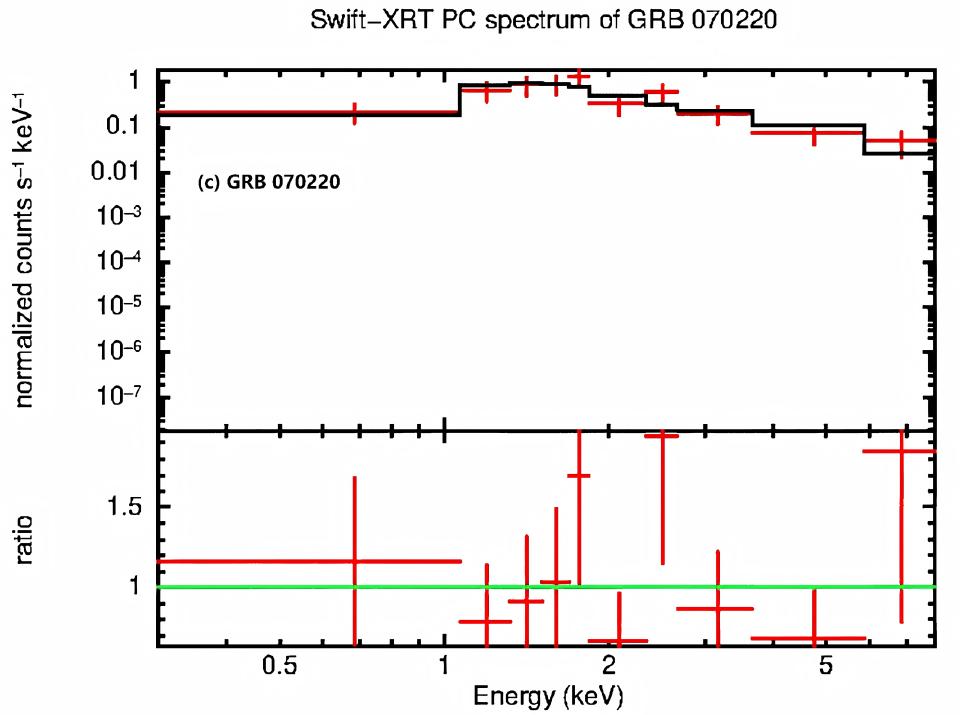}
  \includegraphics[width=0.3\linewidth]{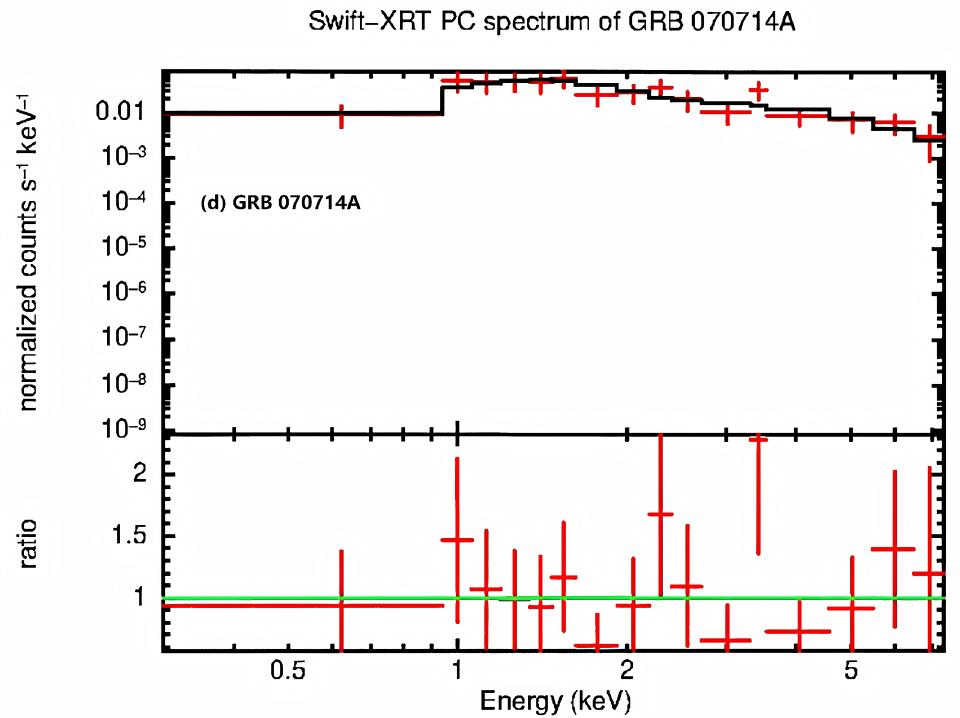}
  \includegraphics[width=0.3\linewidth]{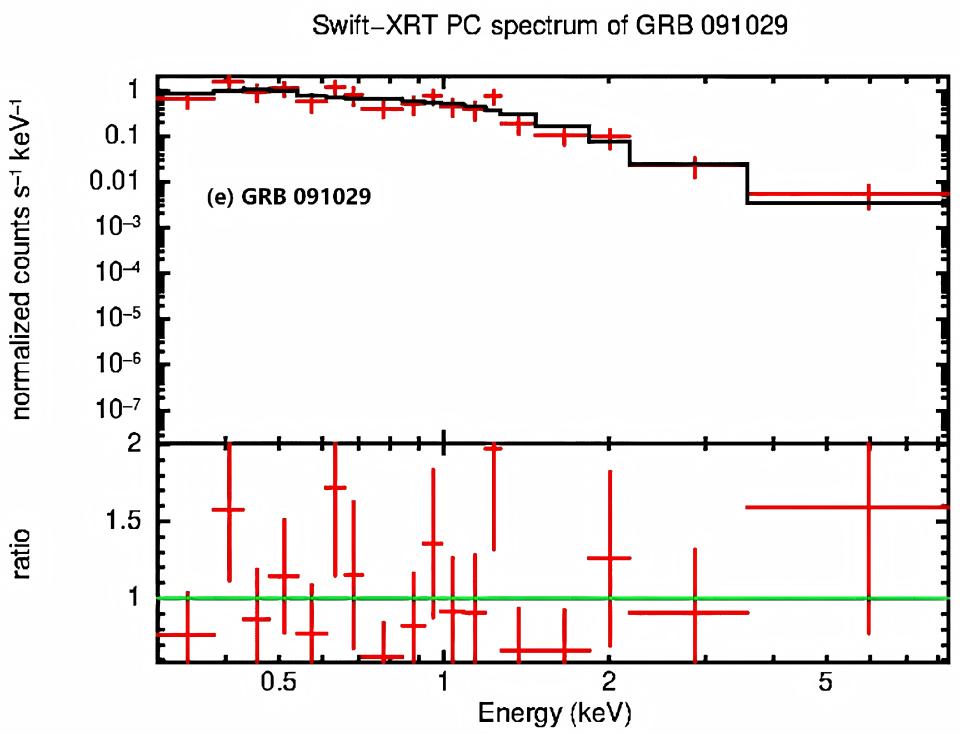}  \includegraphics[width=0.3\linewidth]{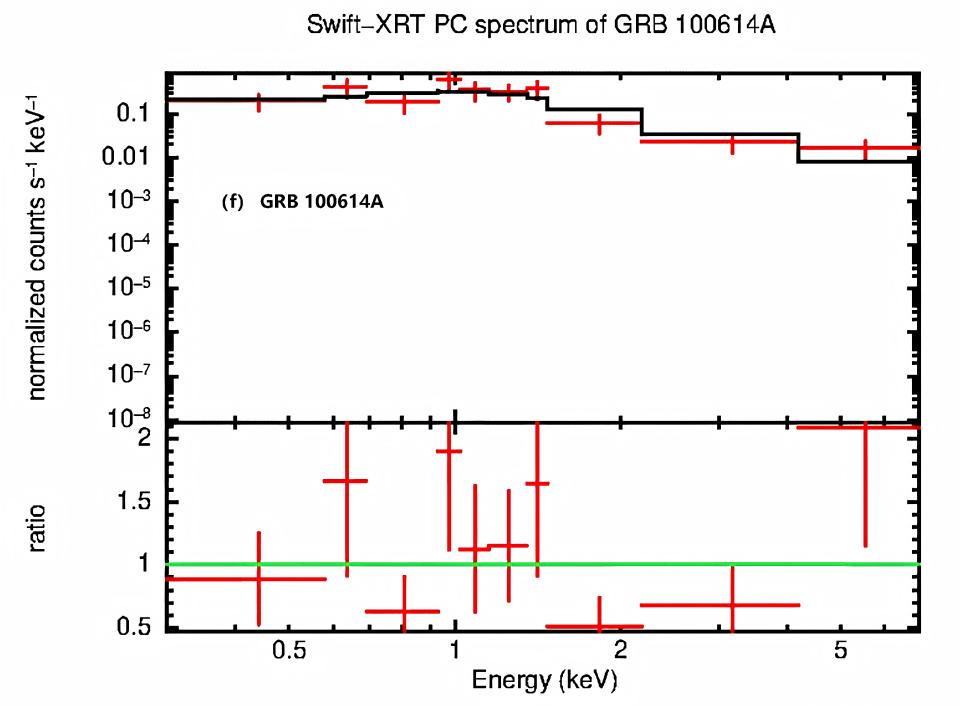}
  \includegraphics[width=0.3\linewidth]{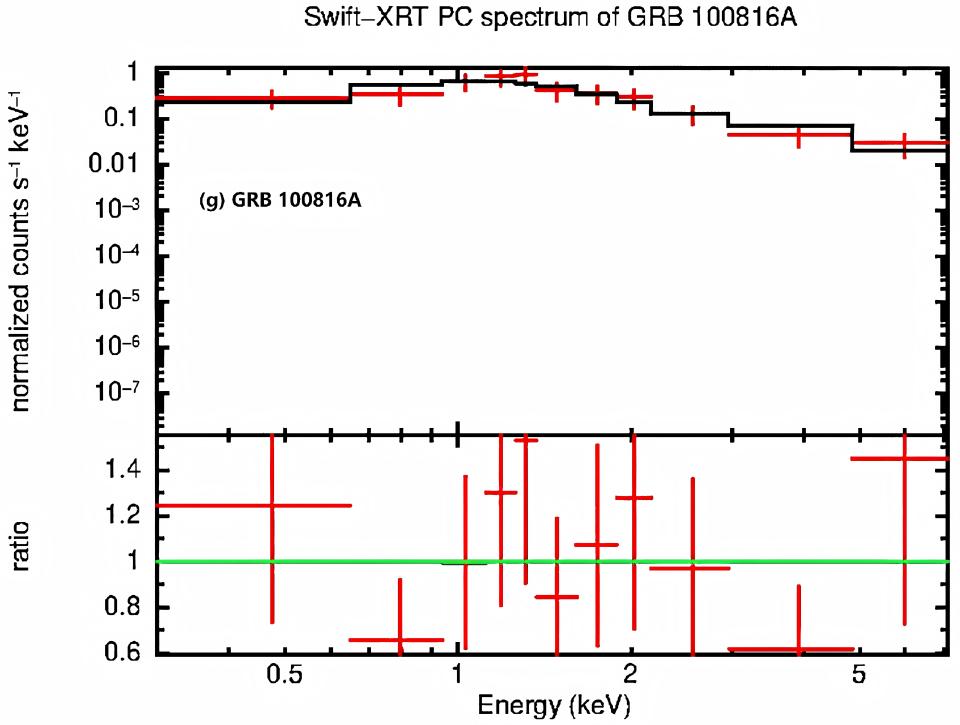}
  \includegraphics[width=0.3\linewidth]{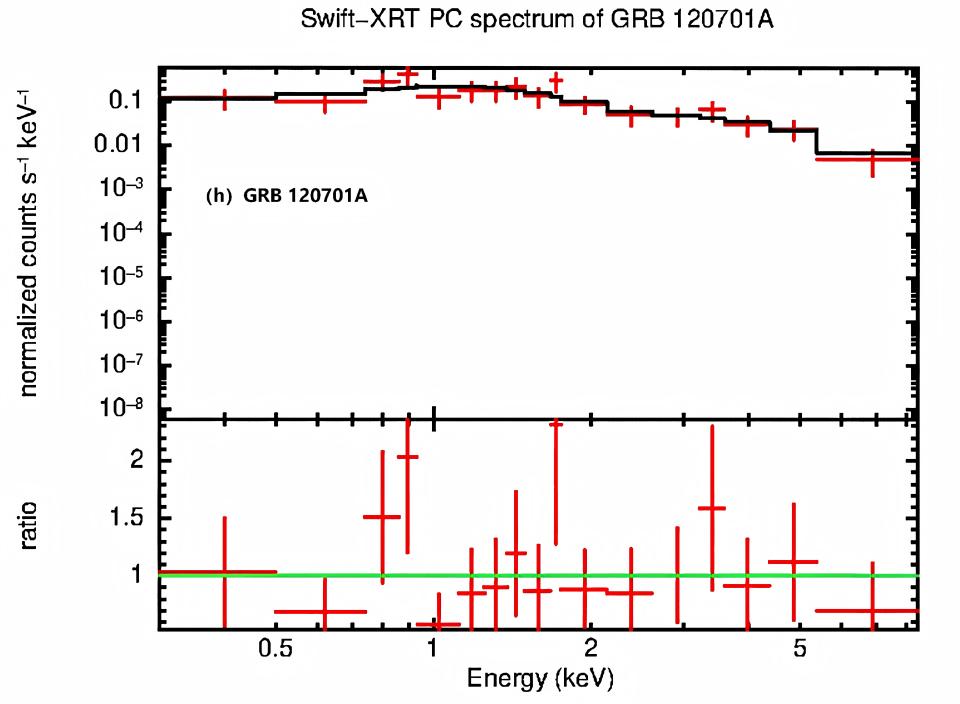}
  \includegraphics[width=0.3\linewidth]{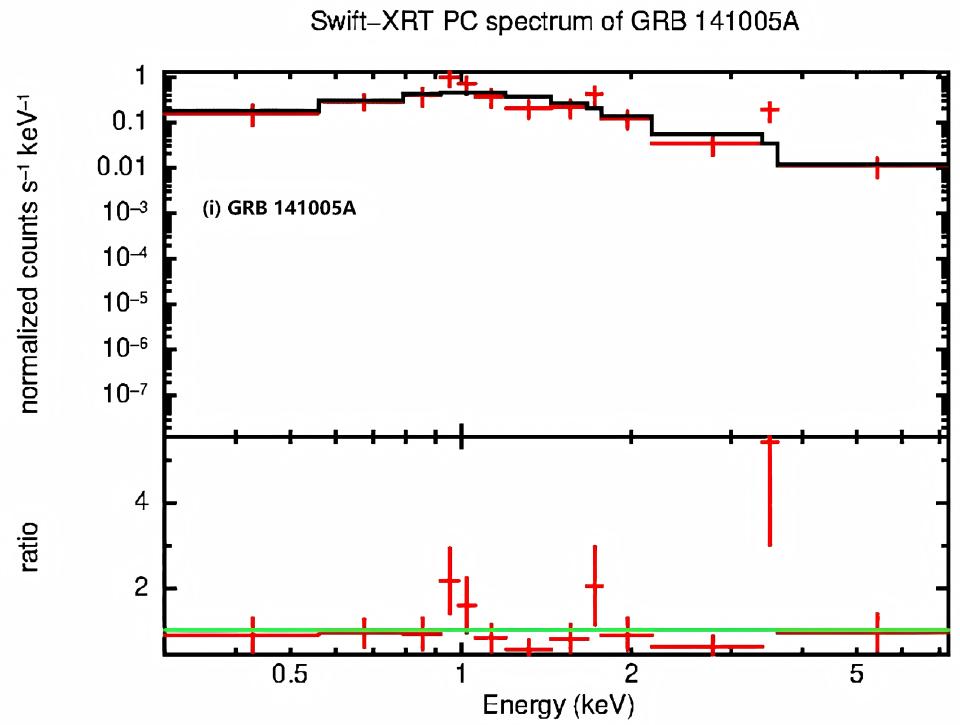}
  \caption{Spectral fit and residual plots for Example 2: constrained $N_{\rm H}$ with relatively large uncertainties. Panels (a)--(j) correspond to GRB 050908, GRB 060512, GRB 070220, GRB 070714A, GRB 091029, GRB 100614A, GRB 100816A, GRB 120701A, and GRB 141005A, respectively. The best-fit parameters are listed in Table A1.}
  \label{fig:A2}
\end{figure*}

\begin{figure*}[!htb]
  \centering
  \includegraphics[width=0.3\linewidth]{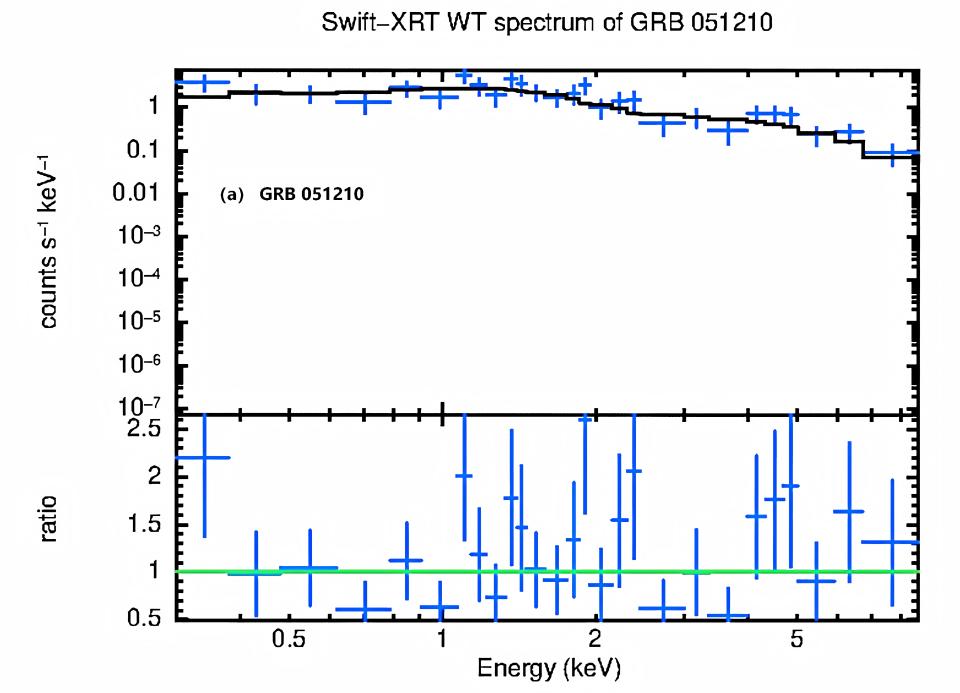}
  \includegraphics[width=0.3\linewidth]{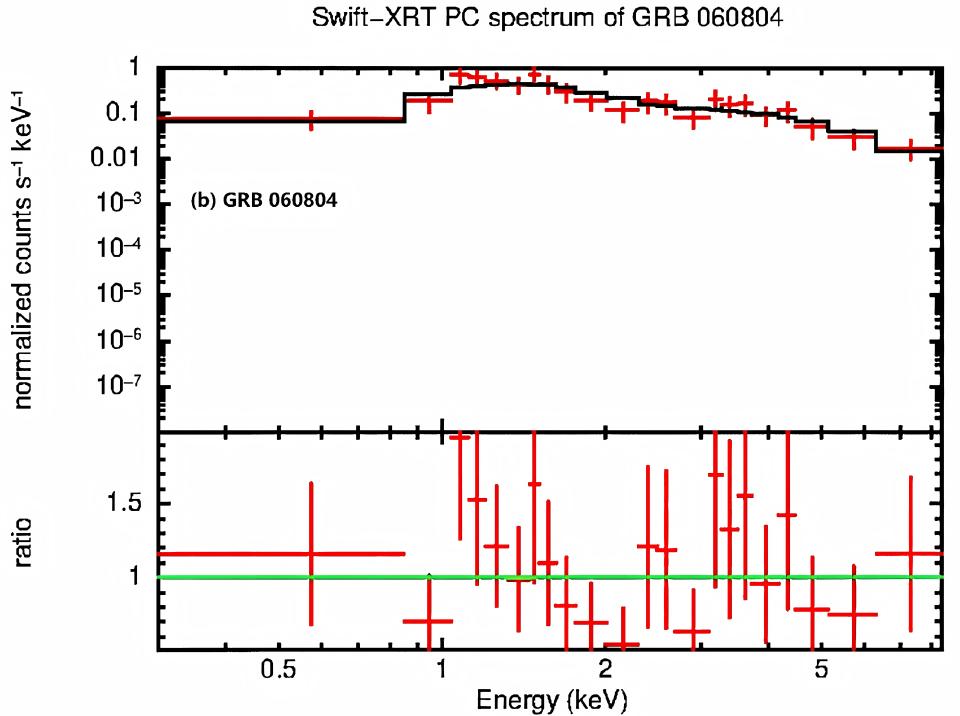}
  \includegraphics[width=0.3\linewidth]{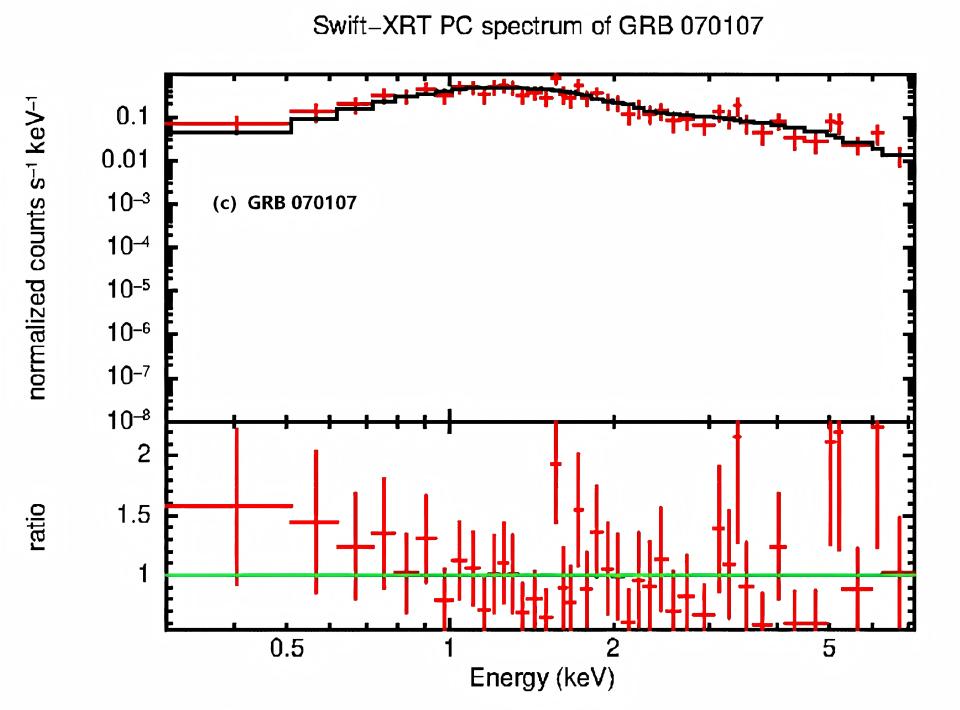}
  \includegraphics[width=0.3\linewidth]{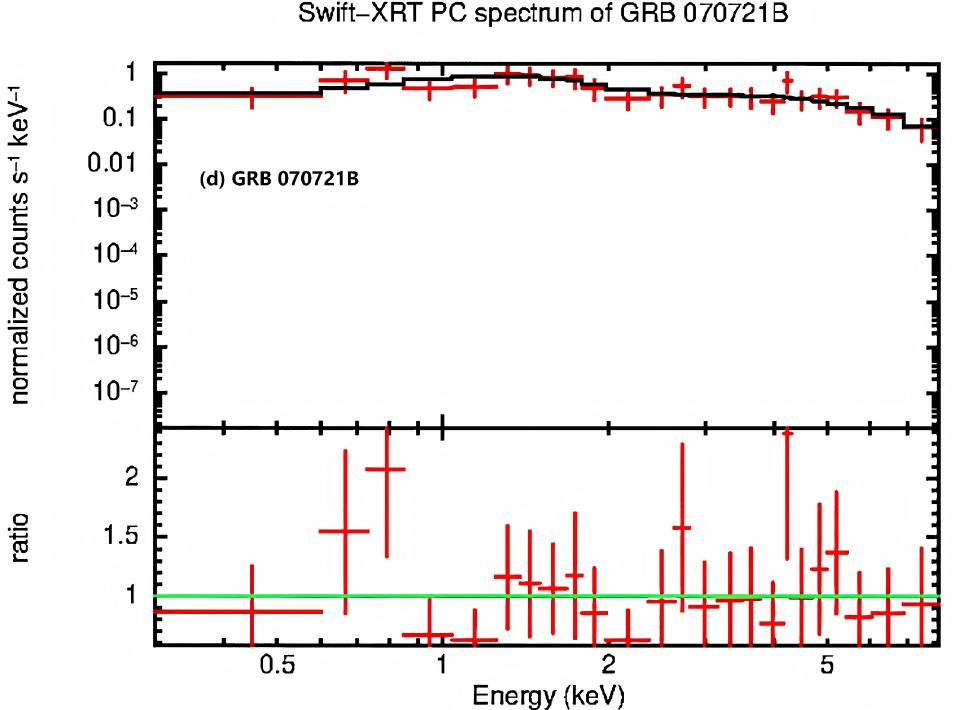}
  \includegraphics[width=0.3\linewidth]{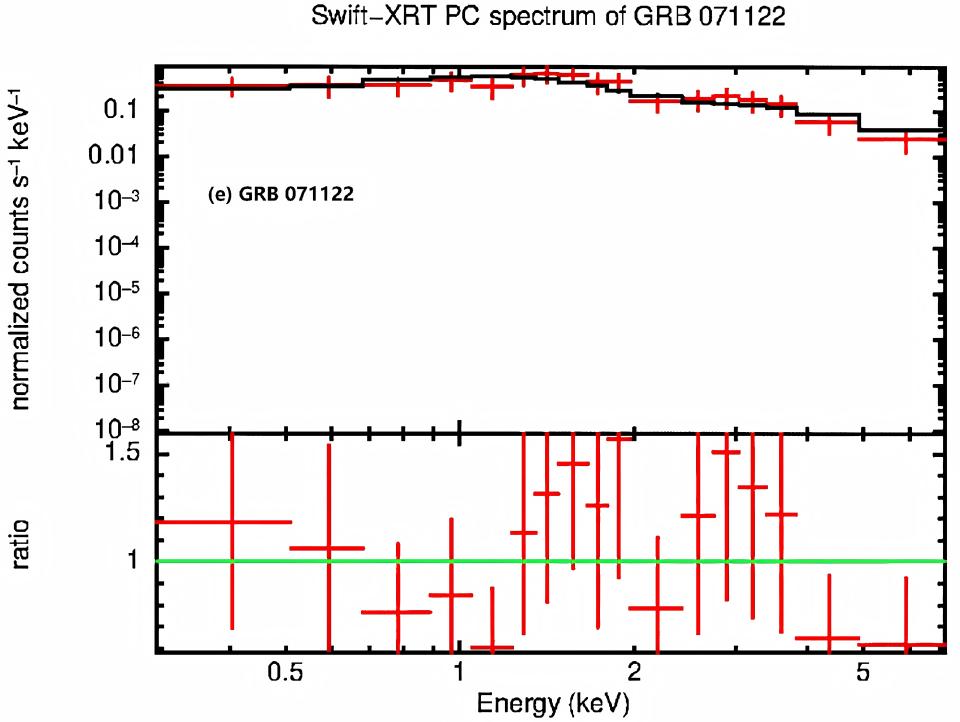}
  \includegraphics[width=0.3\linewidth]{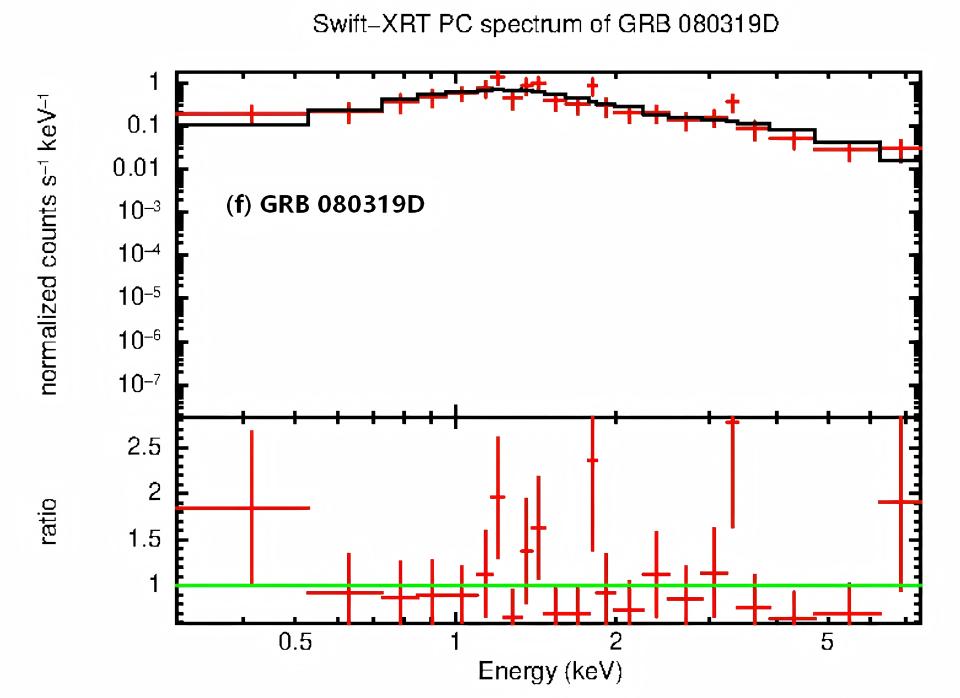}
  \includegraphics[width=0.3\linewidth]{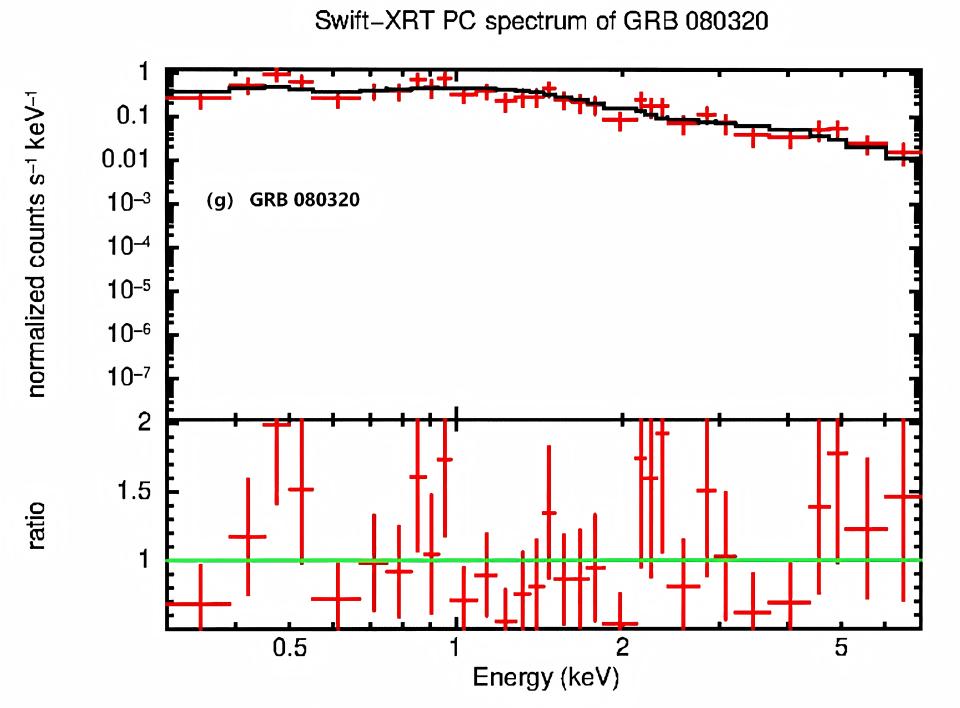}
  \includegraphics[width=0.3\linewidth]{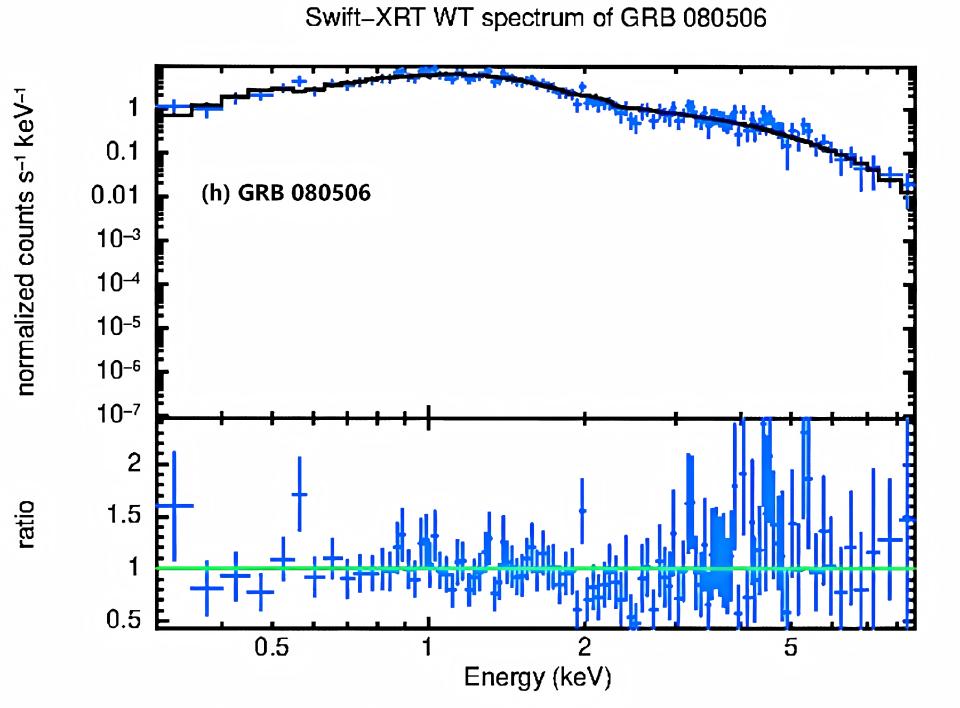}
  \includegraphics[width=0.3\linewidth]{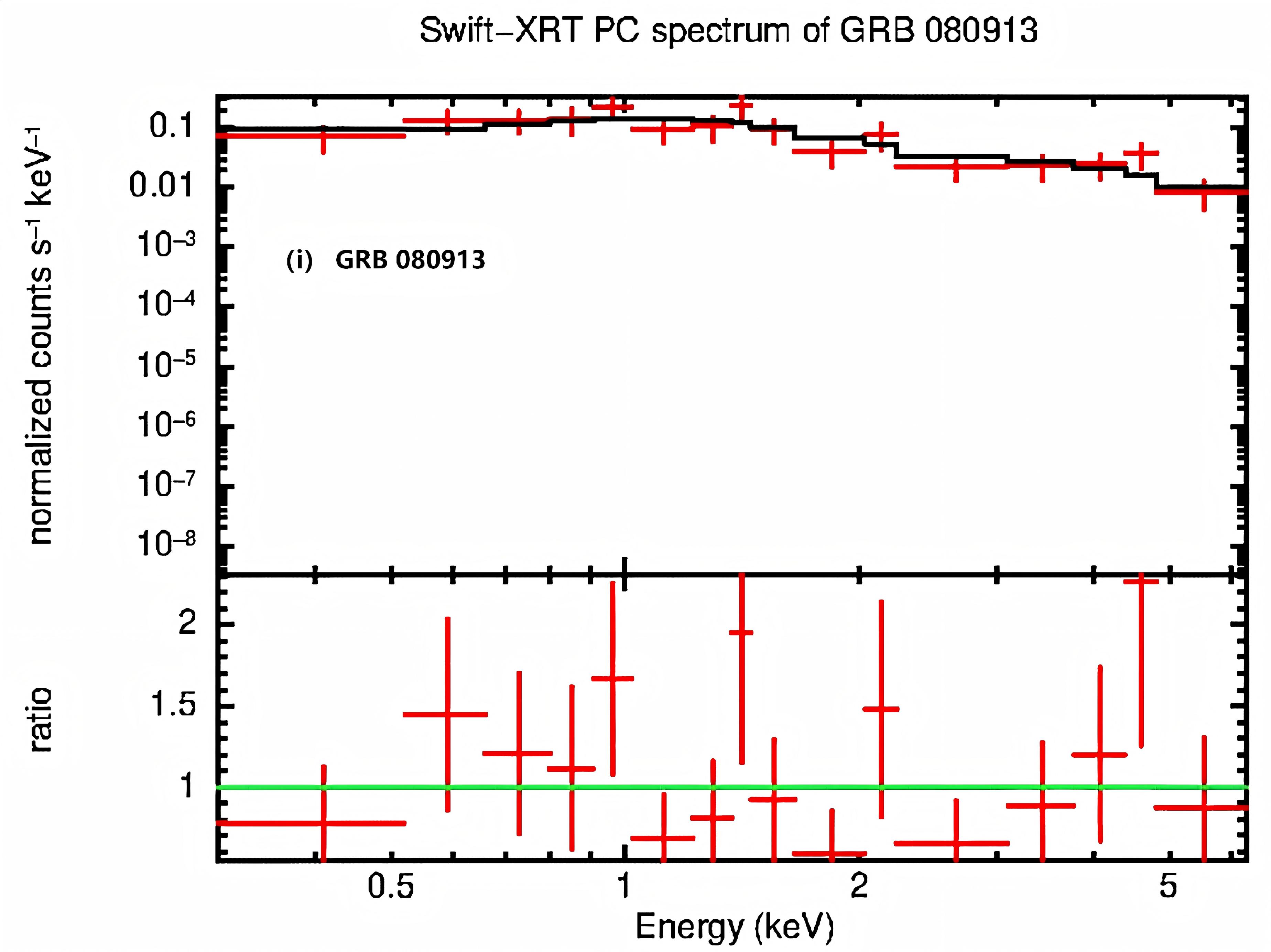}
  \includegraphics[width=0.3\linewidth]{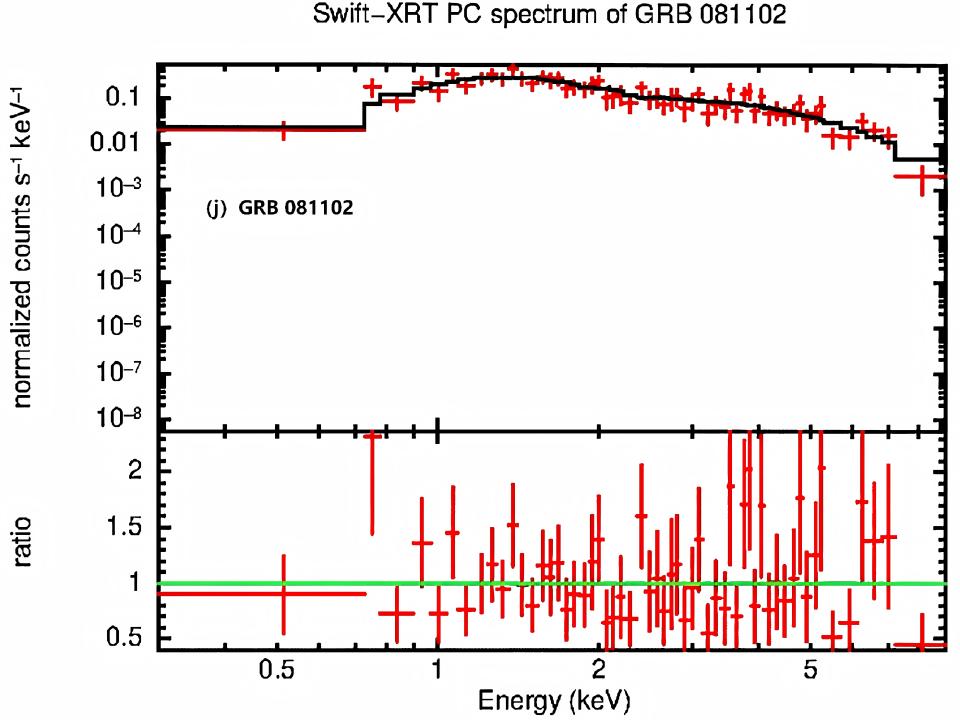}
  \includegraphics[width=0.3\linewidth]{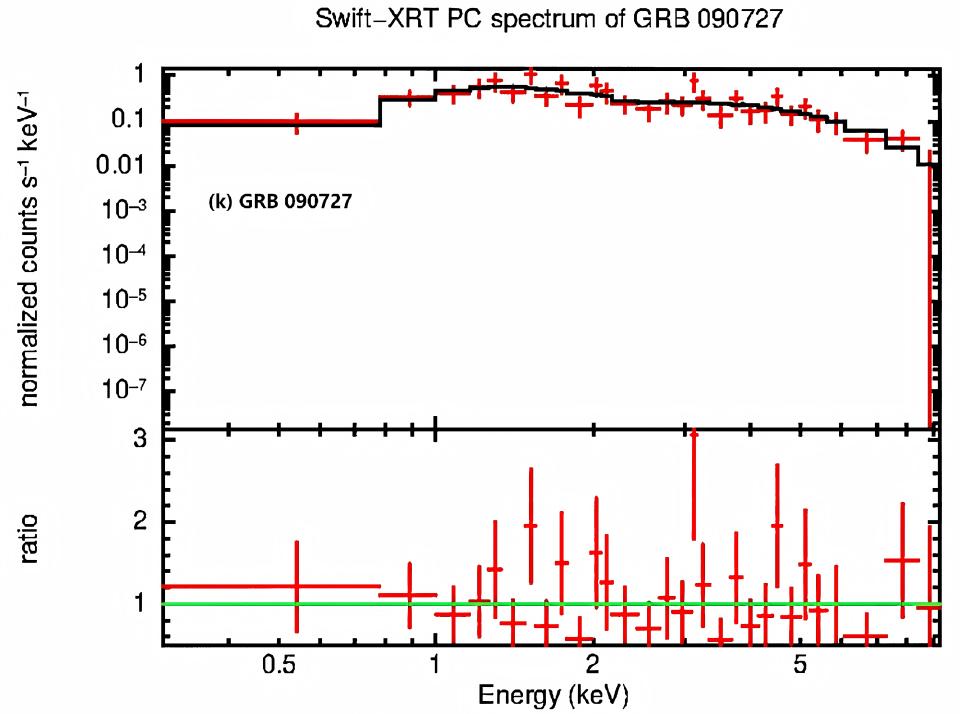}
  \includegraphics[width=0.3\linewidth]{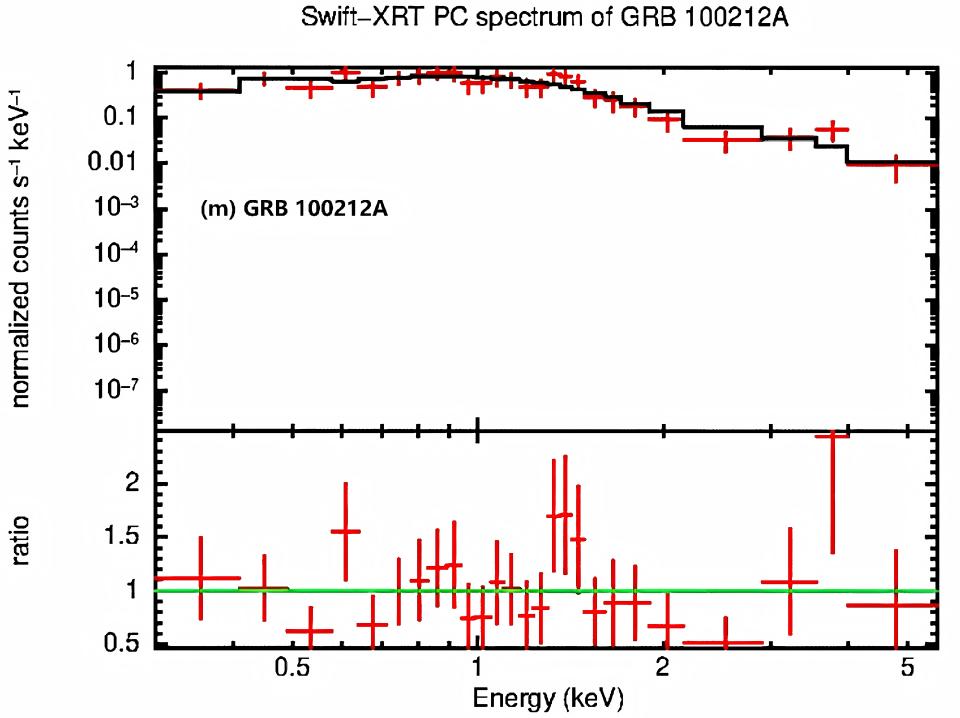}
  \includegraphics[width=0.3\linewidth]{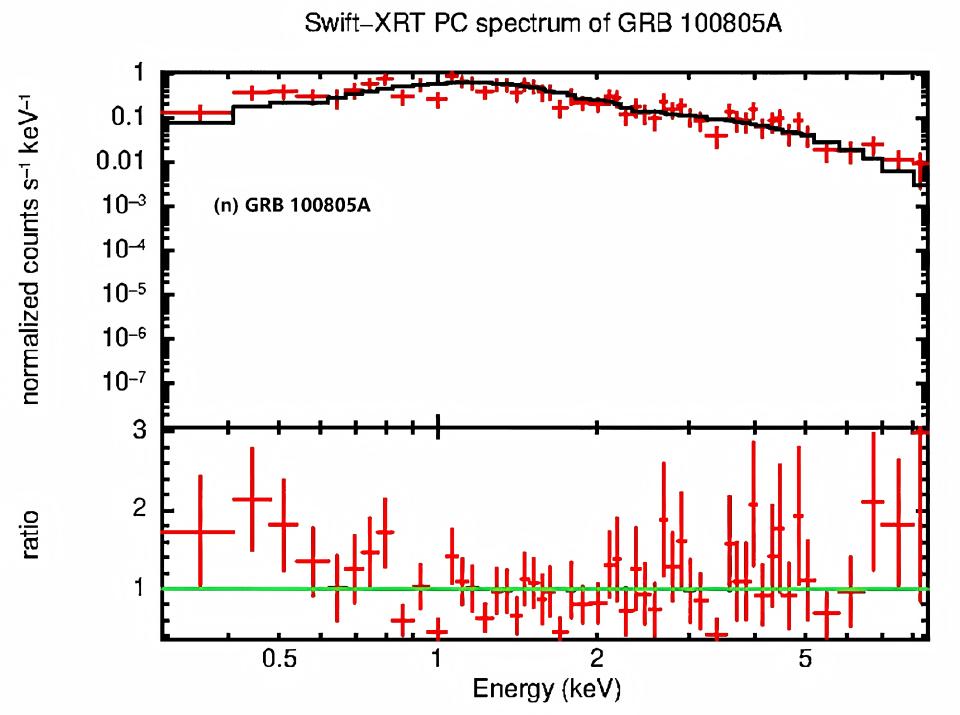}
  \includegraphics[width=0.3\linewidth]{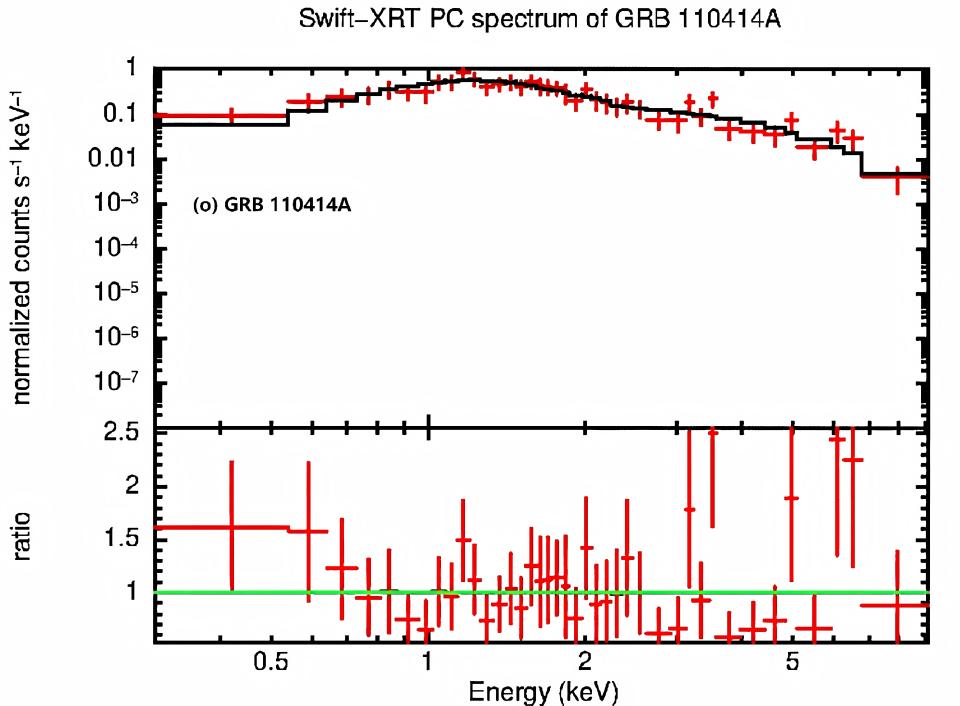}  \includegraphics[width=0.3\linewidth]{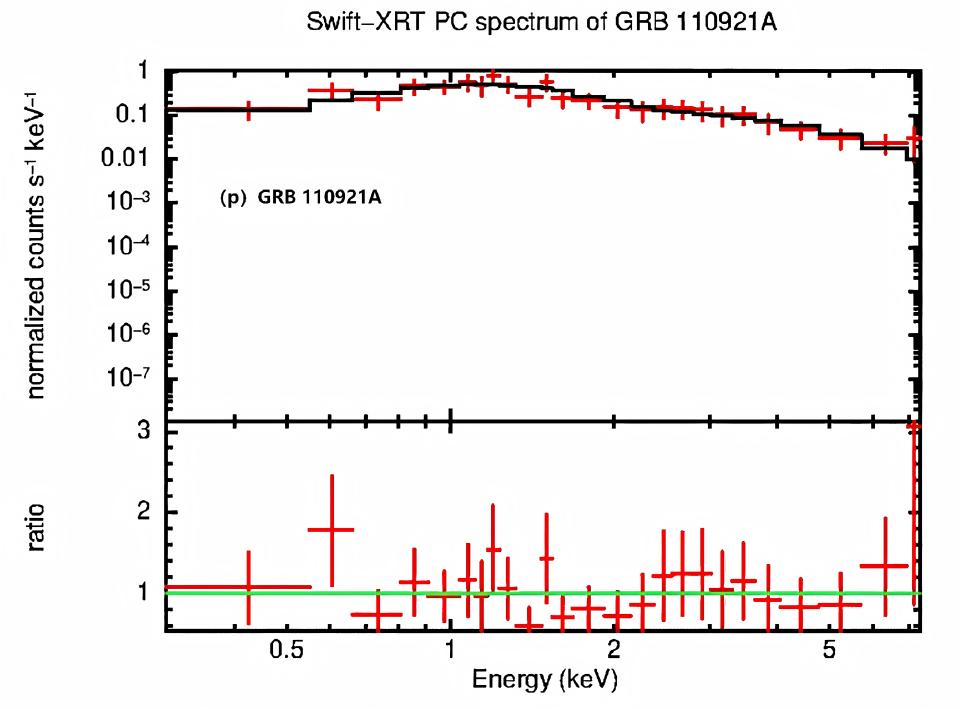}
  \caption{Spectral fit and residual plots for Example 3: only upper limits on $N_{\rm H}$. Panels (a)--(p) correspond to GRB 051210, GRB 060804, GRB 070107, GRB 070721B, GRB 071122, GRB 080319D, GRB 080320, GRB 080506, GRB 080913, GRB 081102, GRB 090727, GRB 100212A, GRB 100805A, GRB 110414A, and GRB 110921A, respectively. The best-fit parameters are listed in Table A1.}
  \label{fig:A3}
\end{figure*}

\begin{deluxetable*}{l c c c c c}
\tablecaption{Best-fit spectral parameters for the representative examples shown in Figures A1--A3.\label{Table:A1}}
\tablewidth{\linewidth}
\tabletypesize{\small}
\setlength{\tabcolsep}{5pt}
\tablehead{ GRB & Interval & Photon Index & $N_{\rm H}$ (\(10^{20}\) $\mathrm{cm}^{-2}$) & $C$-stat/dof & Example }
\startdata
GRB 051117A & 113--209 & 1.83 $\pm$ 0.05 & $14.5^{+1.8}_{-1.7}$ & 647.71(666) & 1 \\
GRB 060111A & 74--443  & 2.20 $\pm$ 0.04 & $22.4^{+1.2}_{-1.2}$ & 794.83(725) & 1 \\
GRB 060604  & 128--187 & 2.10 $\pm$ 0.06 & $355.0^{+43.0}_{-41.0}$ & 496.23(552) &  1 \\
GRB 071031  & 191--235 & 2.13 $\pm$ 0.08 & $190.0^{+37.0}_{-36.0}$ & 448.96(441) &  1 \\
GRB 090407  & 115--192 & 1.76 $\pm$ 0.08 & $42.5^{+5.1}_{-4.8}$ &	546.77(622) &  1 \\
GRB 100802A  & 280--779	& 1.69 $\pm$ 0.03 & $9.9^{+1.4}_{-1.3}$ & 783.10(815) &  1 \\
GRB 100902A  & 350--599 & 2.31 $\pm$ 0.05 & $37.9^{+2.7}_{-2.6}$ & 702.32(691) &  1 \\
GRB 120308A  & 112--250 & 2.97 $\pm$ 0.08 & $13.6^{+1.6}_{-1.6}$ & 429.16(432) &  1 \\
GRB 121123A  & 190--788 & 1.42 $\pm$ 0.02 & $4.3^{+0.8}_{-0.8}$ & 942.93(867) &  1 \\
\cline{1-6}
GRB 050908  & 368--484 &  4.76 $\pm$ 0.78 & $320.0^{+230.0}_{-200.0}$ &  68.25(66) & 2 \\
GRB 060512  & 177--257 &	4.43 $\pm$ 0.65 & $35.0^{+17.0}_{-15.0}$  & 79.77(92) & 2 \\
GRB 070220  & 534--563 & 2.16 $\pm$ 0.79 & $54.0^{+84.0}_{-54.0}$ & 31.09(45) & 2 \\
GRB 070714A & 520--1299 & 1.57 $\pm$ 0.52 & $48.0^{+46.0}_{-34.0}$ & 57.28(74) & 2 \\
GRB 091029  & 278--429 & 2.97 $\pm$ 0.45 & $99.0^{+131.0}_{-99.0}$ & 59.90(86) & 2 \\
GRB 100614A & 986--1093 & 2.18 $\pm$ 0.63 & $970.0^{+1740.0}_{-970.0}$ & 34.83(47) & 2 \\
GRB 100816A & 269--329 & 1.86 $\pm$ 0.50 &	$49.0^{+72.0}_{-49.0}$ & 41.41(51)& 2 \\
GRB 120701A & 250--450 & 1.46 $\pm$ 0.39 & $4.3^{+15.6}_{-4.3}$ & 59.26(80)& 2 \\
GRB 141005A & 389--509 & 2.39 $\pm$ 0.55 & $8.6^{+18.7}_{-8.6}$ & 81.46(62) & 2 \\
\cline{1-6}
GRB 051210  & 123--146  &  1.31 $\pm$ 0.22  & $<7.41$  & 98.11(131) & 3 \\
GRB 060804  & 424--563  & 1.62 $\pm$ 0.31   & $<28.0$  & 78.14(107) & 3 \\
GRB 070107  & 1290--1465 & 2.01 $\pm$ 0.23 & $<7.9$   & 96.27(124)  & 3 \\
GRB 070721B & 256--308 & 0.73 $\pm$ 0.23 & $<431.0$ & 90.14(116) & 3 \\
GRB 071122 & 400--480 & 1.30 $\pm$ 0.28 & $<45.1$ &	66.83(81) & 3 \\
GRB 080319D & 423--533 & 1.74 $\pm$ 0.30 & $<16.1$ &	87.60(111)& 3 \\
GRB 080320 & 700--954 & 1.61 $\pm$ 0.19 & $<4.4$ &	137.43(156) &  3 \\
GRB 080506 & 474--679 & 2.19 $\pm$ 0.09 & $<3.2$ & 384.59(454) &  3 \\
GRB 080913 & 281--599 & 1.39 $\pm$ 0.32 & $<1410.0$ &	66.55(76) &  3 \\
GRB 081102 & 950--1700 & 1.53 $\pm$ 0.17 & $<17.1$ & 	269.00(280)	&  3 \\
GRB 090727 & 150--257 & 0.89 $\pm$ 0.26 & $<29.1$  & 136.84(143) &  3 \\
GRB 100212A & 600--798 & 2.97 $\pm$ 0.32 &  $<9.0$ & 106.18(125)&   3 \\
GRB 100805A & 500--917 & 1.84 $\pm$ 0.12 & $<1.8$ &  	238.56(259) &  3 \\
GRB 110414A & 280--600 & 1.92 $\pm$ 0.17 & $<7.5$ & 	153.90(195)	 &  3 \\
GRB 110921A & 475--640 & 1.67 $\pm$ 0.23 &  $<9.3$	& 	106.93(119) &  3 \\
\enddata
\tablecomments {Example 1: well-constrained $N_{\rm H}$ with small uncertainties; Example 2: constrained $N_{\rm H}$ with relatively large uncertainties; Example 3: only upper limits on $N_{\rm H}$. Interval (Int.) refers to the time interval, in seconds. The symbol $<$ indicates an upper limit on $N_{\rm H}$.}
\end{deluxetable*}

\acknowledgments
We thank the anonymous referee for constructive comments that greatly improved this manuscript. This work made use of data supplied by the UK Swift Science Data Centre at the University of Leicester. This work is supported by the Natural Science Foundation of China 12393813, the National Key R\&D Program of China (2023YFE0101200), CSST grant CMS-CSST-2025-A07, and the Yunnan Revitalization Talent Support Program (YunLing Scholar Project). This work is supported by the Special Basic Cooperative Research Programs of Yunnan Provincial Undergraduate Universities' Association (grant No.~202301BA070001-071).

\clearpage
\setcounter{table}{0}
\renewcommand{\thetable}{\arabic{table}}
%\section*{Long Table (belongs to main text)}
\setlength{\tabcolsep}{0.2pt}
%\begin{center}
\flushleft
\begin{longrotatetable}
\begin{deluxetable*}{llllll llllll  lllll  lllll}
\tablecaption{Spectral Fitting Parameters of the X-ray Flare, Preflare, and Postflare\label{tab1}}
\tablewidth{0pt}
\tabletypesize{\tiny}
\tablehead{
   &   &   &    &   &   & \multicolumn{6}{c}{Flare}& \multicolumn{5}{c}{Preflare} &\multicolumn{5}{c}{Postflare} \\
    \cmidrule(lr){7-12}
    \cmidrule(lr){13-17}
    \cmidrule(lr){18-22}
\colhead{GRB} & \colhead{Type} & \colhead{$z$} & \colhead{$T_{90}$} & \colhead{$\alpha_{\rm PL}$} & \colhead{$\alpha_{\rm CPL}$} & \colhead{Obs.} &\colhead{No.} & \colhead{Int.} & \colhead{$\Gamma$} & \colhead{$N_{\rm H}$} & \colhead{$C$/dof}  &\colhead{Obs.} & \colhead{Int.} & \colhead{$\Gamma$}  &  \colhead{$N_{\rm H}$} & \colhead{$C$/dof}  &\colhead{Obs.}  & \colhead{Int.} & \colhead{$\Gamma$} &  \colhead{$N_{\rm H}$} & \colhead{$C/{\rm dof}$}   }
\startdata
050406 &  XRF & 2.7  &   5.8$\pm$1.4  & 2.40$\pm$0.34 &  -0.24$\pm$2.38 &  WT &  1 & 181-245 & $2.50\pm0.42$  & $5.7^{+7.5}_{-5.7}$ &  117.05(123) &  \nodata & \nodata  & \nodata  & \nodata  & \nodata &   \nodata & \nodata &\nodata  & \nodata & \nodata \\
\cline{1-22}
050502B  &  XRR &   5.2   &  17.7$\pm$1.9 &1.58$\pm$0.13 &  1.15$\pm$0.55 & WT  & tot.   &    501-1296   &  $2.22\pm0.03$  &   $5.8^{+0.7}_{-0.7}$   &  727.90(709) &  \nodata & \nodata & \nodata & \nodata & \nodata &  PC &  1700-19673  &  1.92$\pm$0.19 &  $<3.6$ &  103.99(142) \\
 050502B  &  XRR &   5.2   &  17.7$\pm$1.9 &1.58$\pm$0.13 &  1.15$\pm$0.55 &  WT  &  1  &  501-654  &  $2.19\pm0.07$  &  $11.6 ^{+1.9}_{-1.8}$   &    485.50(549) &  \nodata & \nodata & \nodata & \nodata & \nodata &  \nodata & \nodata & \nodata & \nodata & \nodata \\
050502B  &  XRR &   5.2   &  17.7$\pm$1.9 &1.58$\pm$0.13 &  1.15$\pm$0.55 &  WT  &  2  &   660-700  &   $2.29\pm0.09$  & $12.4 ^{+2.4}_{-2.2}$   &   434.34(470) & \nodata & \nodata & \nodata & \nodata & \nodata &  \nodata & \nodata & \nodata & \nodata & \nodata \\
050502B  &  XRR &   5.2   &  17.7$\pm$1.9 &1.58$\pm$0.13 &  1.15$\pm$0.55 &  WT  &  3   &  724-827  &  $ 2.30\pm0.05$  & $7.9 ^{+1.2}_{-1.2}$   &    560.56(591) &  \nodata & \nodata & \nodata & \nodata & \nodata &  PC &  1700-19673  &  1.92$\pm$0.19 &  $<3.6$ &  103.99(142) \\
\cline{1-22}
050607  & XRF & \nodata  &  48.0$\pm$22.6  & 1.92$\pm$0.18 &  1.80$\pm$0.66 & PC   &  1   &  280-379  &  $ 2.32\pm0.34$ &  $13.0^{+13.0}_{-11.0}$   &   148.87(161) &  \nodata & \nodata & \nodata & \nodata & \nodata & PC & 400-2500 & 1.96$\pm$0.16 & $<3.5$ &  149.56(169) \\
\cline{1-22}
050712   &  C-GRB &  \nodata  & 51.4$\pm$7.9  & 1.50$\pm$0.18 & 1.45$\pm$0.94 &  WT   & tot.  & 212-275  &   $ 2.14\pm0.19$  &  $10.7^{+7.1}_{-6.5}$   &    302.12(327) &  WT & 171-210 &  1.99$\pm$0.32 & $ 0.38^{+1.44}_{-0.38}$ & 128.53(179) & \nodata & \nodata & \nodata & \nodata & \nodata \\
\enddata
\tablecomments{$\alpha_{\rm PL}$ is the power-law photon index in the simple power-law model, and data are derived from \citet{Lien2016}. $\alpha_{\rm CPL}$ is the power-law photon index in the cutoff power-law model, and data are derived from \citet{Lien2016}. {Obs.} denotes observation mode (WT/PC). {No.} denotes the flare number. The label ``tot.'' denotes a combined fit for multiple consecutive flares (total-flare). An asterisk ($*$) marks flares with overlapping WT and PC data that are spectrally fitted separately. Interval (Int.) refers to the time interval in seconds. $N_{\rm H}$ denotes the hydrogen column density, in units of $10^{20}\,\mathrm{cm}^{-2}$. For $N_{\rm H}$, ``<'' denotes an upper limit. $\Gamma$ denotes the photon index. $C/{\rm dof}$ denotes $C$-stat/dof. }
\end{deluxetable*}
\end{longrotatetable}
%\end{center}

\clearpage
\begin{deluxetable*}{clllllllllll}
\tablecaption{Spectral Fitting Parameters Derived in Preflare, Flare, and Postflare Epochs for Bright and Faint X-ray Flares\label{tab2}}
\tablewidth{\linewidth}
\tabletypesize{\small}
\setlength{\tabcolsep}{3pt}
\tablehead{
   &  & \multicolumn{4}{c}{Flare}& \multicolumn{3}{c}{Preflare} &\multicolumn{3}{c}{Postflare} \\
    \cmidrule(lr){3-6}
    \cmidrule(lr){7-9}
    \cmidrule(lr){10-12}
\colhead{Class}& \colhead{GRB}& \colhead{$S_\mathrm{F}$} & \colhead{Int.} & \colhead{$\Gamma$} & \colhead{$N_{\rm H}$} &  \colhead{Int.} & \colhead{$\Gamma$}  &  \colhead{$N_{\rm H}$} &  \colhead{Int.} & \colhead{$\Gamma$} &  \colhead{$N_{\rm H}$}   }
\startdata
\multirow{4}{*}{Bright flare}
   & 060904B	& 200.0$\pm$12.0 & 146--224 & 1.82$\pm$0.06 & $95.0^{+15.0}_{-14.0}$ &  75--139 &	1.57$\pm$0.24 &	$15.0^{+41.0}_{-15.0}$ & 230--315 & 3.25$\pm$0.10 & $67.3^{+8.7}_{-8.4}$  \\
  & 070704 & 100.0$\pm$7.9 & 280--500 & 1.76$\pm$0.07 & $88.0^{+11.0}_{-10.0}$  & 159--270 & 1.98$\pm$0.28 & $100.0^{+43.0}_{-37.0}$ & 550--1078 & 1.47$\pm$0.35 & $47.0^{+53.0}_{-43.0}$  \\
  & 100902A	& 274.1$\pm$8.5 & 350--599 & 2.31$\pm$0.05 & $37.9^{+2.7}_{-2.6}$ & 320--350 & 2.07$\pm$0.25 & $19.0^{+12.0}_{-11.0}$ & 623--2975 & 2.25$\pm$0.24 & $48.0^{+17.0}_{-15.0}$ \\
  & 121123A & 134.3$\pm$2.4 & 190--650 & 1.42$\pm$0.02 & $4.3^{+0.8}_{-0.8}$ & 130--190 & 0.52$\pm$0.12 & $9.5^{+9.0}_{-7.5}$ & 800--1484 & 1.83$\pm$0.14 & $<2.6$  \\
\cline{1-12}
\multirow{6}{*}{Faint flare}
  & 060512 & 4.5$\pm$1.1 & 177--257 & 4.43$\pm$0.65 & $35.0^{+17.0}_{-15.0}$ & 110--155 & 5.13$\pm$0.80 & $29.0^{+16.0}_{-15.0}$ & 260--390 & 4.23$\pm$0.75 & $5.5^{+18.4}_{-5.5}$ \\
  & 091029 & 0.9$\pm$0.3 & 278--429 & 2.97$\pm$0.45 & $99.0^{+131.0}_{-99.0}$ & 85--199 & 2.38$\pm$0.13 & $32.0^{+37.0}_{-32.0}$ 	& 450--1999 & 1.99$\pm$0.19 & $14.0^{+83.0}_{-14.0}$  \\
 & 091104 & 3.0$\pm$0.7 & 191--249 & 2.26$\pm$0.14 & $12.9^{+3.7}_{-3.4}$ & 140--190 & 2.02$\pm$0.14 & $11.7^{+3.9}_{-3.6}$ &	310--800 & 2.31$\pm$0.18 & $<1.7$  \\
  & 110520A & 5.0$\pm$1.0 & 190--360 & 1.57$\pm$0.29 & $37.0^{+19.0}_{-16.0}$ & 105--161 & 1.85$\pm$0.36 & $29.0^{+18.0}_{-14.0}$ & 401--2000 & 1.38$\pm$0.22 & $21.0^{+11.6}_{-9.8}$  \\
\enddata
\tablecomments {$S_\mathrm{F}$ is the fluence of the X-ray flare in units of $10^{-8}\,\mathrm{erg\,cm^{-2}}$. The $S_{\rm F}$ data are taken from \citet{Chincarini2010} and \citet{Yi2016}. Interval (Int.) indicates the time interval adopted for spectral fitting, in units of seconds. $\Gamma$ denotes the photon index. $N_{\rm H}$ denotes the hydrogen column density in units of $10^{20}\,\mathrm{cm^{-2}}$; the symbol $<$ indicates an upper limit on $N_{\rm H}$. Samples are classified as bright flares ($S_\mathrm{F}\gtrsim 100\times10^{-8}\,\mathrm{erg\,cm^{-2}}$) and faint flares ($S_\mathrm{F}<10\times10^{-8}\,\mathrm{erg\,cm^{-2}}$).}
\end{deluxetable*}

\end{document}